\documentclass[twocolumn,showpacs,prd,aps,rsi,floatfix,superscriptaddress,showkeys,citeautoscript,longbibliography,nobalancelastpage]{revtex4-2}
\usepackage{graphicx,wrapfig,epstopdf,booktabs,soul,array,lmodern}
\usepackage[left=1.60cm,right=1.60cm,top=1.90cm,bottom=1.90cm]{geometry}
\usepackage{epigraph}
\usepackage{footmisc}
\usepackage{dcolumn}
\usepackage{dutchcal}
\usepackage{bm,color}
\usepackage{amsmath,amssymb,dsfont,amstext,amsfonts}
\usepackage{extarrows}
\usepackage{lettrine}
\usepackage{times,mathptmx}
\usepackage[english]{babel}
\usepackage{mathtools}
\usepackage{bbold}
\usepackage{wasysym}
\usepackage{mathtools}
\usepackage{bbold} 
\usepackage{braket}
\usepackage{dsfont}
\usepackage{bigints}
\usepackage{soul}
\usepackage{amssymb,amsmath,amsthm,amssymb,multirow,printlen,layouts}
\usepackage{color}
\definecolor{lightgray}{gray}{0.5}
\usepackage[export]{adjustbox}
\usepackage{footnote}
\usepackage{ucs}
\usepackage{braket}
\usepackage{dsfont}
\usepackage{bigints}
\usepackage{multibib}  
\newcites{Main}{References}         
\newcites{Supp}{Supplementary References}  
\usepackage{soul}
\usepackage{url,hyperref}
\usepackage[font=footnotesize,labelfont=bf]{caption}
\usepackage{charter}
\usepackage[T1]{fontenc}
\usepackage[usenames,dvipsnames,svgnames,table]{xcolor}
\definecolor{CustomBlue}{rgb}{0.2, 0.18, 0.65} 

\hypersetup{
    backref=true,
    pagebackref=true,
    colorlinks=true,
    linkcolor=CustomBlue,
    citecolor=CustomBlue,
    urlcolor=CustomBlue
}

\usepackage{float}
\usepackage{fancyhdr}
\usepackage{orcidlink}
\usepackage{fnpos}
\addto{\captionsenglish}{%
  
}
\usepackage[raggedright]{titlesec}
\usepackage{epstopdf}
\definecolor{cream}{RGB}{222,217,201}

\newcommand\ba{\begin{eqnarray}}
\newcommand\ea{\end{eqnarray}}

\definecolor{Nathanblue}{rgb}{0.,0.24,0.51}
\newcommand{\blue}{\color{Nathanblue}}

\def\XXint#1#2#3{{\setbox0=\hbox{$#1{#2#3}{\int}$}
		\vcenter{\hbox{$#2#3$}}\kern-.5\wd0}}

\usepackage{setspace}

\def\AmS{{\protect\the\textfont2
        A\kern-.1667em\lower.5ex\hbox{M}\kern-.125emS}}

\makeatletter
\def\thepage{1-\@arabic\c@page}
\def\@pnumwidth{2em}
\makeatother

\begin{document}
\title{{\blue Dirac Equation Solution with Generalized tanh-Shaped Hyperbolic Potential:\\ Application to Charmonium and Bottomonium Mass Spectra}}

\author{V. H. Badalov{\,}\orcidlink{0000-0002-5468-1978}}
\email{badalovvatan@yahoo.com}
\affiliation{Institute for Physical Problems, Baku State University, AZ-1148 Baku, Azerbaijan}
\affiliation{Center for Theoretical Physics, Khazar University, AZ-1096 Baku, Azerbaijan}

\author{A. I. Ahmadov{\,}\orcidlink{0000-0003-0662-5549}}
\affiliation{Institute for Physical Problems, Baku State University, AZ-1148 Baku, Azerbaijan}
\affiliation{Department of Theoretical Physics, Baku State University, AZ-1148 Baku, Azerbaijan}

\author{E. A. Dadashov{\,}\orcidlink{0000-0002-1835-5833}}
\affiliation{Department of Chemistry and Physics, Lankaran State University, AZ-4200 Lankaran, Azerbaijan}

\author{S. V. Badalov{\,}\orcidlink{0000-0002-8481-4161}}
\email{sabuhi.badalov@uni-bayreuth.de} 
\affiliation{Department of Physics, University of Bayreuth, 95447 Bayreuth, Germany}
\affiliation{Bavarian Center for Battery Technologies, University of Bayreuth, 95448 Bayreuth, Germany}

\noaffiliation

\date{\today}

\begin{abstract}
In this study, we present an analytical solution of the Dirac equation in a generalized tanh-shape hyperbolic potential, which allows us to unify various well-known quantum potentials under a single theoretical framework. This versatile potential model is used to compute the mass spectra for charmonium and bottomonium, with excellent agreement with experimental measurements and does better than some potential models in predicting the several orbital states. Our results not only validate the GTHP as a powerful tool for describing heavy quarkonium systems but also suggest its broader applicability in exploring quantum systems where similar potentials are effective. This work is a stepping stone for new research into fermionic systems with complex interactions, by jointly providing insights into foundational aspects of quantum mechanics as well as applications in particle physics.
\end{abstract}


\maketitle

\section{Introduction}
\lettrine[findent=2pt]{{\textbf{T}}}{}he Dirac equation (DE) is a cornerstone of quantum mechanics and is responsible for revolutionizing theoretical physics, giving rise to what is in essence the first relativistic quantum field theory, leading to the prediction of novel phenomena such as antimatter{\,}\cite{DE1}. The equation has led to advances in many areas in physics, from quantum electrodynamics{\,}\cite{DE2} to the way particles behave in the relativistic limit{\,}\cite{DE3, DE4}. Responsible in its particular nuclear properties studies{\,}\cite{DE5}, which later extended to be used in potent modern effort in a superheavy element{\,}\cite{DE_5_0}, complex quantum systems{\,}\cite{DE_5_1}, chiral symmetric quantum chromodynamics{\,}\cite{DE_5_2} and antiparticles{\,}\cite{DE_5_3}. In addition, DE continues to be relevant for current research and opens the possibility to study electron vortices{\,}\cite{DE10}, coherent control{\,}\cite{DE11} and new theoretical approaches{\,}\cite{DE13, DE14, DE15, DE16, Amruta2024}.

The DE is at the core of the concepts of spin and pseudo-spin symmetries, which are embedded within the Dirac Hamiltonian{\,}\cite{PS,PS1,PS2,PS3,PS4,PS5,PS6,PS7,PS8,PS9}. They explain a variety of physical phenomena associated with the simplicial background and are definitive tools in the interpretation of nuclear structure and reactions{\,}\cite{PS1,PS2,PS3,PS4,PS5,PS6}. Spin symmetry, manifesting as a spin doublet, leads to degeneracy between states characterized by quantum numbers $(n, l, j=l\pm s)${\,}\cite{PS3,PS4,PS5,PS6}. This symmetry is instrumental in elucidating the spectrum of antinucleons in a nucleus and the subtle spin-orbit splitting observed in hadrons{\,}\cite{PS7}. In contrast, pseudo-spin symmetry, another form of degeneracy, arises between states with quantum numbers $(n, l, j=l+\frac{1}{2})$ and $(n-1,l+2, j=l+\frac{3}{2})${\,}\cite{PS8,PS9}. This symmetry has profound implications for understanding nuclear deformation{\,}\cite{PS10,PS11}, identical bands{\,}\cite{PS12,PS13,PS14}, magnetic moments{\,}\cite{PS15,PS16}, magic number shifts{\,}\cite{PS17,PS18,PS19} and effective shell-model structures in nuclear physics{\,}\cite{PS20}.

Pseudo-spin symmetry was initially explored within the non-relativistic framework and later extended into relativistic mean field theory{\,}\cite{PS7,PS8}. Significant strides have been made in this domain, with comprehensive investigations revealing the intricate details of spin and pseudo-spin symmetries{\,}\cite{PS4,PS5,PS21}. Significant advancements have been made in this area, particularly through the analytical solutions of the DE with physical potentials, which are crucial for uncovering the full spectrum of a quantum system's properties{\,}\cite{DEA1,DEA2,DEA3,DEA4}. In this context, our focus is on solving the DE for the generalized tanh-shaped hyperbolic potential (GTHP){\,}\cite{GTHP1,GTHP2}, described as:
\small
\begin{equation}
V(r) = V_1 + V_2 \tanh(\alpha r) + V_3 \tanh^2(\alpha r), \label{a1}
\end{equation}
\normalsize
where $V_1$, $V_2$, and $V_3$ represent potential well depths, and $ \alpha $ characterizes the interaction potential's properties. The GTHP exhibits a minimum value $V_{\text{min}}=V(r_{\text{e}})=V_1-\frac{{V_2}^2}{4V_3}$ at $r_{\text{e}}=\frac{1}{\alpha}\tanh^{-1}\left(-\frac{V_2}{2V_3}\right)$, subject to $\left|V_2\right|<2V_3$ and $V_3>0${\,}\cite{GTHP1,GTHP2}.  This potential encompasses a wide range of physical scenarios, including various standard and generalized potential models, including the standard and generalized Woods-Saxon{\,}\cite{PS,WS1,WS2}, Rosen-Morse{\,}\cite{RM1}, Manning-Rosen type{\,}\cite{MR}, standard and generalized Morse (improved Rosen-Morse){\,}\cite{RM2,RM3}, Schi\"{o}berg{\,}\cite{SCH1,SCH2}, four-parametric exponential type{\,}\cite{FP1,FP2}, Williams-Poulios{\,}\cite{WP1,WP2}, and the sum of the linear and harmonic oscillator potentials in some special cases{\,}\cite{GTHP1,GTHP2}. Notably, considering $V_1\neq 0$ in GTHP facilitates the modeling of these particular cases\cite{GTHP1,GTHP2}. These potentials are particularly relevant in modeling complex quantum systems, where the GTHP's flexibility allows it to capture a broad spectrum of interactions.

While previous studies have significantly advanced our understanding of so-called spin symmetries in relativistic and non-relativistic fields and clarified the critical spin dynamics, there is still a significant gap in applying these insights to complex potentials that more accurately reflect real quantum phenomena. With this study, we aim to fill this gap by advancing our understanding of the hadronic physics and GTHP. This work investigates the bound-state solutions of the DE from the perspective of the GTHP. Employing the Nikiforov-Uvarov (NU) method, reducing the second-order differential equation to a hypergeometric form, we derive exact expressions for the energy eigenvalues and the corresponding radial wave functions articulated through hypergeometric polynomials for various quantum states. Our analysis reveals the sensitivity of these eigenvalues to variations in the potential parameters, providing deep insights into the quantum dynamics encapsulated by the GTHP model. Furthermore, we extend our theoretical framework to studying heavy quark systems, specifically targeting the charmonium and bottomonium mass spectra. Heavy quarkonia, consisting of a heavy quark and antiquark (such as $b\bar{b}$ and $c\bar{c}$), serve as a critical testing ground for quantum chromodynamics (QCD) and various potential models{\,}\cite{Hadron_0, Hadron_1,Hadron_2,Hadron_3,purohit2022,soni2018,bukor2023,Li2009,Ni2022,Molina2027,Qian2024}. These systems are particularly significant owing to their rich spectroscopy, with many states lying below the threshold of open charm or bottom production{\,}\cite{Hadron_3, purohit2022, soni2018, bukor2023, Li2009, Ni2022, Molina2027, Qian2024, Quark_1, Quark_2, Quark_3, Quark_4, Quark_4_1, Quark_5, Sang2023, workman2022, Godfrey2015, Amruta2024}. Employing the GTHP model to these systems, one is able to extract bound-state masses with quite good agreement with experimental results leading to important knowledge on heavy quark interaction dynamics and properties of quark-antiquark potentials on hadronic scales.

To facilitate a comprehensive understanding of the findings, this paper is structured as follows: Section \ref{sec2} presents the bound-state solution of the radial DE for GTHP using the NU method. In Section{\,}\ref{sec3}, we discuss the results for energy levels in specific cases, and the mass spectrum analyses for the $b\bar{b}$ and $c\bar{c}$ systems. Concluding remarks and implications of our study are outlined in Section \ref{sec4}.

\section{Solutions to the Radial Dirac Equation}\label{sec2}
\noindent
The DE serves as a fundamental framework for describing the quantum behavior of fermions in the presence of scalar and vector potentials{\,}\cite{Greiner}. For a particle interacting with an attractive scalar potential $S(\vec{r})$ and a repulsive vector potential $V(\vec{r})$, the DE is given by:
\scriptsize
\begin{equation}
\left[c\vec{\alpha} \cdot \hat{\vec{p}}+\beta\left(Mc^2+S(\vec{r})\right)+V(\vec{r})\right]\psi(\vec{r})=E\psi(\vec{r}),
\label{eq:DE}
\end{equation}
\normalsize
where $E$ is the relativistic total energy, $\hat{\vec{p}} = -i\hbar \vec{\nabla}$ is the momentum operator, and $\vec{\alpha}$ and $\beta$ are the standard Dirac matrices{\,}\cite{Greiner}. To study the system under the influence of the GTHP, we recast the DE into its radial form, exploiting spherical symmetry. In this context, both potentials $S(r)$ and $V(r)$ are functions of the radial coordinate $r$, which allows us to decompose the Dirac spinor into radial and angular parts. Consequently, the DE simplifies to coupled radial second-order linear differential equations for the upper and lower components of the wave function. This formalism is particularly efficient for investigating energy eigenvalues and wave functions in spherically symmetric potentials. Then, we applied NU method{\,}\cite{Nikiforov} to solve this radial second-order linear differential equations, see the derivation which is discussed detailed in Ref.{\,}[\onlinecite{SM}]. The transition to a radial form is crucial as it highlights the symmetries inherent in DE, especially when analyzing the effects of scalar and vector potentials that vary only with radial distance. By effectively isolating the radial components, this approach facilitates a focused study of the impact of potential variations under global symmetries on fermionic behavior.

For spin and pseudo-spin symmetry cases, the corresponding energy eigenvalues $E_{nk}$ and normalized radial wave functions are derived. The details of these derivations are discussed in Ref.{\,}[\onlinecite{SM}], where specific symmetry conditions simplify the DE, such as $\Delta(r) = C_s$ for spin symmetry and $\Sigma(r) = C_{ps}$ for pseudo-spin symmetry.
\\
\\
\begin{description}
\item[\textit{Spin Symmetry Case}]
\end{description}
In the case of spin symmetry, $\frac{d\Delta(r)}{dr} = 0$ leads to the condition $\Delta(r) = C_s$. This simplifies the DE, and the energy eigenvalues $E_{nk}$ are obtained as follows:
\scriptsize
\begin{equation}
\begin{aligned}
&{\left(Mc^2-C_s\right)}^2-E^2_{nk}+\left(Mc^2+E_{nk}-C_s\right)\left(V_1+V_2+V_3\right) \\
&\quad +\frac{{\hslash}^2c^2k(k+1)}{r^2_e}\left(A_0+A_1+A_2\right) \\
&= {\alpha}^2{\hslash}^2c^2\left(\sqrt{\frac{1}{4}+\frac{\left(Mc^2+E_{nk}-C_s\right)V_3}{{\alpha}^2{\hslash}^2c^2}+\frac{k(k+1)}{{\alpha}^2r^2_e}A_2}-n-\frac{1}{2} \right. \\
&\quad \left. +\frac{\frac{\left(Mc^2+E_{nk}-C_s\right)V_2}{2{\alpha }^2{\hslash}^2c^2}+\frac{k(k+1)}{2{\alpha}^2r^2_e}A_1}{\sqrt{\frac{1}{4}+\frac{\left(Mc^2+E_{nk}-C_s\right)V_3}{{\alpha}^2{\hslash}^2c^2}+\frac{k(k+1)}{{\alpha}^2r^2_e}A_2}-n-\frac{1}{2}}\right)^2 .
\end{aligned}\label{a35}
\end{equation}
\normalsize
The information about all parameters in Eq.{\,}\eqref{a35} and the full derivation process to obtain Eq.{\,}\eqref{a35} are provided in Section S1 of Ref.{\,}[\onlinecite{SM}].
\begin{description}
\item[\textit{Pseudospin Symmetry Case}]
\end{description}
For pseudo-spin symmetry, $\frac{d\Sigma(r)}{dr} = 0$ results in $\Sigma(r) = C_{ps}$, which similarly simplifies the DE. The corresponding energy eigenvalues $E_{nk}$ are expressed as:
\scriptsize
\begin{equation}
\begin{aligned}
&{\left(Mc^2+C_{ps}\right)}^2-E^2_{nk}-\left(Mc^2-E_{nk}+C_{ps}\right)\left(V_1+V_2+V_3\right) \\
&\quad +\frac{{\hslash}^2c^2k(k-1)}{r^2_e}(A_0+A_1+A_2) \\
&={\alpha}^2{\hslash}^2c^2{\left(\sqrt{\frac{1}{4}-\frac{(Mc^2-E_{nk}+C_{ps})V_3}{{\alpha}^2{\hslash}^2c^2}+\frac{k(k-1)}{{\alpha}^2r^2_e}A_2}-n-\frac{1}{2} \right.} \\
&\quad \left.+\frac{-\frac{(Mc^2-E_{nk}+C_{ps})V_2}{2{\alpha}^2{\hslash}^2c^2}+\frac{k(k-1)}{2{\alpha}^2r^2_e}A_1}{\sqrt{\frac{1}{4}-\frac{(Mc^2-E_{nk}+C_{ps})V_3}{{\alpha}^2{\hslash}^2c^2}+\frac{k(k-1)}{{\alpha}^2r^2_e}A_2}-n-\frac{1}{2}}\right)^2 .
\end{aligned}
\label{a53}
\end{equation}
\normalsize
The information about all parameters in Eq.{\,}\eqref{a53} and the full derivation process to obtain Eq.{\,}\eqref{a53} are provided in Section S1 of Ref.{\,}[\onlinecite{SM}].

While the current study is based on the NU method, we would like to note that recent developments in gauge theory and quantum integrability offer alternative frameworks for exact spectral analysis. For example, gauge theory approaches using the ordinary differential equation/integrable model correspondences have been applied to solve eigenvalue problems in ${N}=2$ supersymmetric systems{\,}\cite{Fioravanti2020, Fioravanti2023}. Similarly, quantum integrability techniques based on functional Bethe \emph{ansatz} equations have provided exact quantization conditions in contexts such as black hole quasi-normal mode spectra{\,}\cite{Fioravanti2021, Fioravanti2022}. These advances, including new exact solutions in supersymmetric gauge theory and gravity contexts, highlight the power of integrability-based methods and point to promising directions for extending the current model.

\section{Results and Discussion}\label{sec3}
\subsection{Investigation of Spin Symmetry Case in Specific Scenarios}\label{sec3_1}
\noindent
The GTHP is a highly versatile framework that exhibits a remarkable sensitivity to its parameters, making it an effective model to describe several prominent potentials encountered in fundamental quantum systems. The parametric flexibility of GTHP allows it to emulate various well-known potentials, such as the generalized Woods-Saxon{\,}\cite{WS1,WS2}, Rosen-Morse{\,}\cite{RM1}, and Manning-Rosen potentials{\,}\cite{MR}, among others. This adaptability stems from the intricate control of key parameters, such as $V_1$, $V_2$, $V_3$, and $\alpha$, which dictate the potential's shape, depth, and range.

It is well-known that spin symmetry occurs in the DE when the difference between the vector and scalar potentials is constant, $V(r) - S(r) = \text{constant}$. In this regime, the spin-orbit coupling becomes negligible, leading to near-degenerate energy levels and simplifying the relativistic energy spectrum{\,}\cite{PS1,PS2}. The ability to smoothly transition between different potential profiles by tuning these parameters makes GTHP an ideal model for exploring the underlying physics of systems with spin symmetry. Each set of parameters generates a distinct potential landscape, which strongly influences the system's energy levels, the degree of degeneracy, and the manifestation of spin symmetry. This symmetry is significant in high-energy physics and nuclear physics, particularly in understanding mesonic and baryonic spectra, as well as nucleon-nucleon interactions. The sensitivity of GTHP model is also crucial for tailoring specific quantum systems and analyzing the behavior of relativistic particles in various physical environments as well as allowing for the emulation of short-range and long-range interactions across various fields of fundamental physics. We now examine the energy eigenvalue derived from Eq.{\,}\eqref{a35} for different potentials modeled under the GTHP framework.

\noindent
\textbf{i)} For the GTHP, selecting parameters $V_1=-\frac{V_0}{2}-\frac{W}{4}$, $V_2=\frac{V_0}{2}$,
$V_3=\frac{W}{4}$, and $\alpha =\frac{1}{2a}$ yields the following relation for the energy levels equation of the generalized Woods-Saxon potential:
\scriptsize
\begin{equation}
\begin{split}
& \left(Mc^2-C_s\right)^2 -E^2_{nk}+\frac{\hslash^2c^2k(k+1)}{R^2_0}C_0 = \\
& \quad \frac{\hslash^2c^2}{4a^2}\left(\sqrt{\frac{1}{4}+\frac{(Mc^2+E_{nk}-C_s)a^2W}{\hslash^2c^2}+\frac{k(k+1)a^2}{R^2_0}C_2}-n-\frac{1}{2}\right. \\
& \quad \left.+{\frac{\left. \frac{(Mc^2+E_{nk}-C_s)a^2V_0}{\hslash^2c^2}-\frac{k(k+1)a^2}{R^2_0}(C_1+C_2)\right.}{\sqrt{\frac{1}{4}+\frac{(Mc^2+E_{nk}-C_s)a^2W}{\hslash^2c^2}+\frac{k(k+1)a^2}{R^2_0}C_2}-n-\frac{1}{2}}}\right)^2,
\end{split}
\end{equation}
\normalsize
where $n=0, 1, 2, \cdots n_{max}$, and the energy levels are determined by the floor function of the square root expressions within the equation, incorporating the parameters $C_0$, $C_1$, and $C_2$ defined as follows:
\scriptsize
\begin{align}
C_0 &= \frac{A_0+A_1+A_2}{{\left(1+x_e\right)}^2},  \nonumber \\
C_1 &= -\frac{2\left(A_1+2A_2\right)}{{\left(1+x_e\right)}^2}, \\
C_2 &= \frac{4A_2}{{\left(1+x_e\right)}^2},  \nonumber
\end{align}
\normalsize
with $x_e=\frac{r_e-R_0}{R_0}$. In here, $x_e$ represents a dimensionless variable that controls the shape of the potential.

\noindent
\textbf{ii)} Considering the case when $W=0$ and $x_e=0$ for the energy levels equation of the standard Woods-Saxon potential{\,}\cite{Feizi}, we derive:
\scriptsize
\begin{equation}
\begin{aligned}
& \left(Mc^2-C_s\right)^2 - E^2_{nk} + \frac{{\hslash}^2 c^2 k (k+1)}{R^2_0} C_0 \\
&=\frac{{\hslash}^2 c^2}{4a^2}\left(\sqrt{\frac{1}{4}+\frac{k(k+1)a^2}{R^2_0}C_2}-n-\frac{1}{2}+\frac{\frac{(Mc^2+E_{nk}-C_s)a^2V_0}{{\hslash}^2c^2}-\frac{k(k+1)a^2}{R^2_0}(C_1+C_2)}{\sqrt{\frac{1}{4}+\frac{k(k+1)a^2}{R^2_0}C_2}-n-\frac{1}{2}} \right)^2,
\end{aligned}
\label{a58}
\end{equation}
\normalsize
where $n=0, 1, 2, \cdots n_{max}$, and the coefficients $C_0$, $C_1$, and $C_2$ are adjusted accordingly to account for the absence of the $W$ and $x_e$ parameters.

\noindent
\textbf{iii)} For the energy spectrum equation of the Rosen-Morse potential with parameters $\ V_3=-V_1=C\ $ and $V_2=B\ $, the formulation is as follows:
\scriptsize
\begin{equation}
\begin{aligned}
& \left(Mc^2-C_s\right)^2 - E^2_{nk} + (Mc^2+E_{nk}-C_s)B + \frac{{\hslash}^2c^2k(k+1)}{r^2_e}(A_0+A_1+A_2) \\
&= \alpha^2{\hslash}^2c^2 \left( \sqrt{\frac{(Mc^2+E_{nk}-C_s)C}{\alpha^2{\hslash}^2c^2}+\frac{k(k+1)}{\alpha^2r^2_e}A_2+\frac{1}{4}} \right. \\
&\quad \left. -n-\frac{1}{2} + \frac{\frac{(Mc^2+E_{nk}-C_s)B}{2\alpha^2{\hslash}^2c^2}+\frac{k(k+1)}{2\alpha^2r^2_e}A_1}{\sqrt{\frac{(Mc^2+E_{nk}-C_s)C}{\alpha^2{\hslash}^2c^2}+\frac{k(k+1)}{\alpha^2r^2_e}A_2+\frac{1}{4}}-n-\frac{1}{2}} \right)^2,
\end{aligned}
\label{a59}
\end{equation}
\normalsize
where $n=0, 1, 2, \cdots n_{max}$, signifying the quantum number associated with the energy level derived from the equation parameters and conditions.

\noindent
\textbf{iv)} Incorporating the parameters of the GTHP for the Manning-Rosen type-potential, specifically $V_1 = \frac{\beta(\beta - 1) - 2A}{4kb^2}$, $V_2 = -\frac{\beta(\beta - 1) - A}{2kb^2}$, $V_3 = \frac{\beta(\beta - 1)}{4kb^2}$, with $k = \frac{2M}{{\hslash}^2}$ and $2\alpha = \frac{1}{b}$, leads to the following formulation:
\scriptsize
\begin{equation}
\begin{aligned}
& \left(Mc^2-C_s\right)^2 - E^2_{nk} + \frac{{\hslash}^2c^2k(k+1)}{r^2_e}(A_0+A_1+A_2) \\
&= \frac{{\hslash}^2c^2}{4b^2} \left( \sqrt{\frac{1}{4}+\frac{(Mc^2+E_{nk}-C_s)\beta(\beta-1)}{2Mc^2}+\frac{4k(k+1)b^2}{r^2_e}A_2} - n - \frac{1}{2} \right. \\
&\quad \left. + \frac{-\frac{(Mc^2+E_{nk}-C_s)[\beta(\beta-1)-A]}{2{\alpha}^2{\hslash}^2c^2}+\frac{2k(k+1)b^2}{r^2_e}A_1}{\sqrt{\frac{1}{4}+\frac{(Mc^2+E_{nk}-C_s)\beta(\beta-1)}{2Mc^2}+\frac{4k(k+1)b^2}{r^2_e}A_2}-n-\frac{1}{2}} \right)^2,
\end{aligned}
\label{a61}
\end{equation}
\normalsize
where $n=0, 1, 2,\ldots n_{max}$, and the energy levels are evaluated using the complex interplay between the potential parameters and the quantum mechanical properties of the system.

\noindent
\textbf{v)} By meticulously selecting the parameters within the GTHP framework, namely, $V_1=\frac{1}{4}D_e{\left(b-2\right)}^2$, $V_2=-\frac{1}{2}D_eb\left(b-2\right)$, $V_3=\frac{1}{4}D_eb^2$, $2\alpha =\delta $ and $b=e^{\delta r_e}+1>2$, we refine the approach to dissect the energy spectrum equation pertinent to the improved Rosen-Morse potential:
\scriptsize
\begin{align}
& \left(Mc^2-C_s\right)^2 - E^2_{nk} + \left(Mc^2 + E_{nk} - C_s\right)D_e + \frac{\hslash^2c^2k(k+1)}{r^2_e}(A_0 + A_1 + A_2) \nonumber \\
& = \frac{\delta^2 \hslash^2 c^2}{4} \left( \sqrt{ \frac{1}{4} + \frac{(Mc^2 + E_{nk} - C_s)D_e b^2}{\delta^2 \hslash^2 c^2} + \frac{4k(k+1)}{\delta^2 r^2_e}A_2 } - n - \frac{1}{2} \right. \nonumber \\
& \quad \left. - \frac{ \frac{(Mc^2 + E_{nk} - C_s)D_e b(b-2)}{\delta^2 \hslash^2 c^2} - \frac{2k(k+1)}{\delta^2 r^2_e}A_1 }{ \sqrt{ \frac{1}{4} + \frac{(Mc^2 + E_{nk} - C_s)b^2D_e}{\delta^2 \hslash^2 c^2} + \frac{4k(k+1)}{\delta^2 r^2_e}A_2 } - n - \frac{1}{2} } \right)^2. \label{a62}
\end{align}
\normalsize
The quantum numbers are denoted by $n=0, 1, 2,\ldots n_{max}$.

\noindent
\textbf{vi)} Transitioning to the analysis of the Schi\"{o}berg potential, we adopt a nuanced parametrization: $V_{1}={\delta }^2D$, $V_{2}=-2\delta \sigma D$ and $V_{3}={\sigma }^2D$. This setup facilitates the derivation of the energy spectrum equation, articulated as follows:
\scriptsize
\begin{align}
&\left( Mc^2 - C_s \right)^2 - E^2_{nk} + \left( Mc^2 + E_{nk} - C_s \right)(\delta - \sigma)^2 D + \frac{\hslash^2 c^2 k(k + 1)}{r^2_e}C_0 \nonumber \\
&= \alpha^2 \hslash^2 c^2 \left( \sqrt{ \frac{1}{4} + \frac{(Mc^2 + E_{nk} - C_s) \sigma^2 D}{\alpha^2 \hslash^2 c^2} + \frac{k(k + 1)}{4 \alpha^2 r^2_e}C_2 } \right. \nonumber \\
&\quad \left. - n - \frac{1}{2} - \frac{ \frac{(Mc^2 + E_{nk} - C_s)
\delta \sigma D}{\alpha^2 \hslash^2 c^2} - \frac{k(k + 1)}{4
\alpha^2 r^2_e}(C_1 - C_2) }{ \sqrt{ \frac{1}{4} + \frac{(Mc^2 +
E_{nk} - C_s) \sigma^2 D}{\alpha^2 \hslash^2 c^2} + \frac{k(k +
1)}{4 \alpha^2 r^2_e}C_2 } - n - \frac{1}{2} } \right)^2 \ ,
\label{a63}
\end{align}
\normalsize
where $n=0, 1, 2,\ldots n_{max}$. In this context, the coefficients $C_{0}$, $C_{1}$, and $C_{2}$ are explicitly determined as
\scriptsize
\begin{equation}
\begin{aligned}
C_0 &= A_0 + A_1 + A_2 = 1 - \left( \frac{1 + e^{-2\alpha r_e}}{2\alpha r_e} \right)^2 \left[ \frac{8\alpha r_e}{1 + e^{-2\alpha r_e}} - 3 - 2\alpha r_e \right], \\
C_1 &= 2A_1 + 4A_2 = -\left( 1 + e^{2\alpha r_e} \right) \cdot \frac{1 + e^{-2\alpha r_e}}{2\alpha r_e} \left[ 3 - \left( 3 + 2\alpha r_e \right) \left( \frac{1 + e^{-2\alpha r_e}}{2\alpha r_e} \right) \right], \\
C_2 &= 4A_2= \left( 1 + e^{2\alpha r_e} \right)^2 \left( \frac{1 + e^{-2\alpha r_e}}{2\alpha r_e} \right)^2 \left[ 3 + 2\alpha r_e - \frac{4\alpha r_e}{1 + e^{-2\alpha r_e}} \right].
\end{aligned}
\end{equation}
\normalsize

\noindent
\textbf{vii)} Next, we elucidate the energy spectrum equation derived from the GTHP with parameters $V_1=P_1+\frac{P_2}{2}+\frac{P_3}{4}$, $V_2=-\frac{P_2}{2}-\frac{P_3}{2}$, and $V_3=\frac{P_3}{4}$ for the four-parameter exponential-type potential. This culminates in:
\scriptsize
\begin{align}
&\left( Mc^2 - C_s \right)^2 - E^2_{nk} + \left( Mc^2 + E_{nk} - C_s \right) P_1 + \frac{\hslash^2 c^2 k(k + 1)}{r^2_e}(A_0 + A_1 + A_2) \nonumber \\
&= \alpha^2 \hslash^2 c^2 \left( \sqrt{ \frac{1}{4} + \frac{(Mc^2 + E_{nk} - C_s)P_3}{4 \alpha^2 \hslash^2 c^2} + \frac{k(k + 1)}{\alpha^2 r^2_e}A_2 } \right. \nonumber \\
&\quad \left. - n - \frac{1}{2} - \frac{ \frac{(Mc^2 + E_{nk} - C_s)(P_2 + P_3)}{4 \alpha^2 \hslash^2 c^2} - \frac{k(k + 1)}{2 \alpha^2 r^2_e}A_1 }{ \sqrt{ \frac{1}{4} + \frac{(Mc^2 + E_{nk} - C_s)P_3}{4 \alpha^2 \hslash^2 c^2} + \frac{k(k + 1)}{\alpha^2 r^2_e}A_2 } - n - \frac{1}{2} } \right)^2. \label{a65}
\end{align}
\normalsize
The quantum number range is again specified by $n=0, 1, 2, \ldots, n_{\text{max}}$.

\noindent
\textbf{viii)} The parameters for the GTHP, designated as $V_1$, $V_2$, and $V_3$, are defined respectively by the relations $V_1=\frac{W_1+W_2+W_3}{4}$, $V_2=\frac{W_3-W_1}{2}$, and $V_3=\frac{W_1-W_2+W_3}{4}$. These parameters are instrumental in deriving the energy spectrum equation for the Williams-Poulios type-potential. By substituting these definitions, the equation governing the energy spectrum can be expressed as follows:
\scriptsize
\begin{equation}
\begin{aligned}
&{\left(Mc^2-C_s\right)}^2-E^2_{nk}+\left(Mc^2+E_{nk}-C_s\right)W_3+\frac{{\hslash}^2c^2k(k+1)}{r^2_e}(A_0+A_1+A_2)\\
&={\alpha}^2{\hslash}^2c^2\left(\sqrt{\frac{1}{4}+\frac{(Mc^2+E_{nk}-C_s)(W_1-W_2+W_3)}{4{\alpha}^2{\hslash}^2c^2}+\frac{k(k+1)}{{\alpha}^2r^2_e}A_2}-n-\frac{1}{2}\right.\\
&\left.-\frac{\frac{(Mc^2+E_{nk}-C_s)(W_1-W_3)}{4{\alpha}^2{\hslash}^2c^2}-\frac{k(k+1)}{2{\alpha}^2r^2_e}A_1}{\sqrt{\frac{1}{4}+\frac{(Mc^2+E_{nk}-C_s)(W_1-W_2+W_3)}{4{\alpha}^2{\hslash}^2c^2}+\frac{k(k+1)}{{\alpha}^2r^2_e}A_2}-n-\frac{1}{2}}\right)^2 .
\end{aligned}
\label{a66}
\end{equation}
\normalsize
This equation is valid for quantum numbers $n=0, 1, 2,\ldots, n_{max}$.

\noindent
\textbf{ix)} In the regime where $\alpha$ is significantly smaller than unity, the energy levels as described by Eq.{\,}\eqref{a35} simplify to:
\scriptsize
\begin{equation}
\begin{aligned}
&{\left(Mc^2-C_s\right)}^2-E^2_{nk}+\left(V_1-\frac{V^2_2}{4V_3}\right)(Mc^2+E_{nk}-C_s)\\
&+\frac{{\hslash}^2c^2k(k+1)}{r^2_e}+2\alpha\hslash c\left(1-\frac{V^2_2}{4V^2_3}\right)\sqrt{(Mc^2+E_{nk}-C_s)V_3} \\
&\times\left(n+\frac{1}{2}\right)-{\alpha}^2{\hslash}^2c^2\left[\left(1+\frac{3V^2_2}{4V^2_3}\right){\left(n+\frac{1}{2}\right)}^2+\frac{1}{4}\left(1-\frac{V^2_2}{4V^2_3}\right)\right]+o\left({\alpha}^2\right)=0 .
\end{aligned}
\label{a67}
\end{equation}
\normalsize
This approximation holds for small values of $n$ and $l$.

\noindent
\textbf{x)} Setting $C_s=0$ allows us to simplify the potential for the energy level equation of a Klein-Gordon particle, resulting in Ref.{\,}[\onlinecite{GTHP2}]:
\scriptsize
\begin{equation}
\begin{aligned}
&E^2_{nk}-M^2c^4-\left(Mc^2+E_{nk}\right)(V_1+V_2+V_3)-\frac{{\hslash}^2c^2l(l+1)}{r^2_e}(A_0+A_1+A_2)\\
&+{\alpha}^2{\hslash}^2c^2\left(\sqrt{\frac{1}{4}+\frac{(Mc^2+E_{nk})V_3}{{\alpha}^2{\hslash}^2c^2}+\frac{l(l+1)}{{\alpha}^2r^2_e}A_2}-n-\frac{1}{2}\right.\\
&\left.+\frac{\frac{(Mc^2+E_{nk})V_2}{2{\alpha}^2{\hslash}^2c^2}+\frac{l(l+1)}{2{\alpha}^2r^2_e}A_1}{\sqrt{\frac{1}{4}+\frac{(Mc^2+E_{nk})V_3}{{\alpha}^2{\hslash}^2c^2}+\frac{l(l+1)}{{\alpha}^2r^2_e}A_2}-n-\frac{1}{2}}\right)^2=0 .
\end{aligned}
\label{a68}
\end{equation}
\normalsize
Applicable for $n=0, 1, 2,\ldots, n_{max}$.

\noindent
\textbf{xi)} Finally, applying the transformations $E_{nk}-Mc^2+C_s\to E_{nl}$ and $E_{nk}+Mc^2-C_s\to 2\mu c^2$ to Eq.{\,}\eqref{a35}, we derive the energy spectrum for the non-relativistic scenario:
\scriptsize
\begin{equation}
\begin{aligned}
&E_{nl}=(V_1+V_2+V_3)+\frac{{\hslash}^2c^2l(l+1)}{2\mu r^2_e}(A_0+A_1+A_2)\\
&-\frac{{\alpha}^2{\hslash}^2}{2\mu}\left[\sqrt{\frac{1}{4}+\frac{2\mu V_3}{{\alpha}^2{\hslash}^2}+\frac{l(l+1)}{{\alpha}^2r^2_e}A_2}-n-\frac{1}{2}\right.\\
&\left.+\frac{\frac{\mu V_2}{{\alpha}^2{\hslash}^2}+\frac{l(l+1)}{2{\alpha}^2r^2_e}A_1}{\sqrt{\frac{1}{4}+\frac{2\mu V_3}{{\alpha}^2{\hslash}^2}+\frac{l(l+1)}{{\alpha}^2r^2_e}A_2}-n-\frac{1}{2}}\right]^2 .
\end{aligned}
\label{a69}
\end{equation}
\normalsize
This formulation aligns with the findings presented in Ref.{\,}[\onlinecite{GTHP1}], valid for $n=0, 1, 2,\ldots, n_{max}$.

\begin{table*}[htbp]
\caption{\label{tab:charmspectrum} Mass spectra of charmonium \( M_{c \bar{c}} \) in GeV for \( m_{c} = 1.27 \) GeV; \( V_1 = 7.404662 \) GeV; $V_2 = -20.779639$ GeV; $V_3 = 14.959636$ GeV; $\alpha = 0.223612$ GeV; $r_e = 3.830931$ GeV$^{-1}$. The percentage deviation from experimental values is presented below each calculated value.}
\begin{adjustbox}{max width=\textwidth}
\begin{minipage}{\textwidth}
\begin{ruledtabular}
\small
\setlength{\tabcolsep}{2pt}
\renewcommand{\arraystretch}{0.8} 
\begin{tabular}{r|l|cccccccc}
State & Particle & Present Work  & Theory I                   & Theory II                    & Theory III                    & Experiment  \\
      &          & Non-Rel.      & Ref.{\,}[\onlinecite{purohit2022}] & Ref.{\,}[\onlinecite{soni2018}] & Ref.{\,}[\onlinecite{bukor2023}] &  Ref.{\,} [\onlinecite{workman2022}]  \\
\hline \hline
1S  & J/$\psi$(1S)      & 3.096900  & 3.096   & 3.094   & 3.0969  & 3.09690 \\
    &                   & 0.00\%    & -0.03\% & -0.09\% & 0.00\%  &  \\
1P  & $\chi_{c1}$(1P)   & 3.196126  & 3.415   & 3.468   & 3.518   & 3.51067 \\
    &                   & -8.96\%   & -2.71\% & -1.22\% & 0.21\%  &  \\
1D  & -                 & 3.377174  & 3.770   & 3.772   & 3.787   & 3.77370 \\
    &                   &           &         &         &         & - \\
1F  & -                 & 3.617957  & 4.040   & 4.012   & -       & -       \\
    &                   &           &         &         &         &         \\
2S  & $\psi$(2S)        & 3.686100  & 3.733   & 3.681   & 3.686   & 3.68610 \\
    &                   & 0.00\%    & 1.27\%  & -0.14\% & 0.00\%  &  \\
2P  & $\chi_{c1}$(3872) & 3.772214  & 3.894   & 3.938   & 3.823   & 3.87165 \\
    &                   & -2.57\%   & 0.58\%  & 1.71\%  & -1.26\% &  \\
2D  & -                 & 3.930322  & 4.088   & 4.188   & -       & -       \\
    &                   &           &         &         &         &         \\
2F  & -                 & 4.143411  & -       & 4.396   & -       & -       \\
    &                   &           &         &         &         &         \\
3S  & $\psi$(4040)      & 4.039000  & 4.068   & 4.129   & 3.889   & 4.03900 \\
    &                   & 0.00\%    & 0.72\%  & 2.23\%  & -3.71\% &  \\
3P  & $\chi_{c1}$(4140) & 4.106798  & -       & 4.338   & -       & 4.14650 \\
    &                   & -0.96\%   &         & 4.62\%  &         &  \\
3D  & -                 & 4.233365  & -       & 4.557   & -       & -       \\
    &                   &           &         &         &         &         \\
3F  & -                 & 4.409120  & -       & 4.746   & -       & -       \\
    &                   &           &         &         &         &         \\
4S  & $\psi$(4230)      & 4.115000  & 4.263   & 4.514   & 3.982   & 4.22270 \\
    &                   & -2.55\%   & 0.95\%  & 6.90\%  & -5.70\% &  \\
4P  & $\chi_{c1}$(4274) & 4.159579  & -       & 4.696   & -       & 4.28600 \\
    &                   & -2.95\%   &         & 9.57\%  &         &  \\
4D  & -                 & 4.246829  & -       & 4.896   & -       & -       \\
    &                   &           &         &         &         &         \\
\end{tabular}
\end{ruledtabular}
\end{minipage}
\end{adjustbox}
\end{table*}

\begin{table*}[htbp]
\caption{\label{tab:bottomspectrum} Mass spectra of bottomonium $M_{b \bar{b}}$ in GeV for $m_{b} = 4.18$ GeV; $V_1 = 4.883229$ GeV; $V_2 = -12.928267$ GeV; $V_3 = 10.108982$ GeV; $\alpha = 0.412272$ GeV; $r_e = 1.836733$ GeV$^{-1}$. The percentage deviation from experimental values is presented below each calculated value.}
\begin{adjustbox}{max width=\textwidth}
\begin{minipage}{\textwidth}
\begin{ruledtabular}
\small
\setlength{\tabcolsep}{2pt}
\renewcommand{\arraystretch}{0.8}
\begin{tabular}{r|l|cccccccc}
State & Particle & Present Work & Theory I  & Theory II & Theory III & Experiment  \\
&  & Non-Rel. & Ref.{\,}[\onlinecite{purohit2022}] & Ref.{\,}[\onlinecite{soni2018}] & Ref.{\,}[\onlinecite{bukor2023}] &  Ref.{\,}[\onlinecite{workman2022}]          \\
\hline \hline
1S  & $\Upsilon$(1S)   & 9.460302  & 9.460   & 9.463   & 9.460   & 9.46030  \\
    &                  & 0.00\%    & 0.00\%  & 0.03\%  & 0.00\%  &          \\
1P  & $h_{b}$(1P)      & 9.594182  & 9.704   & 9.821   & 9.942   & 9.8993   \\
    &                  & -3.08\%   & -1.97\% & -0.79\% & 0.43\%  &          \\
1D  & $\Upsilon_2$(1D) & 9.827113  & 10.010  & 10.074  & 10.140  & 10.1637  \\
    &                  & -3.31\%   & -1.51\% & -0.88\% & -0.23\% &          \\
1F  & -                & 10.130188 & 10.268  & 10.288  & -       & -        \\
    &                  &           &         &         &         &          \\
2S  & $\Upsilon$(2S)   & 10.023261 & 10.028  & 9.979   & 10.023  & 10.02326 \\
    &                  & 0.00\%    & 0.05\%  & -0.44\% & 0.00\%  &          \\
2P  & $h_{b}$(2P)      & 10.151214 & 10.160  & 10.220  & 10.150  & 10.2598  \\
    &                  & -1.06\%   & -0.97\% & -0.39\% & -1.07\% &          \\
2D  & $\Upsilon_2$(2D) & 10.380036 & 10.332  & 10.424  & -       & -        \\
    &                  &           &         &         &         &          \\
2F  & -                & 10.688957 & -       & 10.607  & -       & -        \\
    &                  &           &         &         &         &          \\
3S  & $\Upsilon$(3S)   & 10.355200 & 10.343  & 10.359  & 10.178  & 10.3552  \\
    &                  & 0.00\%    & -0.12\% & 0.04\%  & -1.71\% &          \\
3P  & $h_{b}$(3P)      & 10.468567 & -       & 10.556  & -       & -        \\
    &                  &           &         &         &         &          \\
3D  & $\Upsilon_2$(3D) & 10.681781 & -       & 10.733  & -       & -        \\
    &                  &           &         &         &         &          \\
3F  & -                & 10.986923 & -       & 10.897  & -       & -        \\
    &                  &           &         &         &         &          \\
4S  & $\Upsilon$(4S)   & 10.405400 & 10.536  & 10.683  & 10.242  & 10.5794  \\
    &                  & -1.64\%   & -0.41\% & 0.98\%  & -3.18\% &          \\
4P  & $h_{b}$(4P)      & 10.497089 & -       & 10.855  & -       & -        \\
    &                  &           &         &         &         &          \\
4D  & $\Upsilon_2$(4D) & 10.686643 & -       & 11.015  & -       & -        \\
    &                  &           &         &         &         &          \\
\end{tabular}
\end{ruledtabular}
\end{minipage}
\end{adjustbox}
\end{table*}

Briefly, in the spin symmetry, the GTHP’s inclusion of Woods--Saxon and Morse potentials means it can be used in nuclear bound-state problems (like nucleon--nucleus potential models) or molecular/atomic systems that employ Morse-type interactions. However, an equally important spin symmetry, pseudospin symmetry, also plays a crucial role in relativistic quantum systems{\,}\cite{PS1, PS2}. Pseudospin symmetry arises when the sum of the vector and scalar potentials satisfies $V(r) + S(r) = \text{constant}${\,}\cite{PS1, PS2}. This symmetry is closely related to the near-degeneracy of nuclear energy levels and offers a complementary perspective to the spin symmetry analysis{\,}\cite{PS8,PS9}. The investigation of pseudospin symmetry is vital for understanding nuclear shell structures, meson spectroscopy, and the underlying physics governing quark-antiquark pairs{\,}\cite{PS10,PS11,PS12,PS13,PS14,PS15,PS16,PS17,PS18,PS19,PS20}. Therefore, we extend our analysis to pseudospin symmetry within the GTHP framework, examining its manifestation across different potential profiles and its impact on the system's energy spectrum in the Section S2 of Ref.{\,}[\onlinecite{SM}]. This transition from spin symmetry to pseudospin symmetry reinforces the unified nature of these symmetries and their relevance to nuclear and particle physics.

\subsection{Analysis of Charmonium and Bottomonium Mass Spectra}\label{sec3_2}
\noindent
In this part, we mainly focus on the non-relativistic modeling of charmonium and bottomonium systems (See detailed modeling information in S3.A section in Ref.{\,}[\onlinecite{SM}]), as it exhibits better agreement with experimental data across both low- and high-energy states. The following analysis delves into specific comparisons and explores the underlying physical implications of these findings.
\begin{figure}[htp!]
    \centering
    \includegraphics[width=1\linewidth]{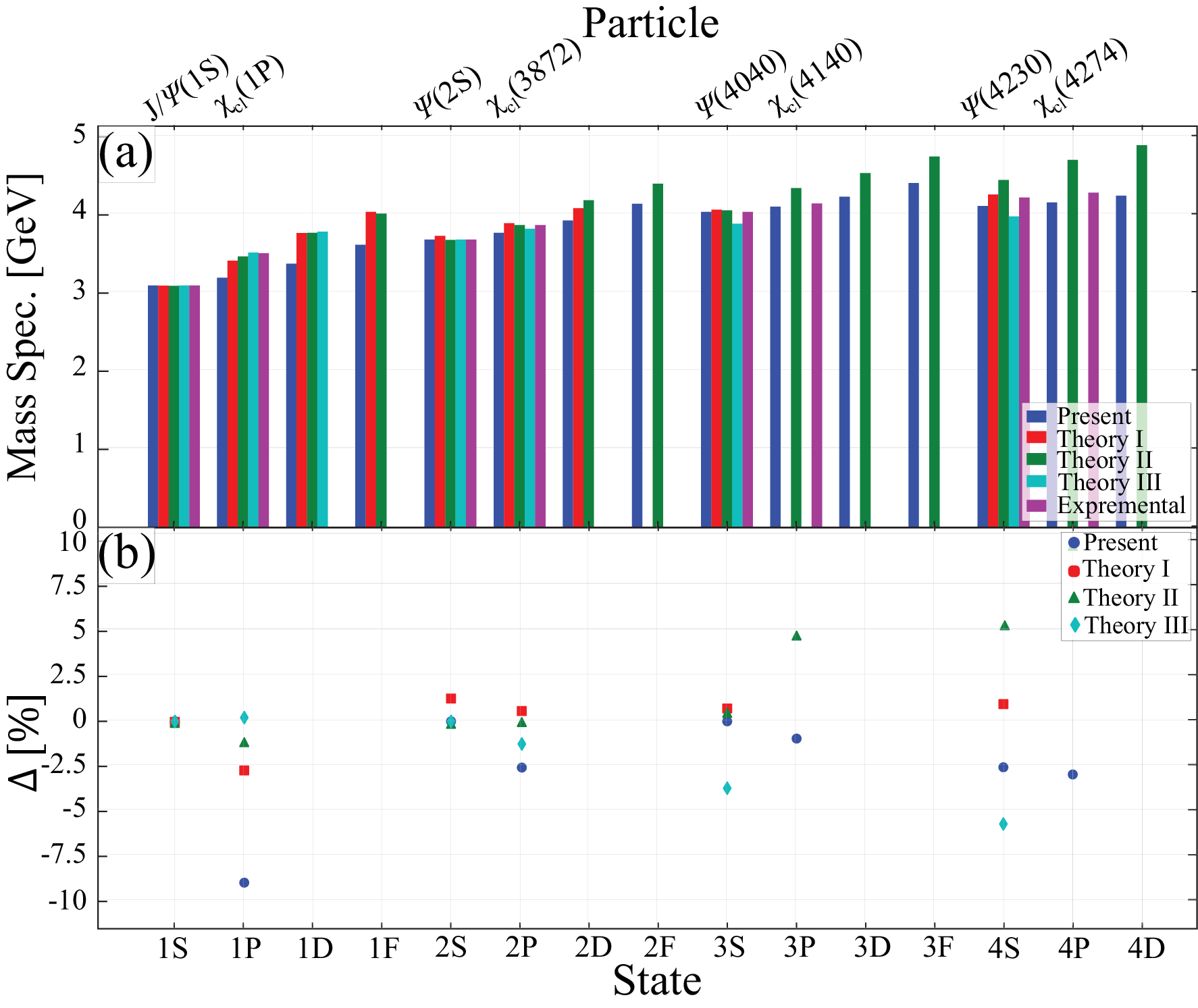}
    \caption{(a) Non-relativistic calculating mass spectra of charmonium $M_{c \tilde{c}}$ in GeV for $m_{c}$ = 1.27 GeV; $V_1$ = 7.404662 GeV; $V_2$ = -20.779639 GeV; $V_3$ = 14.959636 GeV; $\alpha$ = 0.223612 GeV; $r_e$ = 3.830931 GeV$^{-1}$. The experimental data from Ref.{\,}[\onlinecite{workman2022}] and other theoretical predictions: Theory I from Ref.{\,}[\onlinecite{purohit2022}], Theory II from Ref.{\,}[\onlinecite{soni2018}], Theory III from Ref.{\,}[\onlinecite{bukor2023}] are also included for comparison. (b) Corresponding deviations from the experimental data.}
    \label{fig:charmspectrum}
\end{figure}
\begin{figure}[htp!]
\centering
\includegraphics[width=1\linewidth]{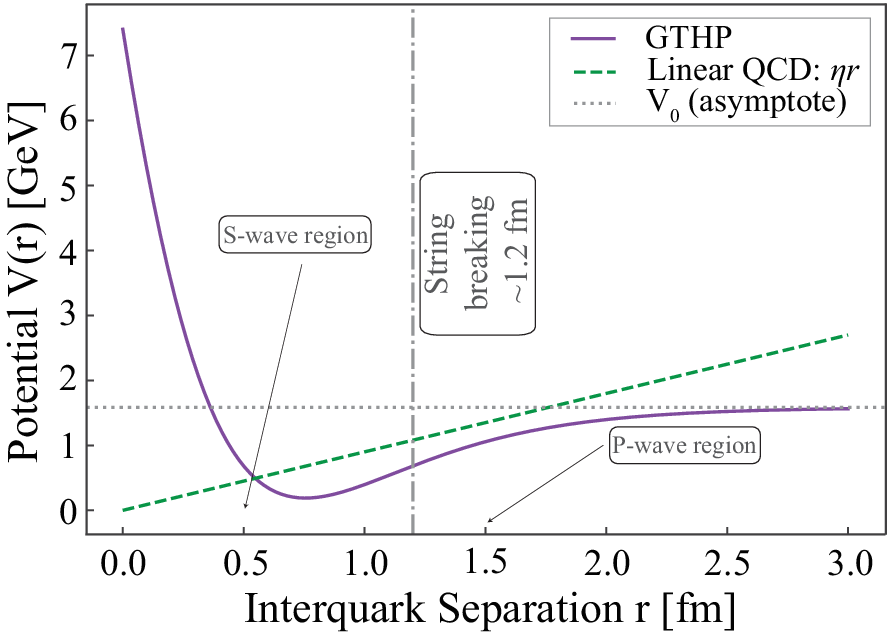}
\caption{Comparison between the GTHP potential (solid) and a linear confinement potential (dashed). The GTHP realistically captures QCD string-breaking by saturating at large distances, accurately describing compact $S$-wave states but reducing confinement for extended $P$-wave states~\cite{Isgur1985, Bali2001, Bali2005, Bulava2019}.}
\label{fig:2}
\end{figure}

The calculated charmonium mass spectrum, as depicted in Figure{\,}\ref{fig:charmspectrum} and detailed in Table{\,}\ref{tab:charmspectrum} and also Table{\,}S1 in Ref.{\,}[\onlinecite{SM}] and, shows a well agreement with experimental results, particularly for the $1S$ ($J/\psi(1S)$), $2S$ ($\psi(2S)$), and $3S$ ($\psi(4040)$) states. Notably, the $J/\psi(1S)$ state matches precisely at $3.096900$ GeV, underscoring the efficacy of the GTHP model in predicting ground state masses where non-relativistic approximations hold strong.

However, the model's performance varies across different orbital states. While $S$-wave states are well-described, higher orbital states, such as the $1P$ ($\chi_{c1}(1P)$) state, exhibit deviations of -8.96\% (see Figure{\,}\ref{fig:charmspectrum} (a)) for non-relativistic case, respectively. Such discrepancies indicate the need for further refinement in modeling higher orbital angular momentum states. The significant deviations observed in these states suggest that the influence of relativistic effects, spin-orbit coupling, and other acceptable structure corrections become more pronounced. Again, in our context of relativistic treatment, using the Eq.{\,}\eqref{a35}, we observe relatively larger deviations from experimental data, especially in the $1P$ and other higher orbital states, primarily due to the overestimation of relativistic contributions without proper spin-orbit interaction corrections.

\begin{figure}[htp!] 
    \centering
\includegraphics[width=1\linewidth]{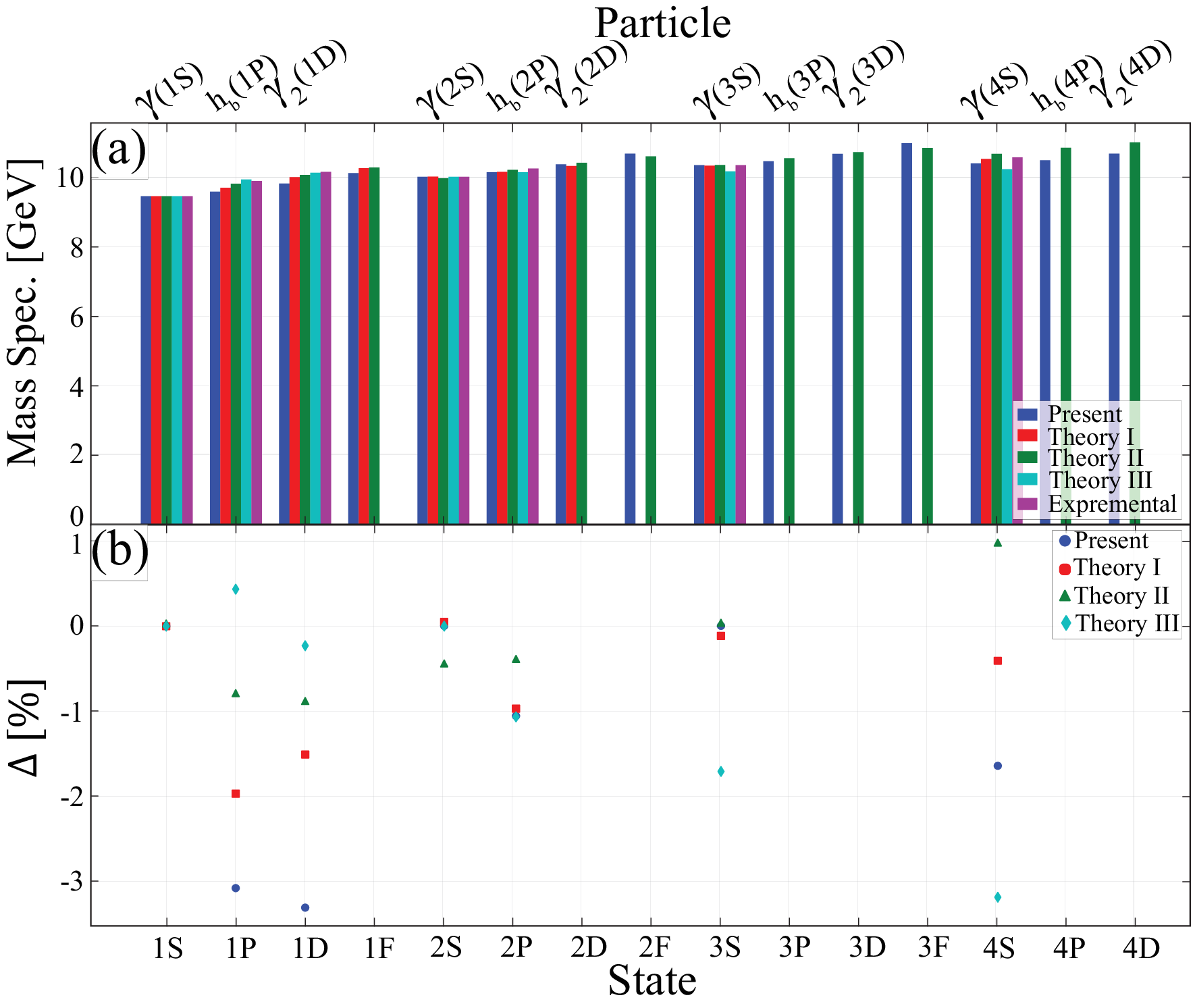}
\caption{(a) Non-relativistic calculating mass spectra of bottomonium $M_{b \tilde{b}}$ in GeV for $m_{b} = 4.18$ GeV; $V_1$ = 4.883229 GeV; $V_2$ = -12.928267 GeV; $V_3$ = 10.108982 GeV; $\alpha$ = 0.412272 GeV; $r_e$ = 1.836733 GeV$^{-1}$. The experimental data from Ref.{\,}[\onlinecite{workman2022}] and other theoretical predictions: Theory I from Ref.{\,}[\onlinecite{purohit2022}], Theory II from Ref.{\,}[\onlinecite{soni2018}], Theory III from Ref.{\,}[\onlinecite{bukor2023}] are also included for comparison. (b) Corresponding deviations from the experimental data.}
\label{fig:bottomspectrum}
\end{figure}

Another reason for the observed deviations can be that the analytic heavy-$Q\bar Q$ potential behaves like a Coulomb+linear ({``Cornell''}) potential at short distances but flattens at large $r$, thereby mimicking QCD string breaking. Figure{\,}\ref{fig:2} compares the GTHP (solid curve) to a conventional linear confining potential. At large separations the GTHP flattens, reflecting the string-breaking screening expected in full QCD{\,}\cite{Isgur1985, Bernard2000, Bali2001, Bali2005, Bulava2019}. In practice this means that beyond roughly $r\sim1.0$-$1.2${\,}fm the effective confining force weakens (approaching the threshold of two non-interacting heavy-light mesons{\,}\cite{Bulava2019}). Consequently, compact $S$-wave quarkonia (with small radii) reside almost entirely within the steep central region of the potential and are well-described by the model. By contrast, $P$-wave states have larger spatial extent and sample the softened tail of the potential; since the confining strength is weaker at these distances, the predicted $P$-wave binding is insufficient. This under-binding of the $P$-waves is therefore a natural consequence of the screened potential choice. Lattice QCD results confirm that the static heavy-quark potential saturates at long range due to light-quark pair creation{\,}\cite{Bali2001, Bulava2019}. We conclude that while the GTHP provides an accurate description of the compact $S$-states, further refinements (e.g. additional interaction terms or coupled-channel effects) will be required to capture the physics of the more extended $P$-wave states.

In order to remove these discrepancies, spin symmetry breaking can be introduced through the parameter $C_s$, which allows for fine-tuning to match the ground state with experimental values. Although this improved the overall agreement, particularly for the charmonium system, the deviations for higher angular momentum states persisted. This is due to the fact that the Eq.{\,}\eqref{a35} are obtained for spin symmetry case. The interaction potential may need to be adjusted to account for quark-antiquark dynamics more comprehensively, particularly in exact spin-orbit correction that naturally arise in systems with non-negligible quark masses. To gain further improvements in the relativistic model, one has to go to higher order corrections like refined GTHP with Tensor potentials or solve two body DE and Bethe-Salpeter formalism with GTHP, as these methods account more accurately for spin-dependent interactions in higher excited states. For a detailed exploration of the relativistic mass spectra and a comprehensive comparison between non-relativistic and relativistic results with the full set of mass spectra tables, analysis of spin symmetry breaking, readers can be found in section S3.B of Ref.{\,}[\onlinecite{SM}].

For bottomonium, the mass spectra shown in Figure{\,}\ref{fig:bottomspectrum} and detailed in Table{\,}\ref{tab:bottomspectrum} and also Table{\,}S2 in Ref.{\,}[\onlinecite{SM}] reveal an even closer agreement with experimental data, particularly for the $1S$ ($\Upsilon(1S)$), $2S$ ($\Upsilon(2S)$), and $3S$ ($\Upsilon(3S)$) states, where deviations from experimental values are around 0.0\%, see Figure . This demonstrates the robustness of the GTHP model in predicting bottomonium states, where the non-relativistic approximation is more justified due to the larger bottom quark mass. For higher states, such as the $1P$ ($h_{b}(1P)$), $1D$ ($\Upsilon_2(1D)$), $2P$ ($h_{b}(2P)$), and $4S$ ($\Upsilon(4S)$) states, the deviations are slightly more significant, ranging from -1.06\% to 3.31\% (see Figure{\,}\ref{fig:bottomspectrum}(b)). These minor discrepancies indicate that while the GTHP model is highly reliable even for excited states, there is room for improvement by incorporating relativistic and spin-orbital corrections. The model's accuracy in these states implies that the essential physics of quark confinement and interaction potentials are well captured. However, spin-spin, spin-orbit, and tensor interactions might need finer adjustments for higher precision. 

While the GTHP model demonstrates good performance across a range of quantum states, a comparison with other theoretical models as Theory I (non-relativistic analyses of the linear plus modified Yukawa potential){\,}\cite{purohit2022}, Theory II (non-relativistic analyses of quark-antiquark Cornell potential){\,}\cite{soni2018}, and Theory III (non-relativistic analyses of Cornell potential){\,}\cite{bukor2023}-reveals varying degrees of agreement. Although Theory II and III show closer results for lower energy states, the GTHP model provides better accuracy overall, especially for higher orbital states. This indicates that the GTHP has provided a more powerful description of quarkonium systems, especially in the high-energy region. Also, our non-relativistic quantum-mechanical treatment yields important insights, especially in the lower-energy region, but we have to be aware of its limitations when considering higher orbital states or systems with significant relativistic effects. This framework includes spin effects semi-inclusively, but a completely relativistic treatment will necessitate in addition the formulation and solution of two-body DE{\,}\cite{Crater,2BDE_2,2BDE_3} as well as the Bethe-Salpeter equation{\,}\cite{Hilger}. Implicitly in this approach would be the relativistic corrections our model now "corrects", providing a more realistic picture of quark-antiquark dynamics, especially regarding higher angular-momentum states. Note that a relativistic treatment could entail more than just this pair term – in particular, we might have important correlations through spin-orbit coupling, which are known to induce fine structure splittings not accounted for by our model{\,}\cite{workman2022}.

Such the two-body DE{\,}\cite{Crater,2BDE_2,2BDE_3} and the Bethe-Salpeter equation{\,}\cite{Molina2027,Hilger} formalism can also open avenues with GTHP for exploring new quarkonium states that may emerge owing to relativistic effects, which are not predicted by conventional non-relativistic models. These heavily bound states may include, e.g., the very tightly-bound ones with large binding energies where tests of the model can be made against experiments; and new states which are created under extremely severe conditions in such things as a heavy-ion collision plasma situations where non-relativistic models might be insufficient. The predictions by this model of possible new states with longevity at extreme conditions can also be significant in the context of heavy-ion collisions, where quark-gluon plasma might be created and where new bound states might be discovered. Expanding this framework and applying it to mixed-flavor systems like bottom-charm ($B_c$) mesons can also yield deeper insights into quark confinement and the dynamics of the strong force. Such predictions, although speculative, highlight the need for closing elliptical states not only in the GTHP model but also by including more sophisticated interactions to study a whole range of quarkonium states. Given these, while hardy in its current form, the GTHP model may serve as a starting block for more advanced models. A more accurate treatment together with incorporating more complex interactions will enable a deeper study of quarkonium states and give us access to a wider domain of the strong interactions that are responsible for these configurations.

\section{Conclusions}\label{sec4}
\noindent
In this study, we derived the analytical solution of the DE in the context of spin and pseudospin symmetries for the GTHP by using NU method and demonstrated its effectiveness in reproducing the mass spectra of charmonium and bottomonium with sufficient accuracy. The flexibility of the GTHP allows its application in different quantum systems, potentially holding its place in our approach to solving complex fermionic systems. For the analysis of the mass spectra of charmonium and bottomonium, our findings for the low $S$-wave states, namely $J/\psi(1S)$, $\psi(2S)$ and $\psi(4040)$, were in good agreement with the reported experimental data. This finding emphasized the accuracy of the model in the non-relativistic regime, especially for heavy quark systems such as bottomonium, where deviations from experimental masses are minimal. However, in the higher orbital states, such as $1P\longmapsto4P$ ($\chi_{c1}$) for charmonium and $1P$ ($h_b(1P)$) and $1D$ ($\Upsilon_2(1D)$) for bottomonium, we observed the relatively larger discrepancy. While the non-relativistic analyses of the GTHP model adequately capture the fundamental dynamics of the heavy quarkonium, it was also shown that the inclusion of relativistic and spin-dependent effects is necessary to advance the description of higher orbital states. For systems with more than one flavour species, such as $B_c$ mesons, a corresponding formulation taking spin-orbit corrections into account can shed more light on the propagation of aberrations owing to quark confinement and the strong force. Its overall performance provided a strong basis for theoretical considerations and reliable predictions with significant potential for further development.

\section*{Acknowledgments}
\noindent
We would like to acknowledge the invaluable contributions of our co-author, A. I. Ahmadov, to beginning of this work. We are deeply saddened by his passing and honor his memory through this publication.

\section*{Data Availability Statement}
\noindent
All data that support the findings of this study are included within the article (and any supplementary files).

\section*{Conflict of Interest}
\noindent
The authors declare no competing interests. All research has been carried out within an appropriate ethical framework.

\clearpage
\newpage

\clearpage
\newpage

\clearpage  
\onecolumngrid  

\begin{center}
    \textcolor{purple}{\LARGE \textbf{Supplemental Material}}  
    \vspace{1em}
    \textcolor{purple}{\rule{\linewidth}{0.7mm}}  
    \vspace{1em}
    \textbf{\large \blue Dirac Equation Solution with Generalized tanh-Shaped Hyperbolic Potential: \\ Application to Charmonium and Bottomonium Mass Spectra}
\end{center}

\twocolumngrid

\setcounter{equation}{0}
\renewcommand{\theequation}{S\arabic{equation}}

\setcounter{figure}{0}
\renewcommand{\thefigure}{S\arabic{figure}}

\setcounter{table}{0}
\renewcommand{\thetable}{S\arabic{table}}

\setcounter{section}{0}
\renewcommand{\thesection}{S\arabic{section}}

\setcounter{page}{1}
\renewcommand{\thepage}{S\arabic{page}}



\section{Detailed Methodology}
\noindent
\lettrine[findent=2pt]{{\color{CustomBlue}{\textbf{W}}}}{}ithin the framework of advanced theoretical physics, the DE provides a profound description of particles possessing mass $M$ under the influence of an attractive scalar potential $S(\vec{{r}})$ and a repulsive vector potential $V(\vec{{r}})$. This formulation is crucial for understanding the quantum behaviors of fermions in various potentials and is given by{\,}\cite{Greiner}:
\scriptsize
\begin{equation}
\left[c\vec{\alpha} \cdot \hat{\vec{p}}+\beta\left(Mc^2+S(\vec{r})\right)+V(\vec{r})\right]\psi(\vec{r})=E\psi(\vec{r}), \label{a2}
\end{equation}
\normalsize
where $E$ represents the relativistic total energy of the particle, and $\hat{\vec{p}}=-i\hslash \vec{\nabla}$ denotes
the momentum operator. The matrices $\vec{\alpha}$ and $\beta$, integral to the equation, are defined as $4{\times}4$ matrices:
\scriptsize
\begin{equation}
\vec{\alpha} = \begin{pmatrix} 0 & \vec{\sigma} \\
\vec{\sigma} & 0 \end{pmatrix}, \quad \beta =
\begin{pmatrix} I & 0 \\ 0 & -I \end{pmatrix},
\end{equation}
\normalsize
with ${\vec{\sigma}}$ denoting the Pauli spin matrices and $I$ being the identity matrix in $2{\times}2$ space. In cases of spherical symmetry, the potentials simplify to radial dependencies, $V(\vec{{r}}) \rightarrow V(r)$ and $S(\vec{{r}}) \rightarrow S(r)$, where $r=|\vec{{r}}|$.

\subsection{Solutions to the Radial Dirac Equation}
\noindent
The eigenfunctions of the Dirac operator, or Dirac spinors, are expressed concerning the radial quantum number $n$ and the spin-orbit coupling quantum number $k$, as follows:
\scriptsize
\begin{equation}
{\psi}_{nk}(r,\theta,\varphi)=\frac{1}{r}\left(
\begin{array}{c}
F_{nk}(r)Y^l_{jm}(\theta,\varphi) \\
iG_{nk}(r)Y^{\tilde{l}}_{jm}(\theta,\varphi)
\end{array} \right), \label{a3}
\end{equation}
\normalsize
where $F_{nk}(r)$ and $G_{nk}(r)$ denote the upper and lower components of the radial wave function, respectively. Here, $Y^l_{jm}(\theta, \varphi)$ and $Y^{\tilde{l}}_{jm}(\theta, \varphi)$ represent the spin spherical harmonic functions, which are intricately coupled to the total angular momentum quantum numbers $j$ and $m$, the latter symbolizing the projection of angular momentum along the $z$-axis. The relationships between the quantum numbers are delineated by $k(k+1) = l(l+1)$ and $k(k-1) = \tilde{l}(\tilde{l}+1)$.

The quantum number $k$ intricately links to the conventional angular momentum quantum number $l$ and the analogous pseudospin angular momentum $\tilde{l}$, as outlined below:
\scriptsize
\begin{equation}
\begin{aligned}
k = \begin{cases}
-\left(l+1\right) = -\left(j+\frac{1}{2}\right), & \text{for aligned spin states}\ (k<0), \\
+l = +\left(j+\frac{1}{2}\right), & \text{for unaligned spin states}\ (k>0),
\end{cases}
\end{aligned}
\label{a4}
\end{equation}
\normalsize
revealing the spin alignment and its influence on the spinor structure. Furthermore, the pseudospin symmetry, manifesting as a quasi-degenerate doublet structure, is described by:
\scriptsize
\begin{equation}
\begin{aligned}
k = \begin{cases}
-\tilde{l} = -\left(j+\frac{1}{2}\right), & \text{for aligned spin states}\ (k<0), \\
+\left(\tilde{l}+1\right) = +\left(j+\frac{1}{2}\right), & \text{for unaligned spin states}\ (k>0),
\end{cases}
\end{aligned}
\label{a5}
\end{equation}
\normalsize
where $k = \pm{1}, \pm{2}, \cdots$, and $\tilde{s}=\frac{1}{2}$ and $\tilde{l}$ represent the pseudospin and pseudo-orbital angular momenta, respectively.

Upon substituting Eq.{\,}\eqref{a3} into Eq.{\,}\eqref{a2}, and leveraging the subsequent algebraic identities involving the Pauli matrices::
\scriptsize
\begin{equation}
\left\{
\begin{array}{l}
(\vec{\sigma} \cdot \vec{a})(\vec{\sigma} \cdot \vec{b}) = (\vec{a} \cdot \vec{b}) + i\vec{\sigma}(\vec{a} \times \vec{b}), \\
\vec{\sigma} \cdot \hat{\vec{p}} = \vec{\sigma} \cdot
\hat{r}(\hat{r} \cdot \hat{\vec{p}} + i\frac{\vec{\sigma} \cdot
\hat{\vec{L}}}{r}), \quad \hat{r} = \frac{\vec{r}}{r}
\end{array}
\right.
\label{a6}
\end{equation}
\normalsize
accompanied by the following spherical harmonic operator relations:
\scriptsize
\begin{equation}
\left\{
\begin{array}{l}
(\vec{\sigma} \cdot \hat{\vec{L}})Y^{\tilde{l}}_{jm}(\theta, \varphi) = \hslash (k-1)Y^{\tilde{l}}_{jm}(\theta, \varphi), \\
(\vec{\sigma} \cdot \hat{\vec{L}})Y^{l}_{jm}(\theta, \varphi) = -\hslash (k+1)Y^{l}_{jm}(\theta, \varphi), \\
(\vec{\sigma} \cdot \hat{r})Y^{\tilde{l}}_{jm}(\theta, \varphi) = -Y^{l}_{jm}(\theta, \varphi), \\
(\vec{\sigma} \cdot \hat{r})Y^{l}_{jm}(\theta, \varphi) =
-Y^{\tilde{l}}_{jm}(\theta, \varphi),
\end{array}
\right.
\label{a7}
\end{equation}
\normalsize
leads us to the derivation of the following two coupled radial DEs for the functions $F_{nk}(r)$ and $G_{nk}(r)$:
\scriptsize
\begin{equation}
\begin{array}{rcl}
\left(\frac{d}{dr} + \frac{k}{r}\right)F_{nk}(r) & = & \frac{Mc^2 + E_{nk} - \Delta(r)}{\hslash c}G_{nk}(r), \\
\left(\frac{d}{dr} - \frac{k}{r}\right)G_{nk}(r) & = & \frac{Mc^2 - E_{nk} + \Sigma(r)}{\hslash c}F_{nk}(r),
\end{array}
\label{a7}
\end{equation}
\normalsize
where $\Delta(r) = V(r) - S(r)$ and $\Sigma(r) = V(r) + S(r)$. Through the elimination of $G_{nk}(r)$ and $F_{nk}(r)$ in the coupled equations, we obtain two distinct second-order differential equations for the upper and lower radial wave functions:
\scriptsize
\begin{equation}
\begin{array}{l}
\left[\frac{d^2}{dr^2} - \frac{k(k+1)}{r^2} - \frac{(Mc^2 + E_{nk} - \Delta(r))(Mc^2 - E_{nk} + \Sigma(r))}{{\hslash}^2c^2} + \frac{\frac{d\Delta(r)}{dr}\left(\frac{d}{dr} + \frac{k}{r}\right)}{Mc^2 + E_{nk} - \Delta(r)}\right]F_{nk}(r) = 0, \\
\\
\left[\frac{d^2}{dr^2} - \frac{k(k-1)}{r^2} - \frac{(Mc^2 + E_{nk} - \Delta(r))(Mc^2 - E_{nk} + \Sigma(r))}{{\hslash}^2c^2} - \frac{\frac{d\Sigma(r)}{dr}\left(\frac{d}{dr} - \frac{k}{r}\right)}{Mc^2 - E_{nk} + \Sigma(r)}\right]G_{nk}(r) = 0.
\end{array}
\label{a8}
\end{equation}
\normalsize
These equations highlight the distinct spin and pseudospin symmetries scenarios within the relativistic quantum mechanical description of fermions under spherical potentials.

\subsubsection{Spin symmetry case}
\noindent
The regime of spin symmetry is delineated by the condition $\frac{d\triangle(r)}{dr}=0$, leading to a constant $\triangle(r)=C_s$. This pivotal assumption simplifies the analysis significantly, as delineated in the seminal works cited herein{\,}\cite{PS4, PS5}. Within this framework, the modified potential terms, $S(r)=V(r)-C_s$ and $\Sigma(r)=2V(r)-C_s$, are instrumental in reformulating the second equation of our foundational Eq.{\,}\eqref{a7} set as follows:
\scriptsize
\begin{equation}
\left[\frac{d^2}{dr^2}-\frac{k\left(k+1\right)}{r^2}-\frac{{\left(Mc^2-C_s\right)}^2-E^2_{nk}+2\left(Mc^2+E_{nk}-C_s\right)V\left(r\right)}{{\hslash}^2c^2}\right]F_{nk}(r)=0\ , \label{a12}
\end{equation}
\normalsize
where $k=l$ for positive $k$ values and $k=-(l+1)$ for negative ones. The energy levels $E_{nk}$ are functions of the quantum numbers $n$ and $l$, intricately linked to the spin symmetry quantum number. It is noteworthy that non-zero $\tilde{l}$ values engender degenerate states characterized by $j=l\pm {1}/{2}$, epitomizing $SU(2)$ spin symmetry.

The potential function $2V(r)$ is specified in Eq.{\,}\eqref{a12}, taken as:
\scriptsize
\begin{equation}
2V(r)=V_{GTHP}(r)=V_1+V_2{\tanh \left(\alpha r\right)}+V_3{{\rm tanh}}^2\left(\alpha r\right)\ . \label{a13}
\end{equation}
\normalsize
Then, substituting Eq.{\,}\eqref{a13} into Eq.{\,}\eqref{a12} yields:
\scriptsize
\begin{equation}
\begin{split}
&\left(\frac{d^2}{dr^2}-\frac{\left(Mc^2-C_s\right)^2-E^2_{nk}}{\hslash^2c^2}-\frac{k\left(k+1\right)}{r^2}\right.\\
&\quad-\left.\frac{\left(Mc^2+E_{nk}-C_s\right)}{\hslash^2c^2}\left[V_1+V_2
\tanh \left(\alpha r\right)+V_3 \tanh^2\left(\alpha
r\right)\right]\right)F_{nk}(r)=0. \label{a14}
\end{split}
\end{equation}
\normalsize
Given the spin-orbit centrifugal term $V_{soc}(r)=\frac{k(k+1)}{r^2}$, analytical solutions to Eq.{\,}\eqref{a14} are elusive except for the cases where $k=0$ or $k=-1$. Thus, the widely embraced Pekeris approximation{\,}\cite{Pekeris} is adopted for the pragmatic resolution of the equation, facilitating the treatment of the spin-orbit centrifugal term{\,}\cite{GTHP1,GTHP2}.
\scriptsize
\begin{equation}
\frac{1}{r^2}\approx \frac{1}{r^2_e}\left[A_0+A_1{\tanh \left(\alpha r\right)}+A_2{{\rm tanh}}^2\left(\alpha r\right)\right]\ ,\label{a15}
\end{equation}
\normalsize
The coefficients $A_0$, $A_1$, and $A_2$ are determined through the relations presented below, as derived in the Ref.{\,}[\onlinecite{GTHP1,GTHP2}]:
\scriptsize
\begin{equation}
\left\{ \begin{array}{l}
A_0=1+\frac{\cosh^4\left(\alpha r_e\right)}{\alpha^2 r_e^2}\left[3\tanh^2\left(\alpha r_e\right)+2\alpha r_e\tanh \left(\alpha r_e\right)\left(1-2\tanh^2\left(\alpha r_e\right)\right)\right], \\
A_1=-\frac{2\cosh^4\left(\alpha r_e\right)}{\alpha^2 r_e^2}\left[3\tanh \left(\alpha r_e\right)+\alpha r_e\left(1-3\tanh^2\left(\alpha r_e\right)\right)\right], \\
A_2=\frac{\cosh^4\left(\alpha r_e\right)}{\alpha^2 r_e^2}\left(3-2\alpha r_e\tanh \left(\alpha r_e\right)\right)
\end{array} \right.
\label{a16}
\end{equation}
\normalsize
Integrating Eq.{\,}\eqref{a15} into Eq.{\,}\eqref{a14} facilitates the following reformulation:
\scriptsize
\begin{equation}
\begin{aligned}
& \frac{d^2F_{nk}(r)}{dr^2}-\left(\frac{{\left(Mc^2-C_s\right)}^2-E^2_{nk}}{{\hslash}^2c^2} + \frac{\left(Mc^2+E_{nk}-C_s\right)}{{\hslash}^2c^2}V_1 + \frac{k(k+1)}{r^2_e}A_0 \right. \\
& \quad \left. + \left[\frac{\left(Mc^2+E_{nk}-C_s\right)}{{\hslash}^2c^2}V_2 + \frac{k(k+1)}{r^2_e}A_1\right]{\tanh \left(\alpha r\right)} \right. \\
& \quad \left. + \left[\frac{\left(Mc^2+E_{nk}-C_s\right)}{{\hslash}^2c^2}V_3 + \frac{k(k+1)}{r^2_e}A_2\right]{{\rm tanh}}^2\left(\alpha r\right)\right)F_{nk}(r) = 0.
\end{aligned}
\label{a17}
\end{equation}
\normalsize
Adopting a transformation to a new variable $z={\tanh \left(\alpha r\right)}$, with $0 \leq z \leq 1$, simplifies Eq.{\,}\eqref{a17} to:
\scriptsize
\begin{equation}
F''_{nk}(z)-\frac{2z}{1-z^2}F'_{nk}(z)-\frac{\varepsilon +\beta
z+\gamma z^2}{{\left(1-z^2\right)}^2}F_{nk}(z)=0\ ,\label{a18}
\end{equation}
\normalsize
where the coefficients $\varepsilon$, $\beta$, and $\gamma$ are defined as:
\scriptsize
\begin{equation}
\left\{
\begin{array}{l}
\frac{{\left(Mc^2-C_s\right)}^2-E^2_{nk}+\left(Mc^2+E_{nk}-C_s\right)V_1}{{\alpha }^2{\hslash }^2c^2}+\frac{k\left(k+1\right)}{{\alpha }^2r^2_e}A_0=\varepsilon \\
\frac{\left(Mc^2+E_{nk}-C_s\right)V_2}{{\alpha }^2{\hslash }^2c^2}+\frac{k(k+1)}{{\alpha }^2r^2_e}A_1=\beta \\
\frac{\left(Mc^2+E_{nk}-C_s\right)V_3}{{\alpha }^2{\hslash}^2c^2}+\frac{k(k+1)}{{\alpha }^2r^2_e}A_2=\gamma
\end{array}
\right.
\label{a19}
\end{equation}
\normalsize
The boundary conditions are set as $F_{nk}(0)=0$ and $F_{nk}(1)=0$, facilitating the application of the NU method for the solution. Accordingly, Eq.{\,}\eqref{a18} is recast into the form of a hypergeometric-type equation:
\scriptsize
\begin{equation}
u''(z)+\frac{\widetilde{\tau }(z)}{\sigma(z)}u'(z)+\frac{\widetilde{\sigma }(z)}{{\sigma}^2(z)}u(z)=0\ ,\label{a20}
\end{equation}
\normalsize
Upon a thorough comparison between Eqs.{\,}\eqref{a18} and \eqref{a20}, the following relationships emerge:
\scriptsize
\begin{equation}
\sigma(z) = 1 - z^2; \quad \widetilde{\tau}(z) = -2z; \quad \widetilde{\sigma}(z) = -\varepsilon - \beta z - \gamma z^2. \label{a21}
\end{equation}
\normalsize
Utilizing these expressions and following the methodology outlined in Ref.{\,}[\onlinecite{Nikiforov}], we deduce the form of the function $\pi(z)$ by substituting from Eq.{\,}\eqref{a21} and considering the derivative $\sigma'(z) = -2z$, thus obtaining:
\scriptsize
\begin{equation}
\pi(z) = \pm \sqrt{(\gamma - k)z^2 + \beta z + \varepsilon + k}. \label{a22}
\end{equation}
\normalsize
To ascertain the constant parameter $k$, we impose the condition that the discriminant of the quadratic form under the square root in Eq.{\,}\eqref{a22} vanishes, which leads to:
\scriptsize
\begin{equation}
k_1 = D^2 - \varepsilon, \quad k_2 = P^2 - \varepsilon, \label{a23}
\end{equation}
\normalsize
where $D$ and $P$ are defined as:
\scriptsize
\begin{equation}
D = \sqrt{\frac{\varepsilon + \gamma + \sqrt{(\varepsilon + \gamma)^2 - \beta^2}}{2}}, \quad P = \sqrt{\frac{\varepsilon + \gamma - \sqrt{(\varepsilon + \gamma)^2 - \beta^2}}{2}}, \label{a24}
\end{equation}
\normalsize
with the relationships $D > P$, $2DP = |\beta|$, and $D^2 + P^2 = \varepsilon + \gamma$ ensuring a hierarchical structure in the solutions. Substituting the values of $k$ from Eq.{\,}\eqref{a23} into Eq.{\,}\eqref{a22}, eight possible configurations of $\pi(z)$ are derived as follows:
\scriptsize
\begin{equation}
\pi(z) = \pm \left\{
\begin{array}{ll}
Pz \pm D, & \text{for } k = D^2 - \varepsilon, \\
Dz \pm P, & \text{for } k = P^2 - \varepsilon.
\end{array}
\right.
\label{a25}
\end{equation}
\normalsize
Among these eight formulations of $\pi(z)$, dictated by the NU method, we select the variant for which the derivative of $\tau(z) = \widetilde{\tau}(z) + 2\pi(z)$ is strictly negative, that is, $\tau'(z) = -2(D + 1) < 0$, and $\tau(z) = 0$ within the interval $z = \frac{P}{D + 1} \in (0, 1)$, dismissing the rest as physically irrelevant. Thus, the suitable choice for $\pi(z)$ is:
\scriptsize
\begin{equation}
\pi(z) = -Dz + P, \label{a26}
\end{equation}
\normalsize
and consequently,
\scriptsize
\begin{equation}
\begin{aligned}
\tau(z) &= -2(D + 1)z + 2P, \\
k &= P^2 - \varepsilon.
\end{aligned} \label{a28}
\end{equation}
\normalsize
Further, by employing the relations for $\lambda = k + \pi'(z)$ and the quantized eigenvalue condition $\lambda_n = -n\tau'(z) - \frac{n(n - 1)}{2}\sigma''(z)$, where $n = 0, 1, 2, \ldots$, as prescribed in Ref.[\onlinecite{Nikiforov}], we acquire:
\scriptsize
\begin{equation}
\lambda = P^2 - \varepsilon - D, \label{a29}
\end{equation}
\begin{equation}
\lambda_n = 2nD + n(n + 1), \label{a30}
\end{equation}
\normalsize
where $n$ denotes the radial quantum number. Aligning Eq.{\,}\eqref{a29} with Eq.{\,}\eqref{a30} and leveraging the sum of squares identity $D^2 + P^2 = \varepsilon + \gamma$, we deduce:
\scriptsize
\begin{equation}
D + n + \frac{1}{2} \pm \sqrt{\gamma + \frac{1}{4}} = 0, \label{a31}
\end{equation}
\normalsize
yielding the condition:
\scriptsize
\begin{equation}
D = n' > 0, \label{a32}
\end{equation}
\normalsize
where $n'$ is defined as:
\scriptsize
\begin{equation}
n' = \sqrt{\gamma + \frac{1}{4}} - n - \frac{1}{2}, \quad n = 0, 1, 2, \ldots. \label{a33}
\end{equation}
\normalsize
Next, incorporating Eq.{\,}\eqref{a24} into Eq.{\,}\eqref{a32}, we arrive at:
\scriptsize
\begin{equation}
\varepsilon + \beta + \gamma = \left(n' + \frac{\beta}{2n'}\right)^2. \label{a34}
\end{equation}
\normalsize
After substituting the expressions from Eqs.{\,}\eqref{a19} and \eqref{a33} into Eq.{\,}\eqref{a34}, which represents the energy levels equation, we arrive at the following expression:
\scriptsize
\begin{equation}
\begin{split}
&{\left(Mc^2-C_s\right)}^2-E^2_{nk}+\left(Mc^2+E_{nk}-C_s\right)\left(V_1+V_2+V_3\right)\\
&\quad+\frac{{\hslash}^2c^2k(k+1)}{r^2_e}\left(A_0+A_1+A_2\right) \\
&\quad = {\alpha}^2{\hslash}^2c^2\left(\sqrt{\frac{1}{4}+\frac{\left(Mc^2+E_{nk}-C_s\right)V_3}{{\alpha}^2{\hslash}^2c^2}+\frac{k(k+1)}{{\alpha}^2r^2_e}A_2}-n-\frac{1}{2}\right. \\
&\quad \left. +\frac{\frac{\left(Mc^2+E_{nk}-C_s\right)V_2}{2{\alpha }^2{\hslash}^2c^2}+\frac{k(k+1)}{2{\alpha}^2r^2_e}A_1}{\sqrt{\frac{1}{4}+\frac{\left(Mc^2+E_{nk}-C_s\right)V_3}{{\alpha}^2{\hslash}^2c^2}+\frac{k(k+1)}{{\alpha}^2r^2_e}A_2}-n-\frac{1}{2}}\right)^2 .
\end{split}
\label{a35}
\end{equation}
\normalsize
Here, \(n\) takes values from \(0\) to \(n_{max}\) defined as:
\scriptsize
\begin{equation}
\begin{aligned}
n_{max} = \left\lfloor \sqrt{\frac{\left(Mc^2+E_{nk}-C_s\right)V_3}{{\alpha }^2{\hslash}^2c^2} + \frac{k\left(k+1\right)}{{\alpha }^2r^2_e}A_2 + \frac{1}{4}} \right. \\
\left. - \frac{1}{2} -  \sqrt{-\frac{\left(Mc^2+E_{nk}-C_s\right)V_2}{2{\alpha }^2{\hslash}^2c^2} - \frac{k\left(k+1\right)}{2{\alpha }^2r^2_e}A_1} \right\rfloor .
\end{aligned}
\label{a36}
\end{equation}
\normalsize
Utilizing the NU method, we can determine the radial eigenfunctions. By substituting the expressions for \(\pi(z)\) and \(\sigma(z)\) into the differential equations \(\frac{{\Phi'}(z)}{\Phi(z)}=\frac{\pi(z)}{\sigma(z)}\) and \(\frac{{\rho'}(z)}{\rho(z)}+\frac{{\sigma'}(z)}{\sigma(z)}=\frac{\tau(z)}{\sigma(z)}\), we can solve the higher-order differential equation to find the finite functions \(\Phi(z)\) and \(\rho(z)\) within the interval \((0,1)\). These functions take the form:
\scriptsize
\begin{equation}
\Phi \left(z\right)={\left(1-z\right)}^{\frac{\eta
}{2}}{\left(1+z\right)}^{\frac{\nu }{2}}\ \ ,\label{a37}
\end{equation}
\begin{equation}
\rho \left(z\right)={(1-z)}^{\eta }{\left(1+z\right)}^{\nu }\
,\label{a38}
\end{equation}
\normalsize
where \(\eta =D-P=\sqrt{\varepsilon +\gamma -\left|\beta \right|}>0\) and \(\nu =D+P=\sqrt{\varepsilon +\gamma +\left|\beta \right|}>0\).

The other part of the wave function \(y_n\left(z\right)\) takes on a hypergeometric-type function whose polynomial solutions are given by the Rodrigues relation{\,}\cite{Nikiforov}. It can be expressed as:
\scriptsize
\begin{equation}
y_n\left(z\right)=P^{(\eta ,\ \nu)}_n(z) . \label{a41}
\end{equation}
\normalsize
Here $P^{(\eta ,\ \nu)}_n(z)$ is the Jacobi polynomial:
\scriptsize
\begin{equation}
P^{(\eta ,\ \nu)}_n(z)=\frac{{(-1)}^n}{2^nn!}{(1-z)}^{-\eta
}{(1+z)}^{-\nu }\frac{d^n}{dz^n}\left[{(1-z)}^{n+\eta
}{(1+z)}^{n+\nu }\right]\ .
\end{equation}
\normalsize
According to the relation \(F_{nk}(z)=\Phi(z)y_n(z)\), we obtain the radial wave functions as:
\scriptsize
\begin{equation}
F_{nk}\left(z\right)=C_{nk}{\left(1-z\right)}^{\frac{\eta
}{2}}{\left(1+z\right)}^{\frac{\nu }{2}}P^{\left(\eta ,\ \nu
\right)}_n\left(z\right)\ ,\label{a42}
\end{equation}
\normalsize
where \(C_{nk}\) represents the normalization constant \cite{Abramowitz}.

Finally, the normalization constant \(C_{nk}\) is determined by utilizing the orthogonality condition:
\scriptsize
\begin{equation}
\int^{\infty
}_0{{\left|F_{nk}\left(r\right)\right|}^2dr}=\frac{1}{\alpha
}\int^1_0{\frac{{\left|F_{nk}\left(z\right)\right|}^2dz}{1-z^2}}=1\
.\label{a43}
\end{equation}
\normalsize
From the first equation presented in Eq.{\,}\eqref{a7} set, the lower-spinor component can be expressed as:
\scriptsize
\begin{equation}
G_{nk}\left(r\right)=\frac{\hslash
c}{Mc^2+E_{nk}-C_s}\left(\frac{d}{dr}+\frac{k}{r}\right)F_{nk}\left(r\right)\
,\label{a44}
\end{equation}
\normalsize
Here, it is essential to note that for \(C_s=0\) (exact spin symmetry), the energy spectrum consists only of real positive values as \(E_{nk}\ne -Mc^2\).

\subsubsection{Pseudospin symmetry case}
\noindent Pseudospin symmetry is characterized by the condition $\frac{d\Sigma(r)}{dr}=0$, resulting in $\Sigma(r)=C_{ps}=const${\,}\cite{PS4,PS5}. This leads to the simplification of the second equation presented in Eq.{\,}\eqref{a8} set with $S(r)=C_{ps}-V(r)$ and $\triangle(r)=2V(r)-C_{ps}$, yielding:
\scriptsize
\begin{equation}
\left[\frac{d^2}{dr^2}-\frac{k(k-1)}{r^2}-\frac{{\left(Mc^2+C_{ps}\right)}^2-E^2_{nk}-2\left(Mc^2-E_{nk}+C_{ps}\right)V(r)}{{\hslash}^2c^2}\right]G_{nk}(r)=0,
\label{a45}
\end{equation}
\normalsize
where $k=-\tilde{l}$ for $k<0$ and $k=\tilde{l}+1$ for $k>0$. The energy $E_{nk}$ depends on quantum numbers $n$ and $\tilde{l}$, associated with pseudospin symmetry. Notably, for $\tilde{l}\neq 0$, degenerate states with $j=\tilde{l}\pm\frac{1}{2}$ are produced, categorized under $SU(1,1)$ pseudospin symmetry.

Furthermore, if we express $2V(r)$ as a generalized hyperbolic potential in the form of $\tanh$ functions, Eq.{\,}\eqref{a15} becomes:
\scriptsize
\begin{equation}
\begin{aligned}
&\left\{
\frac{d^2}{dr^2} -
\frac{{\left(Mc^2+C_{ps}\right)}^2-E^2_{nk}}{{\hslash}^2c^2} -
\frac{k(k-1)}{r^2} \right. \\
&\left. + \frac{\left(Mc^2-E_{nk}+C_{ps}\right)}{{\hslash}^2c^2}
\left[V_1+V_2\tanh\left(\alpha r\right) + V_3\tanh^2\left(\alpha r\right)\right]
\right\}G_{nk}(r) = 0.
\end{aligned}
\label{a46}
\end{equation}
\normalsize
In cases where the pseudo-centrifugal term ($V_{pc}(r)=\frac{k(k-1)}{r^2}$) cannot be solved analytically for arbitrary $k$, the Pekeris approximation scheme{\,}\cite{Pekeris} Eq.{\,}\eqref{a15} is applied to it, resulting in the following equation:
\scriptsize
\begin{equation}
\begin{aligned}
\frac{d^2G_{nk}(r)}{dr^2} &- \left\{ \frac{(Mc^2+C_{ps})^2-E^2_{nk}}{{\hslash}^2c^2} \right. \\
&- \frac{(Mc^2-E_{nk}+C_{ps})}{{\hslash}^2c^2} \left[V_1 + V_2\tanh(\alpha r) + V_3\tanh^2(\alpha r)\right] \\
&\left. + \frac{k(k-1)}{r^2_e} \left[A_0 + A_1\tanh(\alpha r) + A_2\tanh^2(\alpha r)\right] \right\} G_{nk}(r) = 0.
\end{aligned}
\label{a47}
\end{equation}
\normalsize
To simplify Eq.{\,}\eqref{a47} further, we introduce a new variable $z=\tanh(\alpha r)$, where $z\in[0,1]$, resulting in:
\scriptsize
\begin{equation}
G''_{nk}(z)-\frac{2z}{1-z^2}G'_{nk}(z)-\frac{\varepsilon+\beta z+\gamma z^2}{(1-z^2)^2}G_{nk}(z)=0,
\label{a48}
\end{equation}
\normalsize
where the coefficients $\varepsilon$, $\beta$, and $\gamma$ are defined as:
\scriptsize
\begin{align}
\frac{(Mc^2+C_{ps})^2-E^2_{nk}-(Mc^2-E_{nk}+C_{ps})V_1}{{\alpha}^2{\hslash}^2c^2} + \frac{k(k-1)}{{\alpha}^2r^2_e}A_0 &= \varepsilon, \nonumber \\
-\frac{(Mc^2-E_{nk}+C_{ps})V_2}{{\alpha}^2{\hslash}^2c^2} + \frac{k(k-1)}{{\alpha}^2r^2_e}A_1 &= \beta, \label{a49} \\
-\frac{(Mc^2-E_{nk}+C_{ps})V_3}{{\alpha}^2{\hslash}^2c^2} + \frac{k(k-1)}{{\alpha}^2r^2_e}A_2 &= \gamma. \nonumber
\end{align}
\normalsize

\noindent
In the investigation of the quantum mechanical system described by Eq.{\,}\eqref{a48}, the imposed boundary conditions stipulate that the wave functions $G_{nk}(0) = 0$ and $G_{nk}(1) = 0$, ensuring the physical realizability of the system within the defined spatial domain. Upon the application of a series of transformations, the system undergoes a significant reparameterization: the radial wave function $G_{nk}(r)$ transformed into $F_{nk}(r)$, the quantum number $k$ is inverted to $-k$, the energy levels $E_{nk}$ are negated to $-E_{nk}$, and the constants and potentials are modified accordingly ($C_{ps}$ $\rightarrow$ $-C_s$, $V_1$ $\rightarrow$ $-V_1$, $V_2$ $\rightarrow$ $-V_2$, $V_3$ $\rightarrow$ $-V_3$). As a consequence of these transformations, Eq.{\,}\eqref{a48} and its associated expressions, denoted by Eq.{\,}\eqref{a49}, evolve into Eqs.{\,}\eqref{a18} and \eqref{a19}, respectively. This approach underscores the inherent symmetry between the pseudospin and spin symmetry frameworks, enabling a unified perspective on the energy spectrum and wavefunction solutions within the quantum mechanical paradigm.

In the context of pseudospin symmetry, the determination of the energy spectrum and the solutions to Eq.{\,}\eqref{a48} hinges on Eqs.{\,}\eqref{a34} and \eqref{a42}. The relation encapsulates the energy eigenvalues:
\scriptsize
\begin{equation}
\varepsilon +\beta +\gamma ={\left(n'+\frac{\beta}{2n'}\right)}^2, \label{a50}
\end{equation}
\normalsize
where the quantum number $n'$ is defined as:
\scriptsize
\begin{equation}
n'=\sqrt{\gamma +\frac{1}{4}}-n-\frac{1}{2}, \quad (n=0, 1, 2, \ldots). \label{a51}
\end{equation}
\normalsize
Furthermore, the functional form of $G_{nk}(z)$ is specified by:
\scriptsize
\begin{equation}
G_{nk}(z) = C_{nk}{\left(1-z\right)}^{\frac{\eta}{2}}{\left(1+z\right)}^{\frac{\nu}{2}}P^{\left(\eta, \nu\right)}_n(z), \label{a52}
\end{equation}
\normalsize
where $\eta = D-P = \sqrt{\varepsilon +\gamma -|\beta|} > 0$, $\nu = D+P = \sqrt{\varepsilon +\gamma +|\beta|} > 0$, and $C_{nk}$ represents the normalization constant.

Incorporating the expressions from Eqs.{\,}\eqref{a49} and \eqref{a51} into Eq.{\,}\eqref{a50} for the derivation of the energy levels yields:
\scriptsize
\begin{equation}
\begin{aligned}
&{\left(Mc^2+C_{ps}\right)}^2-E^2_{nk}-\left(Mc^2-E_{nk}+C_{ps}\right)\left(V_1+V_2+V_3\right) \\
&+\frac{{\hslash}^2c^2k(k-1)}{r^2_e}(A_0+A_1+A_2) \\
&={\alpha}^2{\hslash}^2c^2{\left(\sqrt{\frac{1}{4}-\frac{(Mc^2-E_{nk}+C_{ps})V_3}{{\alpha}^2{\hslash}^2c^2}+\frac{k(k-1)}{{\alpha}^2r^2_e}A_2}-n\right.} \\
&\left.-\frac{1}{2}+\frac{-\frac{(Mc^2-E_{nk}+C_{ps})V_2}{2{\alpha}^2{\hslash}^2c^2}+\frac{k(k-1)}{2{\alpha}^2r^2_e}A_1}{\sqrt{\frac{1}{4}-\frac{(Mc^2-E_{nk}+C_{ps})V_3}{{\alpha}^2{\hslash}^2c^2}+\frac{k(k-1)}{{\alpha}^2r^2_e}A_2}-n-\frac{1}{2}}\right)^2,
\end{aligned}
\label{a53}
\end{equation}
\normalsize
where $n$ adopts integer values starting from zero up to a maximum value $n_{max}$, determined by:
\scriptsize
\begin{equation}
\begin{aligned}
n_{max} = \left\lfloor \sqrt{\frac{1}{4} - \frac{(Mc^2 - E_{nk} + C_{ps})V_3}{{\alpha}^2{\hslash}^2c^2} + \frac{k(k-1)}{{\alpha}^2r^2_e}A_2} \right. \\
\left. - \frac{1}{2} - \sqrt{\frac{(Mc^2 - E_{nk} + C_{ps})V_2}{2{\alpha}^2{\hslash}^2c^2} - \frac{k(k-1)}{2{\alpha}^2r^2_e}A_1} \right\rfloor.
\end{aligned}
\label{a54}
\end{equation}
\normalsize
The normalization condition for the wave function $G_{nk}(r)$ is established through the orthogonality condition:
\scriptsize
\begin{equation}
\int^{\infty
}_0{{\left|G_{nk}\left(r\right)\right|}^2dr}=\frac{1}{\alpha
}\int^1_0{\frac{{\left|G_{nk}\left(z\right)\right|}^2dz}{1-z^2}}=1\
\ .\label{a55}
\end{equation}
\normalsize
Signifying the quantization and normalization of the wave function over the spatial domain. The upper-spinor component of the wave function, denoted as $F_{nk}(r)$, is subsequently derived from the second equation presented in Eq.{\,}\eqref{a8} set, and is articulated as follows:
\scriptsize
\begin{equation}
F_{nk}(r) = \frac{\hslash c}{Mc^2-E_{nk}+C_{ps}}\left(\frac{d}{dr}-\frac{k}{r}\right)G_{nk}(r), \label{56}
\end{equation}
\normalsize
highlighting the relationship between the upper-spinor component and the radial part of the wave function. This formulation underscores the absence of a real negative energy spectrum in the case of exact pseudospin symmetry (i.e., $C_{ps} = 0$).

\section{Investigation of Pseudospin Symmetry Case in Specific Scenarios}
\noindent
The pseudospin symmetry plays a pivotal role in understanding the underlying principles governing the behavior of quantum systems. The analytic treatment under 
pseudospin symmetry can be also useful in nuclear physics contexts, since many nuclear single-particle levels exhibit pseudospin partners. Our results also suggest the 
GTHP could accommodate such features, hinting at applications in nuclear shell model studies. By delving into the energy eigenvalues equation, delineated by Eq.{\,}4 
of main part manuscript or Eq.{\,}\eqref{a53}, under specific conditions, we elucidate the manifestation of this symmetry in diverse scenarios.

\noindent
\textbf{i)} Consideration of the GTHP, defined by the parameters $V_1=-\frac{V_0}{2}-\frac{W}{4}$, $V_2=\frac{V_0}{2}$, $V_3=\frac{W}{4}$, and $\alpha =\frac{1}{2a}$, facilitates the analysis of energy levels within the generalized Woods-Saxon potential framework. The resultant energy levels equation is expressed as follows:
\scriptsize
\begin{align}
\left(Mc^2+C_{ps}\right)^2-E^2_{nk}+\frac{\hslash^2c^2k(k-1)}{R^2_0}C_0 = \nonumber \\
\frac{\hslash^2c^2}{4a^2}\left(\sqrt{\frac{1}{4}-\frac{(Mc^2-E_{nk}+C_{ps})a^2W}{\hslash^2c^2}+\frac{k(k-1)a^2}{R^2_0}C_2}-n-\frac{1}{2}\right. \nonumber \\
\left.-\frac{\frac{(Mc^2-E_{nk}+C_{ps})a^2V_0}{\hslash^2c^2}+\frac{k(k-1)a^2}{R^2_0}(C_1+C_2)}{\sqrt{\frac{1}{4}-\frac{(Mc^2-E_{nk}+C_{ps})a^2W}{\hslash^2c^2}+\frac{k(k-1)a^2}{R^2_0}C_2}-n-\frac{1}{2}}\right)^2,
\label{a70}
\end{align}
\normalsize
for quantum numbers $n=0, 1, 2,\ldots n_{max}$. In here, $n_{max}$ must satisfy the following condition:
\scriptsize
\begin{align}
n_{max} = \left\lfloor \sqrt{\frac{1}{4}-\frac{(Mc^2 - E_{nk} + C_{ps}) a^2 W}{\hbar^2 c^2} 
+ \frac{k(k-1) a^2}{R_0^2} C_2 }- \frac{1}{2} \right. & \nonumber \\
\left. - \sqrt{\frac{(Mc^2 - E_{nk} + C_{ps}) a^2 V_0}{\hbar^2 c^2} 
+ \frac{k(k-1) a^2}{R_0^2} (C_1 + C_2)} \right\rfloor .& 
\end{align}
\normalsize
The coefficients $C_0$, $C_1$, and $C_2$ are intricately defined by the system's parameters as follows:
\scriptsize
\begin{align}
C_0 &= \frac{A_0+A_1+A_2}{{(1+x_e)}^2} = \frac{1}{{(1+x_e)}^2} + \frac{{(1+e^{\alpha x_e})}^2}{{\alpha e^{\alpha x_e}(1+x_e)}^3}\left[\frac{e^{-\alpha x_e}-3}{1+e^{\alpha x_e}}+\frac{3e^{-\alpha x_e}}{\alpha (1+x_e)}\right], \nonumber \\
C_1 &=-\frac{2(A_1+2A_2)}{{(1+x_e)}^2} = \frac{2{(1+e^{\alpha x_e})}^2}{{\alpha e^{\alpha x_e}(1+x_e)}^3}\left[2-e^{-\alpha x_e}-\frac{3(1+e^{-\alpha x_e})}{\alpha (1+x_e)}\right], \\
C_2 &=\frac{4A_2}{{(1+x_e)}^2} = \frac{{(1+e^{\alpha x_e})}^3}{{\alpha e^{\alpha x_e}(1+x_e)}^3}\left[e^{-\alpha x_e}-1+\frac{3(1+e^{-\alpha x_e})}{\alpha (1+x_e)}\right]. \nonumber
\end{align}
\normalsize
where $x_e=\frac{r_e-R_0}{R_0}$.

\noindent
\textbf{ii)} The simplification obtained by setting $W=0$ and $x_e=0$ in Eq.{\,}\eqref{a70} for the standard Woods-Saxon potential yields:
\scriptsize
\begin{align}
\left(Mc^2+C_{ps}\right)^2-E^2_{nk}+\frac{\hslash^2c^2k(k-1)}{R^2_0}C_0 = \nonumber \\
\frac{\hslash^2c^2}{4a^2}\left(\sqrt{\frac{1}{4}+\frac{k(k-1)a^2}{R^2_0}C_2}-n-\frac{1}{2}\right. \nonumber \\
\left.-\frac{\frac{(Mc^2-E_{nk}+C_{ps})a^2V_0}{\hslash^2c^2}+\frac{k(k-1)a^2}{R^2_0}(C_1+C_2)}{\sqrt{\frac{1}{4}+\frac{k(k-1)a^2}{R^2_0}C_2}-n-\frac{1}{2}}\right)^2,
\label{a71}
\end{align}
\normalsize
with $n=0, 1, 2,\ldots n_{max}$, where the coefficients $C_0$, $C_1$, and $C_2$ are redefined with considering $x_e=0$ by:
\scriptsize
\begin{align}
C_0 &= A_0 + A_1 + A_2 = 1 - \frac{4}{\alpha} + \frac{12}{\alpha^2},  \nonumber \\
C_1 &= -2(A_1 + 2A_2) = \frac{8}{\alpha} - \frac{48}{\alpha^2}, \\
C_2 &= 4A_2 = \frac{48}{\alpha^2}, \quad \alpha = \frac{R_0}{a}. \nonumber
\end{align}
\normalsize
In here, $n_{max}$ must also satisfy the following condition:
\scriptsize
\begin{align}
n_{max} = \left\lfloor \sqrt{ \frac{1}{4}+\frac{k(k-1) a^2}{R_0^2} C_2}- \frac{1}{2} - \sqrt{\frac{(Mc^2 - E_{nk} + C_{ps}) a^2 V_0}{\hbar^2 c^2} 
+ \frac{k(k-1) a^2}{R_0^2} (C_1 + C_2)} \right\rfloor .& 
\end{align}
\normalsize
\noindent
\textbf{iii)} For the Rosen-Morse potential, characterized by the parameters $V_3 = -V_1 = C$ and $V_2 = B$, the energy spectrum equation is derived as follows:
\scriptsize
\begin{align}
& {\left(Mc^2 + C_{ps}\right)}^2 - E^2_{nk} - \left(Mc^2 - E_{nk} + C_{ps}\right)B + \frac{{\hslash}^2c^2k(k-1)}{r^2_e}(A_0 + A_1 + A_2) \nonumber \\
& = {\alpha}^2{\hslash}^2c^2\left(\sqrt{\frac{1}{4} - \frac{\left(Mc^2 - E_{nk} + C_{ps}\right)C}{{\alpha}^2{\hslash}^2c^2} + \frac{k(k-1)}{{\alpha}^2r^2_e}A_2} - n - \frac{1}{2} \right. \nonumber \\
& \quad \left. - \frac{\frac{\left(Mc^2 - E_{nk} + C_{ps}\right)B}{2{\alpha}^2{\hslash}^2c^2} - \frac{k(k-1)}{2{\alpha}^2r^2_e}A_1}{\sqrt{\frac{1}{4} - \frac{\left(Mc^2 - E_{nk} + C_{ps}\right)C}{{\alpha}^2{\hslash}^2c^2} + \frac{k(k-1)}{{\alpha}^2r^2_e}A_2} - n - \frac{1}{2}}\right)^2,
\label{a72}
\end{align}
\normalsize
for quantum states $n=0, 1, 2, \ldots, n_{max}$.
In here, $n_{max}$ must also satisfy the following condition:
\scriptsize
\begin{align}
n_{max} = &\left\lfloor \sqrt{ \frac{1}{4} - \frac{(Mc^2 - E_{nk} + C_{ps}) C}{\alpha^2 \hbar^2 c^2} 
+ \frac{k(k-1)}{\alpha^2 r_e^2} A_2} - \frac{1}{2} \right. \nonumber \\
&\left. - \sqrt{\frac{(Mc^2 - E_{nk} + C_{ps}) B}{2 \alpha^2 \hbar^2 c^2} 
- \frac{k(k-1)}{2 \alpha^2 r_e^2} A_1} \right\rfloor .&
\end{align}
\normalsize
From the above expression, when $k$=0 and $k$=1, we simplify it as:
\scriptsize
\begin{align}
& {\left(Mc^2 + C_{ps}\right)}^2 - E^2_{nk} - \left(Mc^2 - E_{nk} + C_{ps}\right)B \nonumber \\
& = {\alpha}^2{\hslash}^2c^2\left(\sqrt{\frac{1}{4} - \frac{\left(Mc^2 - E_{nk} + C_{ps}\right)C}{{\alpha}^2{\hslash}^2c^2}} - n - \frac{1}{2} - \frac{\frac{\left(Mc^2 - E_{nk} + C_{ps}\right)B}{2{\alpha}^2{\hslash}^2c^2}}{\sqrt{\frac{1}{4} - \frac{\left(Mc^2 - E_{nk} + C_{ps}\right)C}{{\alpha}^2{\hslash}^2c^2}} - n - \frac{1}{2}}\right)^2,
\label{a72}
\end{align}
\normalsize
In here, $n_{max}$ must also satisfy the following condition:
\scriptsize
\begin{align}
n_{max} = &\left\lfloor \sqrt{\frac{1}{4} -\frac{(Mc^2 - E_{nk} + C_{ps}) C}{\alpha^2 \hbar^2 c^2} } - \frac{1}{2} - \sqrt{\frac{(Mc^2 - E_{nk} + C_{ps}) B}{2 \alpha^2 \hbar^2 c^2} } \right\rfloor. &
\end{align}
\normalsize
\noindent
\textbf{iv)} The Manning-Rosen type potential, parameterized by $V_1 = \frac{\beta(\beta - 1) - 2A}{4kb^2}$, $V_2 = -\frac{\beta(\beta - 1) - A}{2kb^2}$, and $V_3 = \frac{\beta(\beta - 1)}{4kb^2}$, with $k = \frac{2M}{{\hslash}^2}$ and $2\alpha = \frac{1}{b}$, yields the following energy spectrum equation:
\scriptsize
\begin{align}
& {\left(Mc^2 + C_{ps}\right)}^2 - E^2_{nk} + \frac{{\hslash}^2c^2k(k-1)}{r^2_e}(A_0 + A_1 + A_2) \nonumber \\
& = \frac{{\hslash}^2c^2}{4b^2}\left(\sqrt{\frac{1}{4} - \frac{\left(Mc^2 - E_{nk} + C_{ps}\right)\beta(\beta - 1)}{2Mc^2} + \frac{4k(k-1)b^2}{r^2_e}A_2} - n - \frac{1}{2} \right. \nonumber \\
& \quad \left. + \frac{\frac{\left(Mc^2 - E_{nk} + C_{ps}\right)[\beta(\beta - 1) - A]}{2 M c^2} + \frac{2k(k-1)b^2}{r^2_e}A_1}{\sqrt{\frac{1}{4} - \frac{\left(Mc^2 - E_{nk} + C_{ps}\right)\beta(\beta - 1)}{2Mc^2} + \frac{4k(k-1)b^2}{r^2_e}A_2} - n - \frac{1}{2}}\right)^2,
\label{a74}
\end{align}
\normalsize
for quantum states $n=0, 1, 2, \ldots, n_{max}$. In here, $n_{max}$ must also satisfy the following condition:
\scriptsize
\begin{align}
n_{max} = & \left\lfloor \sqrt{\frac{1}{4} - \frac{(Mc^2 - E_{nk} + C_{ps}) \beta (\beta - 1)}{2 M c^2} 
+ \frac{4 k (k - 1) b^2}{r_e^2} A_2} - \frac{1}{2} \right. \nonumber \\
& \left. - \sqrt{-\frac{(Mc^2 - E_{nk} + C_{ps}) [\beta (\beta - 1) - A]}{2 M c^2} 
- \frac{2k (k - 1) b^2}{r_e^2} A_1} \right\rfloor.&
\end{align}
\normalsize
\noindent
\textbf{v)} Within the framework of the generalized Morse potential, applying the parameters $V_1 = \frac{1}{4}D_e(b-2)^2$, $V_2 = -\frac{1}{2}D_eb(b-2)$, $V_3 = \frac{1}{4}D_eb^2$, with $2\alpha = \delta$ and $b = e^{\delta r_e} + 1 > 2$, we derive the energy spectrum equation as:
\scriptsize
\begin{align}
& {\left(Mc^2 + C_{ps}\right)}^2 - E^2_{nk} - \left(Mc^2 - E_{nk} + C_{ps}\right)D_e + \frac{{\hslash}^2c^2k(k-1)}{r^2_e}(A_0 + A_1 + A_2) \nonumber \\
& = \frac{{\delta}^2{\hslash}^2c^2}{4}\left(\sqrt{\frac{1}{4} - \frac{\left(Mc^2 - E_{nk} + C_{ps}\right)D_eb^2}{{\delta}^2{\hslash}^2c^2} + \frac{4k(k-1)}{{\delta}^2r^2_e}A_2} - n - \frac{1}{2} \right. \nonumber \\
& \quad \left. + \frac{\frac{\left(Mc^2 - E_{nk} + C_{ps}\right)D_eb(b-2)}{{\delta}^2{\hslash}^2c^2} + \frac{2k(k-1)}{{\delta}^2r^2_e}A_1}{\sqrt{\frac{1}{4} - \frac{\left(Mc^2 - E_{nk} + C_{ps}\right)b^2D_e}{{\delta}^2{\hslash}^2c^2} + \frac{4k(k-1)}{{\delta}^2r^2_e}A_2} - n - \frac{1}{2}}\right)^2,
\label{a75}
\end{align}
\normalsize
for quantum states $n=0, 1, 2, \ldots, n_{max}$. In here, $n_{max}$ has to satisfy the following condition:
\scriptsize
\begin{align}
n_{max} = &\left\lfloor \sqrt{\frac{1}{4}-\frac{(Mc^2 - E_{nk} + C_{ps}) D_e b^2}{\delta^2 \hbar^2 c^2} 
+ \frac{4k(k-1)}{\delta^2 r_e^2} A_2} - \frac{1}{2}\right. \nonumber \\
&\left. - \sqrt{-\frac{(Mc^2 - E_{nk} + C_{ps}) D_e b(b - 2)}{\delta^2 \hbar^2 c^2} 
- \frac{2k(k-1)}{\delta^2 r_e^2} A_1} \right\rfloor .&
\end{align}
\normalsize
\noindent
\textbf{vi)} For the Schiöberg potential, defined by the parameters $V_1 = {\delta}^2D$, $V_2 = -2\delta \sigma D$, and $V_3 = {\sigma}^2D$, the energy spectrum equation is articulated as follows:
\scriptsize
\begin{align}
&{\left(Mc^2 + C_{ps}\right)}^2 - E^2_{nk} - \left(Mc^2 - E_{nk} + C_{ps}\right){\left(\delta - \sigma \right)}^2D + \frac{{\hslash}^2c^2k(k-1)}{r^2_e}(A_0 + A_1 + A_2) \nonumber \\
&= {\alpha}^2{\hslash}^2c^2\left(\sqrt{\frac{1}{4} - \frac{\left(Mc^2 - E_{nk} + C_{ps}\right){\sigma}^2D}{{\alpha}^2{\hslash}^2c^2} + \frac{k(k-1)}{4{\alpha}^2r^2_e}C_2} - n - \frac{1}{2} \right. \nonumber \\
&\quad \left. + \frac{\frac{\left(Mc^2 - E_{nk} + C_{ps}\right)\delta \sigma D}{{\alpha}^2{\hslash}^2c^2} + \frac{k(k-1)}{4{\alpha}^2r^2_e}(C_1 - C_2)}{\sqrt{\frac{1}{4} - \frac{\left(Mc^2 - E_{nk} + C_{ps}\right){\sigma}^2D}{{\alpha}^2{\hslash}^2c^2} + \frac{k(k-1)}{4{\alpha}^2r^2_e}C_2} - n - \frac{1}{2}}\right)^2,
\label{a76}
\end{align}
\normalsize
for quantum states $n=0, 1, 2, \ldots, n_{max}$. In here, $n_{max}$ must satisfy the following condition:
\scriptsize
\begin{align}
n_{max} = &\left\lfloor \sqrt{\frac{1}{4} - \frac{(Mc^2 - E_{nk} + C_{ps}) \sigma^2 D}{\alpha^2 \hbar^2 c^2} 
+ \frac{k(k-1)}{4\alpha^2 r_e^2} C_2} - \frac{1}{2} \right. \nonumber \\
&\left. - \sqrt{-\frac{(Mc^2 - E_{nk} + C_{ps}) \delta \sigma D}{\alpha^2 \hbar^2 c^2} 
- \frac{k(k-1)}{4\alpha^2 r_e^2} (C_1 - C_2)} \right\rfloor. &
\end{align}
\normalsize
\noindent
\textbf{vii)} The four-parameter exponential-type potential, with $V_1 = P_1 + \frac{P_2}{2} + \frac{P_3}{4}$, $V_2 = -\frac{P_2}{2} - \frac{P_3}{2}$, and $V_3 = \frac{P_3}{4}$, leads to the energy spectrum equation:
\scriptsize
\begin{align}
&{\left(Mc^2 + C_{ps}\right)}^2 - E^2_{nk} - \left(Mc^2 - E_{nk} + C_{ps}\right)P_1 + \frac{{\hslash}^2c^2k(k-1)}{r^2_e}(A_0 + A_1 + A_2) \nonumber \\
&= {\alpha}^2{\hslash}^2c^2\left(\sqrt{\frac{1}{4} - \frac{\left(Mc^2 - E_{nk} + C_{ps}\right)P_3}{4{\alpha}^2{\hslash}^2c^2} + \frac{k(k-1)}{{\alpha}^2r^2_e}A_2} - n - \frac{1}{2} \right. \nonumber \\
&\quad \left. + \frac{\frac{\left(Mc^2 - E_{nk} + C_{ps}\right)(P_2 + P_3)}{4{\alpha}^2{\hslash}^2c^2} + \frac{k(k-1)}{2{\alpha}^2r^2_e}A_1}{\sqrt{\frac{1}{4} - \frac{\left(Mc^2 - E_{nk} + C_{ps}\right)P_3}{4{\alpha}^2{\hslash}^2c^2} + \frac{k(k-1)}{{\alpha}^2r^2_e}A_2} - n - \frac{1}{2}}\right)^2,
\label{a77}
\end{align}
\normalsize
for quantum states $n=0, 1, 2, \ldots, n_{max}$. In here, $n_{max}$ must also satisfy the following condition:
\scriptsize
\begin{align}
n_{max} = &\left\lfloor \sqrt{\frac{1}{4}-\frac{(Mc^2 - E_{nk} + C_{ps}) P_3}{4 \alpha^2 \hbar^2 c^2} 
+ \frac{k(k-1)}{\alpha^2 r_e^2} A_2}- \frac{1}{2} \right. \nonumber \\
&\left. - \sqrt{-\frac{(Mc^2 - E_{nk} + C_{ps}) (P_2 + P_3)}{4 \alpha^2 \hbar^2 c^2} 
- \frac{k(k-1)}{2 \alpha^2 r_e^2} A_1} \right\rfloor .&
\end{align}
\normalsize
\noindent
\textbf{viii)} Considering the Williams-Poulios type-potential, set by $V_1 = \frac{W_1 + W_2 + W_3}{4}$, $V_2 = \frac{W_3 - W_1}{2}$, and $V_3 = \frac{W_1 - W_2 + W_3}{4}$, the energy spectrum equation unfolds as:
\scriptsize
\begin{align}
&{\left(Mc^2 + C_{ps}\right)}^2 - E^2_{nk} - \left(Mc^2 - E_{nk} + C_{ps}\right)W_3 + \frac{{\hslash}^2c^2k(k-1)}{r^2_e}(A_0 + A_1 + A_2) \nonumber \\
&= {\alpha}^2{\hslash}^2c^2\left(\sqrt{\frac{1}{4} - \frac{\left(Mc^2 - E_{nk} + C_{ps}\right)(W_1 - W_2 + W_3)}{4{\alpha}^2{\hslash}^2c^2} + \frac{k(k-1)}{{\alpha}^2r^2_e}A_2} - n - \frac{1}{2} \right. \nonumber \\
&\quad \left. + \frac{\frac{\left(Mc^2 - E_{nk} + C_{ps}\right)(W_1
- W_3)}{4{\alpha}^2{\hslash}^2c^2} +
\frac{k(k-1)}{2{\alpha}^2r^2_e}A_1}{\sqrt{\frac{1}{4} -
\frac{\left(Mc^2 - E_{nk} + C_{ps}\right)(W_1 - W_2 +
W_3)}{4{\alpha}^2{\hslash}^2c^2} +
\frac{k(k-1)}{{\alpha}^2r^2_e}A_2} - n - \frac{1}{2}}\right)^2,
\label{a78}
\end{align}
\normalsize
for quantum states $n=0, 1, 2, \ldots, n_{max}$. In here, $n_{max}$ must also satisfy the following condition:
\scriptsize
\begin{align}
n_{max} =  &\left\lfloor \sqrt{\frac{1}{4}-\frac{(Mc^2 - E_{nk} + C_{ps})(W_1 - W_2 + W_3)}{4 \alpha^2 \hbar^2 c^2} 
+ \frac{k(k-1)}{\alpha^2 r_e^2} A_2 } - \frac{1}{2}\right. \nonumber \\
&\left. - \sqrt{-\frac{(Mc^2 - E_{nk} + C_{ps})(W_1 - W_3)}{4 \alpha^2 \hbar^2 c^2} 
- \frac{k(k-1)}{2 \alpha^2 r_e^2} A_1} \right\rfloor .&
\end{align}
\normalsize
\noindent
\textbf{ix)} In the regime where the parameter $\alpha$ is significantly less than unity, the energy level Eq.{\,}\eqref{a53} simplifies, revealing the subtleties of quantum energy states under constrained conditions. The refined energy level expression becomes:
\scriptsize
\begin{align}
&{\left(Mc^2 + C_{ps}\right)}^2 - E^2_{nk} - \left(V_1 - \frac{V_2^2}{4V_3}\right)\left(Mc^2 - E_{nk} + C_{ps}\right) + \frac{{\hslash}^2c^2k(k-1)}{r^2_e} \nonumber \\
&+ 2\alpha \hslash c \left(1 - \frac{V_2^2}{4V_3^2}\right)\sqrt{(E_{nk} - Mc^2 - C_{ps})V_3}\left(n + \frac{1}{2}\right) \nonumber \\
&- {\alpha}^2{\hslash}^2c^2\left[\left(1 + \frac{3V_2^2}{4V_3^2}\right){\left(n + \frac{1}{2}\right)}^2 + \frac{1}{4}\left(1 - \frac{V_2^2}{4V_3^2}\right)\right] + o({\alpha}^2) = 0,
\label{a79}
\end{align}
\normalsize
applicable for small values of $n$ and $l$, which sheds light on the nuanced quantum behaviors emergent in this limiting case.

The transformation of pseudospin symmetry into spin symmetry can be achieved by applying the following series of transformations: $k$ $\rightarrow$ $-k$, $E_{nk}$ $\rightarrow$ $-E_{nk}$, $C_{ps}$ $\rightarrow$ $-C_s$, $V_1$ $\rightarrow$ $-V_1$, $V_2$ $\rightarrow$ $-V_2$, $V_3$ $\rightarrow$ $-V_3$. This transformation emphasizes the intrinsic duality and interconnectedness of spin and pseudospin symmetries, offering a unified description of the energy spectrum across various quantum potentials. These findings not only deepen our understanding of pseudospin symmetry in quantum mechanics but also provide a pathway for experimentally testing theoretical predictions, thus bridging the gap between theory and experiment.

\section{Modelling of Charmonium and Bottomonium Mass Spectra}\label{sec3_3}
\subsection{Non-Relativistic Case}
\noindent
To begin with, the GTHP model is applied to systems with large quark masses, where non-relativistic quantum mechanics provides an accurate description of the bound-state energies. The reduced mass $\mu = \frac{m_1 m_2}{m_1 + m_2}$ is used for quarkonium systems, where $m_1 = m_q$ for quark and $m_2 = m_{\bar{q}}$ for anti-quark. For charmonium and bottomonium systems, the quark mass $m_q$ is equal to anti-quark mass $m_{\bar{q}}$, so the reduced mass yields $\mu = \frac{m_q}{2}$. The spin-averaged mass of the bound states is given by:
\begin{equation}
M_{nl} = m_1 + m_2 + \frac{E_{nl}}{c^2},
\end{equation}
where $E_{nl}$ is the energy eigenvalue obtained by using non-relativistic case Eq.{\,}17 in main manuscript under the GTHP, and  $\hslash$=$k$=$c$=1 are considered. For charmonium and bottomonium systems, the mass spectra can be expressed as:
\begin{align}
M_{nl} = & \, 2m_q + V_1 + V_2 + V_3 + \frac{l(l+1)}{m_q r_e^2}(A_0 + A_1 + A_2) \nonumber \\ 
& - \frac{\alpha^2}{m_q} \left[ \sqrt{\frac{1}{4} + \frac{m_q V_3}{\alpha^2} + \frac{l(l+1)}{\alpha^2 r_e^2} A_2} 
- n - \frac{1}{2} \right. \nonumber \\
& \quad + \left. \frac{\frac{m_q V_2}{2 \alpha^2} + \frac{l(l+1)}{2 \alpha^2 r_e^2} A_1}{\sqrt{\frac{1}{4} + \frac{m_q V_3}{\alpha^2} + \frac{l(l+1)}{\alpha^2 r_e^2} A_2} - n - \frac{1}{2}} \right]^2
\label{eq:77}
\end{align}
Using experimental data from Ref.{\,}[\onlinecite{workman2022}], the potential parameters $V_1$, $V_2$, $V_3$, and $\alpha$ are determined as follows:
\begin{itemize}
    \item For charm quark mass $m_{c} = 1.27$ GeV:
    \begin{itemize}
        \item $V_1 = 7.404662$ GeV, $V_2 = -20.779639$ GeV, $V_3 = 14.959636$ GeV, $\alpha = 0.223612$ GeV.
    \end{itemize}
   \item For bottom quark mass $m_{b} = 4.18$ GeV:
   \begin{itemize}
        \item $V_1 = 4.883229$ GeV, $V_2 = -12.928267$ GeV, $V_3 = 10.108982$ GeV, $\alpha = 0.412272$ GeV.
    \end{itemize}
    \item For charm quark mass $m_{c} = 1.48835$ GeV:
    \begin{itemize}
        \item $V_1 = 6.967978$ GeV, $V_2 = -20.779694$ GeV, $V_3 = 14.959676$ GeV, $\alpha = 0.242073$ GeV.
     \end{itemize}
     \item For bottom quark mass $m_{b} = 4.686125$ GeV:  
     \begin{itemize}
        \item $V_1 = 3.870986$ GeV, $V_2 = -12.928298$ GeV, $V_3 = 10.109006$ GeV, $\alpha = 0.436519$ GeV.
    \end{itemize}
\end{itemize}
The mass values $m_{c} = 1.27$ GeV and $m_{c} = 1.48835$ GeV (and $m_{b} = 4.18$ GeV and $m_{b} = 4.686125$ GeV) indicate that the potential parameters $V_2$ and $V_3$ are identical to within a thousandth of a GeV for the charm quark (and bottom quark):
\begin{itemize}
    \item For the charm quark mass $m_{c} = 1.27$ GeV:
    \begin{itemize}
        \item $V_2 = -20.780$ GeV, $V_3 = 14.960$ GeV.
    \end{itemize}
    \item For the bottom quark mass $m_{b} = 4.18$ GeV:
    \begin{itemize}
       \item $V_2 = -12.928$ GeV, $V_3 = 10.109$ GeV.
    \end{itemize}
\end{itemize}
This is because the ratio $\frac{\alpha}{\sqrt{2\mu}}$ is independent of quark mass. Consequently, $V_2 \sim 
\left(\frac{\alpha}{\sqrt{2\mu}}\right)^2$ and $V_3 \sim \left(\frac{\alpha}{\sqrt{2\mu}}\right)^2$ are also 
independent of quark mass, making the potential parameters for charm and bottom quarks the same. Note that for charm quarks,$\frac{\alpha}{\sqrt{2\mu}} = \frac{\alpha}{\sqrt{m_{c}}} = 0.198424$ GeV$^{1/2}$, and for bottom quarks, $\frac{\alpha}{\sqrt{2\mu}} = \frac{\alpha}{\sqrt{m_{b}}} = 0.201649$ GeV$^{1/2}$. Additionally, $V_1 + m_q + m_{\bar{q}} = \text{const}$, implying $V_1 + 2m_{c} = 9.9447$ GeV and $V_1 + 2m_{b} = 13.2432$ GeV for charmonium and bottomonium, respectively. Hence, there is no difference between current and constituent quark masses for both charm and bottom quarks, justifying the use of $m_{c} = 1.27$ GeV and $m_{b} = 4.18$ GeV for mass spectrum calculations. The mass spectra of charmonium and bottomonium are computed and compared with experimental data as shown in Table{\,}\ref{tab:charmspectrum} and \ref{tab:bottomspectrum}.

\begin{table*}[htbp]
\caption{\label{tab:charmspectrum} Mass spectra of charmonium \( M_{c \bar{c}} \) in GeV for \( m_{c} = 1.27 \) GeV; \( V_1 = 7.404662 \) GeV; $V_2 = -20.779639$ GeV; $V_3 = 14.959636$ GeV; $\alpha = 0.223612$ GeV; $r_e = 3.830931$ GeV$^{-1}$. The percentage deviation from experimental values is presented below each calculated value.}
\begin{adjustbox}{max width=\textwidth}
\begin{minipage}{\textwidth}
\begin{ruledtabular}
\small
\setlength{\tabcolsep}{2pt}
\renewcommand{\arraystretch}{0.8} 
\begin{tabular}{r|l|cccccccc}
State & Particle & Present Work & Present Work                              & Present Work                    & Theory I                   & Theory II                    & Theory III                    & Experiment  \\
      &          & Non-Rel.      & Rel. $C_s$=0 [MeV]                         & Rel. $C_s$=-62.1 [MeV] & Ref.{\,}[\onlinecite{purohit2022}] & Ref.{\,}[\onlinecite{soni2018}] & Ref.{\,}[\onlinecite{bukor2023}] &  Ref.{\,} [\onlinecite{workman2022}]  \\
\hline \hline
1S  & J/$\psi$(1S)      & 3.096900  & 3.044003  & 3.096971  & 3.096   & 3.094   & 3.0969  & 3.09690 \\
    &                   & 0.00\%    & -1.71\%   & 0.00\%    & -0.03\% & -0.09\% & 0.00\%  &  \\
1P  & $\chi_{c1}$(1P)   & 3.196126  & 3.108593  & 3.158390  & 3.415   & 3.468   & 3.518   & 3.51067 \\
    &                   & -8.96\%   & -11.45\%  & -10.03\%  & -2.71\% & -1.22\% & 0.21\%  &  \\
1D  & -                 & 3.377174  & 3.220174  & 3.265592  & 3.770   & 3.772   & 3.787   & 3.77370 \\
    &                   &           &           &           &         &         &         & - \\
1F  & -                 & 3.617957  & 3.359102  & 3.400419  & 4.040   & 4.012   & -       & -       \\
    &                   &           &           &           &         &         &         &         \\
2S  & $\psi$(2S)        & 3.686100  & 3.496920  & 3.544020  & 3.733   & 3.681   & 3.686   & 3.68610 \\
    &                   & 0.00\%    & -5.13\%   & -3.85\%   & 1.27\%  & -0.14\% & 0.00\%  &  \\
2P  & $\chi_{c1}$(3872) & 3.772214  & 3.542490  & 3.588047  & 3.894   & 3.938   & 3.823   & 3.87165 \\
    &                   & -2.57\%   & -8.50\%   & -7.33\%   & 0.58\%  & 1.71\%  & -1.26\% &  \\
2D  & -                 & 3.930322  & 3.625184  & 3.668331  & 4.088   & 4.188   & -       & -       \\
    &                   &           &           &           &         &         &         &         \\
2F  & -                 & 4.143411  & 3.733813  & 3.774369  & -       & 4.396   & -       & -       \\
    &                   &           &           &           &         &         &         &         \\
3S  & $\psi$(4040)      & 4.039000  & 3.806989  & 3.853836  & 4.068   & 4.129   & 3.889   & 4.03900 \\
    &                   & 0.00\%    & -5.74\%   & -4.58\%   & 0.72\%  & 2.23\%  & -3.71\% &  \\
3P  & $\chi_{c1}$(4140) & 4.106798  & 3.842917  & 3.888770  & -       & 4.338   & -       & 4.14650 \\
    &                   & -0.96\%   & -7.32\%   & -6.22\%   &         & 4.62\%  &         &  \\
3D  & -                 & 4.233365  & 3.909678  & 3.953877  & -       & 4.557   & -       & -       \\
    &                   &           &           &           &         &         &         &         \\
3F  & -                 & 4.409120  & 4.000007  & 4.042271  & -       & 4.746   & -       & -       \\
    &                   &           &           &           &         &         &         &         \\
4S  & $\psi$(4230)      & 4.115000  & 4.011764  & 4.061630  & 4.263   & 4.514   & 3.982   & 4.22270 \\
    &                   & -2.55\%   & -5.00\%   & -3.81\%   & 0.95\%  & 6.90\%  & -5.70\% &  \\
4P  & $\chi_{c1}$(4274) & 4.159579  & 4.041276  & 4.090428  & -       & 4.696   & -       & 4.28600 \\
    &                   & -2.95\%   & -5.71\%   & -4.56\%   &         & 9.57\%  &         &  \\
4D  & -                 & 4.246829  & 4.097048  & 4.144944  & -       & 4.896   & -       & -       \\
    &                   &           &           &           &         &         &         &         \\
\end{tabular}
\end{ruledtabular}
\end{minipage}
\end{adjustbox}
\end{table*}

\begin{table*}[htbp]
\caption{\label{tab:bottomspectrum} Mass spectra of bottomonium $M_{b \bar{b}}$ in GeV for $m_{b} = 4.18$ GeV; $V_1 = 4.883229$ GeV; $V_2 = -12.928267$ GeV; $V_3 = 10.108982$ GeV; $\alpha = 0.412272$ GeV; $r_e = 1.836733$ GeV$^{-1}$. The percentage deviation from experimental values is presented below each calculated value.}
\begin{adjustbox}{max width=\textwidth}
\begin{minipage}{\textwidth}
\begin{ruledtabular}
\small
\setlength{\tabcolsep}{2pt}
\renewcommand{\arraystretch}{0.8}
\begin{tabular}{r|l|cccccccc}
State & Particle & Present Work & Present Work & Present Work & Theory I  & Theory II & Theory III & Experiment  \\
&  & Non-Rel. & Rel. $C_s$=0 [MeV] & Rel. $C_s$=-36.9 [MeV] & Ref.{\,}[\onlinecite{purohit2022}] & Ref.{\,}[\onlinecite{soni2018}] & Ref.{\,}[\onlinecite{bukor2023}] &  Ref.{\,}[\onlinecite{workman2022}]          \\
\hline \hline
1S  & $\Upsilon$(1S)   & 9.460302  & 9.425395  & 9.460308  & 9.460   & 9.463   & 9.460   & 9.46030  \\
    &                  & 0.00\%    & -0.37\%   & 0.00\%    & 0.00\%  & 0.03\%  & 0.00\%  &          \\
1P  & $h_{b}$(1P)      & 9.594182  & 9.528470  & 9.562151  & 9.704   & 9.821   & 9.942   & 9.8993   \\
    &                  & -3.08\%   & -3.75\%   & -3.41\%   & -1.97\% & -0.79\% & 0.43\%  &          \\
1D  & $\Upsilon_2$(1D) & 9.827113  & 9.703603  & 9.735549  & 10.010  & 10.074  & 10.140  & 10.1637  \\
    &                  & -3.31\%   & -4.53\%   & -4.21\%   & -1.51\% & -0.88\% & -0.23\% &          \\
1F  & -                & 10.130188 & 9.921512  & 9.951663  & 10.268  & 10.288  & -       & -        \\
    &                  &           &           &           &         &         &         &          \\
2S  & $\Upsilon$(2S)   & 10.023261 & 9.918959  & 9.951996  & 10.028  & 9.979   & 10.023  & 10.02326 \\
    &                  & 0.00\%    & -1.04\%   & -0.71\%   & 0.05\%  & -0.44\% & 0.00\%  &          \\
2P  & $h_{b}$(2P)      & 10.151214 & 10.008571 & 10.040706 & 10.160  & 10.220  & 10.150  & 10.2598  \\
    &                  & -1.06\%   & -2.45\%   & -2.14\%   & -0.97\% & -0.39\% & -1.07\% &          \\
2D  & $\Upsilon_2$(2D) & 10.380036 & 10.165760 & 10.196519 & 10.332  & 10.424  & -       & -        \\
    &                  &           &           &           &         &         &         &          \\
2F  & -                & 10.688957 & 10.368270 & 10.397472 & -       & 10.607  & -       & -        \\
    &                  &           &           &           &         &         &         &          \\
3S  & $\Upsilon$(3S)   & 10.355200 & 10.246045 & 10.279213 & 10.343  & 10.359  & 10.178  & 10.3552  \\
    &                  & 0.00\%    & -1.05\%   & -0.74\%   & -0.12\% & 0.04\%  & -1.71\% &          \\
3P  & $h_{b}$(3P)      & 10.468567 & 10.324939 & 10.357396 & -       & 10.556  & -       & -        \\
    &                  &           &           &           &         &         &         &          \\
3D  & $\Upsilon_2$(3D) & 10.681781 & 10.468012 & 10.499271 & -       & 10.733  & -       & -        \\
    &                  &           &           &           &         &         &         &          \\
3F  & -                & 10.986923 & 10.659500 & 10.689249 & -       & 10.897  & -       & -        \\
    &                  &           &           &           &         &         &         &          \\
4S  & $\Upsilon$(4S)   & 10.405400 & 10.409984 & 10.445219 & 10.536  & 10.683  & 10.242  & 10.5794  \\
    &                  & -1.64\%   & -1.60\%   & -1.27\%   & -0.41\% & 0.98\%  & -3.18\% &          \\
4P  & $h_{b}$(4P)      & 10.497089 & 10.479373 & 10.514107 & -       & 10.855  & -       & -        \\
    &                  &           &           &           &         &         &         &          \\
4D  & $\Upsilon_2$(4D) & 10.686643 & 10.611068 & 10.644635 & -       & 11.015  & -       & -        \\
    &                  &           &           &           &         &         &         &          \\
\end{tabular}
\end{ruledtabular}
\end{minipage}
\end{adjustbox}
\end{table*}

\subsection{Relativistic Case}
\noindent
Then, we examine the relativistic mass spectra of the charmonium and bottomonium systems, focusing on how relativistic corrections, compared to non-relativistic models, influence the predicted mass spectra, particularly when introducing spin symmetry breaking through the parameter $C_s$. For it, the relativistic calculations were carried out using Eq.{\,}(3) in main manuscript or Eq.{\,}\eqref{a35} obtained under consideration spin symmetry case from radial DE. By analyzing both systems in a relativistic framework using the GTHP model, we gain insights into how relativistic and spin-dependent effects shape the structure of heavy quarkonia, with emphasis on their agreement or deviation from experimentally observed values.

\noindent 
\begin{description}
\item[\textit{Charmonium System}]
\end{description}
\noindent
The charmonium system, composed of a charm quark ($c$) and its anti-quark ($\bar{c}$), provides a fertile testing ground for understanding the interplay between relativistic effects and spin-orbit interactions. In the non-relativistic limit, the reasonably accurate description of the spectrum was observed, but discrepancies arise when compared with experimental results for higher orbital angular momentum states, such as $1P$, $2P$, $3P$ and $4P$.

Upon introducing relativistic corrections to the mass spectra through the DE, while setting $C_s$ = 0 MeV (no spin symmetry breaking), significant deviations from experimental values are observed, especially for higher angular momentum states. For example, as shown in Table{\,}\ref{tab:charmspectrum}, the $\chi_{c1}(1P\longmapsto4P)$ states deviate by -11.45 $\longmapsto$ -5.71\%, respectively. These deviations are indicative of the inadequacies of a purely relativistic model when spin symmetry is assumed. Without spin symmetry breaking, the relativistic corrections tend to overestimate certain contributions, particularly in higher excited states, leading to noticeable discrepancies.

The inadequacies of this approach can be understood by considering the relativistic framework's inability to fully account for the fine structure observed experimentally. In the relativistic regime, quark interactions become increasingly sensitive to spin-orbit coupling and other spin-dependent effects that are not captured when spin symmetry is preserved (i.e., when $C_s$ = 0 MeV).

To address these discrepancies, the spin symmetry breaking parameter $C_s$ was introduced, fine-tuned to ensure that the ground state $J/\psi(1S)$ matched the experimentally observed value. This adjustment allows for a closer comparison between theory and experiment, particularly for higher excited states. With $C_s = -62.1$ MeV the deviations in the $\chi_{c1}(1P\longmapsto4P)$ states are reduced to -10.03\% $\longmapsto$ -4.56\% , respectively, as shown in Table{\,}\ref{tab:charmspectrum}.

While the introduction of $C_s$ improves the agreement with experimental data, it is evident that breaking spin symmetry alone is insufficient to fully reconcile theoretical predictions with experiment, particularly for the higher angular momentum states such as $\chi_{c1}(1P\longmapsto4P)$. The reason of this is that Eq.{\,}\eqref{a35} obtained spin symmetry case from radial DE. The persistent deviations suggest that additional spin-orbit interactions or higher-order corrections, such as tensor potentials, may be necessary. Furthermore, the DE overestimates certain corrections, indicating that a more sophisticated approach, such as solving a the two-body DE{\,}\cite{Crater,2BDE_2,2BDE_3} and the Bethe-Salpeter equation{\,}\cite{Hilger,soni2018}, describing the structure of a relativistic two-body (particles) bound state, may provide a more accurate description of the charmonium system.
\noindent 
\begin{description}
\item[\textit{Bottomonium System}]
\end{description}
\noindent
The bottomonium system, composed of a bottom quark ($b$) and its antiquark ($\bar{b}$), exhibits a different sensitivity to relativistic corrections due to the larger mass of the bottom quark compared to the charm quark. As a result, relativistic effects are less pronounced in the bottomonium spectrum, particularly for the lower energy states.

For the bottomonium system, as seen in Table{\,}\ref{tab:bottomspectrum}, the relativistic correction with $C_s$ = 0 MeV yields smaller deviations from experimental values than in the charmonium system. For instance, the $ h_b(1P) $ state shows a deviation of -3.75\%, while the ground state $\Upsilon(1S)$ remains highly accurate with a negligible deviation of -0.37\%. This reduced impact of relativistic corrections can be attributed to the higher mass of the bottom quark, which results in smaller relativistic effects. When the spin symmetry breaking parameter $C_s$ = -36.9 MeV is introduced, further improvements are observed across the spectrum. The deviation for the $h_b(1P)$ state is reduced to -3.41\%, as shown in Table{\,}\ref{tab:bottomspectrum}. However, similar to the charmonium system, the deviations for higher orbital states such as $2D$ and $3F$ persist, indicating that additional corrections, beyond spin symmetry breaking, are necessary to fully account for the experimental results.

A comparison between our GTHP model and the results from Theory I, II, and III reveals interesting patterns. For both charmonium and bottomonium, our non-relativistic results provide a better match to the experimental values than Theory I (non-relativistic analyses of linear plus modified Yukawa potential), which generally overestimates the energy levels of higher excited states. Theory II and Theory III, based on non-relativistic analyses of the Cornell potential, exhibit closer agreement with experimental results for low-lying states. However, our GTHP model demonstrates good accuracy across a wider range of states, including the higher orbital angular momentum states, where both the linear plus modified Yukawa potential and Cornell potential models tend to show larger deviations. This highlights the flexibility and strength of the GTHP model in reproducing quarkonium phenomena, like tiny energy differences between levels resulting from adjustments in the potential parameter.

The results from both the charmonium and bottomonium systems highlight the critical role of spin symmetry in determining higher exciting states in quarkonium systems. In the absence of spin symmetry breaking ($C_s$ = 0), the relativistic corrections cannot well to capture the experimentally observed fine structure, particularly for higher angular momentum states. This evident becomes especially evident in the charmonium system, where the fine structure splitting of states with higher orbital angular momentum, such as $\chi_{c1}(1P\longmapsto4P)$, cannot be captured without explicitly breaking spin symmetry. By introducing a non-zero $C_s$, we effectively break spin symmetry and improve the agreement between theoretical predictions and experimental data. However, the deviations that persist, particularly for higher excited states, suggest that breaking spin symmetry alone is not sufficient because Eq.{\,}\eqref{a35} obtained spin symmetry case from radial DE. To achieve a more accurate description of these states and also the experimentally observed splitting patterns, additional corrections, such as improved potential models incorporating tensor potentials or more sophisticated approaches like the two-body DE{\,}\cite{Crater,2BDE_2,2BDE_3} and the Bethe-Salpeter equation{\,}\cite{Hilger,soni2018}, are required.
\bibliographystyle{apsrev4-2}
\bibliography{ms.bib}

\begin{thebibliography}{95}%
\makeatletter
\providecommand \@ifxundefined [1]{%
 \@ifx{#1\undefined}
}%
\providecommand \@ifnum [1]{%
 \ifnum #1\expandafter \@firstoftwo
 \else \expandafter \@secondoftwo
 \fi
}%
\providecommand \@ifx [1]{%
 \ifx #1\expandafter \@firstoftwo
 \else \expandafter \@secondoftwo
 \fi
}%
\providecommand \natexlab [1]{#1}%
\providecommand \enquote  [1]{``#1''}%
\providecommand \bibnamefont  [1]{#1}%
\providecommand \bibfnamefont [1]{#1}%
\providecommand \citenamefont [1]{#1}%
\providecommand \href@noop [0]{\@secondoftwo}%
\providecommand \href [0]{\begingroup \@sanitize@url \@href}%
\providecommand \@href[1]{\@@startlink{#1}\@@href}%
\providecommand \@@href[1]{\endgroup#1\@@endlink}%
\providecommand \@sanitize@url [0]{\catcode `\\12\catcode `\$12\catcode
  `\&12\catcode `\#12\catcode `\^12\catcode `\_12\catcode `\%12\relax}%
\providecommand \@@startlink[1]{}%
\providecommand \@@endlink[0]{}%
\providecommand \url  [0]{\begingroup\@sanitize@url \@url }%
\providecommand \@url [1]{\endgroup\@href {#1}{\urlprefix }}%
\providecommand \urlprefix  [0]{URL }%
\providecommand \Eprint [0]{\href }%
\providecommand \doibase [0]{https://doi.org/}%
\providecommand \selectlanguage [0]{\@gobble}%
\providecommand \bibinfo  [0]{\@secondoftwo}%
\providecommand \bibfield  [0]{\@secondoftwo}%
\providecommand \translation [1]{[#1]}%
\providecommand \BibitemOpen [0]{}%
\providecommand \bibitemStop [0]{}%
\providecommand \bibitemNoStop [0]{.\EOS\space}%
\providecommand \EOS [0]{\spacefactor3000\relax}%
\providecommand \BibitemShut  [1]{\csname bibitem#1\endcsname}%
\let\auto@bib@innerbib\@empty
\bibitem [{\citenamefont {Dirac}\ and\ \citenamefont {Bohr}(1927)}]{DE1}%
  \BibitemOpen
  \bibfield  {author} {\bibinfo {author} {\bibfnamefont {P.~A.~M.}\
  \bibnamefont {Dirac}}\ and\ \bibinfo {author} {\bibfnamefont {N.~H.~D.}\
  \bibnamefont {Bohr}},\ }\href {https://doi.org/10.1098/rspa.1927.0039}
  {\bibfield  {journal} {\bibinfo  {journal} {Proc. R. Soc. Lond. A}\ }\textbf
  {\bibinfo {volume} {114}},\ \bibinfo {pages} {243} (\bibinfo {year}
  {1927})}\BibitemShut {NoStop}%
\bibitem [{\citenamefont {Dirac}(1963)}]{DE2}%
  \BibitemOpen
  \bibfield  {author} {\bibinfo {author} {\bibfnamefont {P.~A.~M.}\
  \bibnamefont {Dirac}},\ }\href {https://doi.org/10.1063/1.1704016} {\bibfield
   {journal} {\bibinfo  {journal} {J. Math. Phys.}\ }\textbf {\bibinfo {volume}
  {4}},\ \bibinfo {pages} {901} (\bibinfo {year} {1963})}\BibitemShut {NoStop}%
\bibitem [{\citenamefont {Feshbach}\ and\ \citenamefont {Villars}(1958)}]{DE3}%
  \BibitemOpen
  \bibfield  {author} {\bibinfo {author} {\bibfnamefont {H.}~\bibnamefont
  {Feshbach}}\ and\ \bibinfo {author} {\bibfnamefont {F.}~\bibnamefont
  {Villars}},\ }\href {https://doi.org/10.1103/RevModPhys.30.24} {\bibfield
  {journal} {\bibinfo  {journal} {Rev. Mod. Phys.}\ }\textbf {\bibinfo {volume}
  {30}},\ \bibinfo {pages} {24} (\bibinfo {year} {1958})}\BibitemShut {NoStop}%
\bibitem [{\citenamefont {Dirac}(1949)}]{DE4}%
  \BibitemOpen
  \bibfield  {author} {\bibinfo {author} {\bibfnamefont {P.~A.~M.}\
  \bibnamefont {Dirac}},\ }\href {https://doi.org/10.1103/RevModPhys.21.392}
  {\bibfield  {journal} {\bibinfo  {journal} {Rev. Mod. Phys.}\ }\textbf
  {\bibinfo {volume} {21}},\ \bibinfo {pages} {392} (\bibinfo {year}
  {1949})}\BibitemShut {NoStop}%
\bibitem [{\citenamefont {Chirgwin}\ and\ \citenamefont {Flint}(1945)}]{DE5}%
  \BibitemOpen
  \bibfield  {author} {\bibinfo {author} {\bibfnamefont {B.~H.}\ \bibnamefont
  {Chirgwin}}\ and\ \bibinfo {author} {\bibfnamefont {H.~T.}\ \bibnamefont
  {Flint}},\ }\href {https://doi.org/10.1038/155724a0} {\bibfield  {journal}
  {\bibinfo  {journal} {Nature}\ }\textbf {\bibinfo {volume} {155}},\ \bibinfo
  {pages} {724} (\bibinfo {year} {1945})}\BibitemShut {NoStop}%
\bibitem [{\citenamefont {Giuliani}\ \emph {et~al.}(2019)\citenamefont
  {Giuliani}, \citenamefont {Matheson}, \citenamefont {Nazarewicz},
  \citenamefont {Olsen}, \citenamefont {Reinhard}, \citenamefont {Sadhukhan},
  \citenamefont {Schuetrumpf}, \citenamefont {Schunck},\ and\ \citenamefont
  {Schwerdtfeger}}]{DE_5_0}%
  \BibitemOpen
  \bibfield  {author} {\bibinfo {author} {\bibfnamefont {S.~A.}\ \bibnamefont
  {Giuliani}}, \bibinfo {author} {\bibfnamefont {Z.}~\bibnamefont {Matheson}},
  \bibinfo {author} {\bibfnamefont {W.}~\bibnamefont {Nazarewicz}}, \bibinfo
  {author} {\bibfnamefont {E.}~\bibnamefont {Olsen}}, \bibinfo {author}
  {\bibfnamefont {P.-G.}\ \bibnamefont {Reinhard}}, \bibinfo {author}
  {\bibfnamefont {J.}~\bibnamefont {Sadhukhan}}, \bibinfo {author}
  {\bibfnamefont {B.}~\bibnamefont {Schuetrumpf}}, \bibinfo {author}
  {\bibfnamefont {N.}~\bibnamefont {Schunck}},\ and\ \bibinfo {author}
  {\bibfnamefont {P.}~\bibnamefont {Schwerdtfeger}},\ }\href
  {https://doi.org/10.1103/RevModPhys.91.011001} {\bibfield  {journal}
  {\bibinfo  {journal} {Rev. Mod. Phys.}\ }\textbf {\bibinfo {volume} {91}},\
  \bibinfo {pages} {011001} (\bibinfo {year} {2019})}\BibitemShut {NoStop}%
\bibitem [{\citenamefont {Li}\ \emph {et~al.}(2016)\citenamefont {Li},
  \citenamefont {Shi}, \citenamefont {Guo}, \citenamefont {Niu},\ and\
  \citenamefont {Liang}}]{DE_5_1}%
  \BibitemOpen
  \bibfield  {author} {\bibinfo {author} {\bibfnamefont {N.}~\bibnamefont
  {Li}}, \bibinfo {author} {\bibfnamefont {M.}~\bibnamefont {Shi}}, \bibinfo
  {author} {\bibfnamefont {J.-Y.}\ \bibnamefont {Guo}}, \bibinfo {author}
  {\bibfnamefont {Z.-M.}\ \bibnamefont {Niu}},\ and\ \bibinfo {author}
  {\bibfnamefont {H.}~\bibnamefont {Liang}},\ }\href
  {https://doi.org/10.1103/PhysRevLett.117.062502} {\bibfield  {journal}
  {\bibinfo  {journal} {Phys. Rev. Lett.}\ }\textbf {\bibinfo {volume} {117}},\
  \bibinfo {pages} {062502} (\bibinfo {year} {2016})}\BibitemShut {NoStop}%
\bibitem [{\citenamefont {Ding}\ \emph {et~al.}(2021)\citenamefont {Ding},
  \citenamefont {Li}, \citenamefont {Mukherjee}, \citenamefont {Tomiya},
  \citenamefont {Wang},\ and\ \citenamefont {Zhang}}]{DE_5_2}%
  \BibitemOpen
  \bibfield  {author} {\bibinfo {author} {\bibfnamefont {H.-T.}\ \bibnamefont
  {Ding}}, \bibinfo {author} {\bibfnamefont {S.-T.}\ \bibnamefont {Li}},
  \bibinfo {author} {\bibfnamefont {S.}~\bibnamefont {Mukherjee}}, \bibinfo
  {author} {\bibfnamefont {A.}~\bibnamefont {Tomiya}}, \bibinfo {author}
  {\bibfnamefont {X.-D.}\ \bibnamefont {Wang}},\ and\ \bibinfo {author}
  {\bibfnamefont {Y.}~\bibnamefont {Zhang}},\ }\href
  {https://doi.org/10.1103/PhysRevLett.126.082001} {\bibfield  {journal}
  {\bibinfo  {journal} {Phys. Rev. Lett.}\ }\textbf {\bibinfo {volume} {126}},\
  \bibinfo {pages} {082001} (\bibinfo {year} {2021})}\BibitemShut {NoStop}%
\bibitem [{\citenamefont {Agarwal}(2019)}]{DE_5_3}%
  \BibitemOpen
  \bibfield  {author} {\bibinfo {author} {\bibfnamefont {A.}~\bibnamefont
  {Agarwal}},\ }\href {https://doi.org/10.1103/PhysRevLett.123.211601}
  {\bibfield  {journal} {\bibinfo  {journal} {Phys. Rev. Lett.}\ }\textbf
  {\bibinfo {volume} {123}},\ \bibinfo {pages} {211601} (\bibinfo {year}
  {2019})}\BibitemShut {NoStop}%
\bibitem [{\citenamefont {Barnett}(2017)}]{DE10}%
  \BibitemOpen
  \bibfield  {author} {\bibinfo {author} {\bibfnamefont {S.~M.}\ \bibnamefont
  {Barnett}},\ }\href {https://doi.org/10.1103/PhysRevLett.118.114802}
  {\bibfield  {journal} {\bibinfo  {journal} {Phys. Rev. Lett.}\ }\textbf
  {\bibinfo {volume} {118}},\ \bibinfo {pages} {114802} (\bibinfo {year}
  {2017})}\BibitemShut {NoStop}%
\bibitem [{\citenamefont {Campos}\ \emph {et~al.}(2017)\citenamefont {Campos},
  \citenamefont {Cabrera}, \citenamefont {Rabitz},\ and\ \citenamefont
  {Bondar}}]{DE11}%
  \BibitemOpen
  \bibfield  {author} {\bibinfo {author} {\bibfnamefont {A.~G.}\ \bibnamefont
  {Campos}}, \bibinfo {author} {\bibfnamefont {R.}~\bibnamefont {Cabrera}},
  \bibinfo {author} {\bibfnamefont {H.~A.}\ \bibnamefont {Rabitz}},\ and\
  \bibinfo {author} {\bibfnamefont {D.~I.}\ \bibnamefont {Bondar}},\ }\href
  {https://doi.org/10.1103/PhysRevLett.119.173203} {\bibfield  {journal}
  {\bibinfo  {journal} {Phys. Rev. Lett.}\ }\textbf {\bibinfo {volume} {119}},\
  \bibinfo {pages} {173203} (\bibinfo {year} {2017})}\BibitemShut {NoStop}%
\bibitem [{\citenamefont {Bialynicki-Birula}(2004)}]{DE13}%
  \BibitemOpen
  \bibfield  {author} {\bibinfo {author} {\bibfnamefont {I.}~\bibnamefont
  {Bialynicki-Birula}},\ }\href {https://doi.org/10.1103/PhysRevLett.93.020402}
  {\bibfield  {journal} {\bibinfo  {journal} {Phys. Rev. Lett.}\ }\textbf
  {\bibinfo {volume} {93}},\ \bibinfo {pages} {020402} (\bibinfo {year}
  {2004})}\BibitemShut {NoStop}%
\bibitem [{\citenamefont {Kaminer}\ \emph {et~al.}(2015)\citenamefont
  {Kaminer}, \citenamefont {Nemirovsky}, \citenamefont {Rechtsman},
  \citenamefont {Bekenstein},\ and\ \citenamefont {Segev}}]{DE14}%
  \BibitemOpen
  \bibfield  {author} {\bibinfo {author} {\bibfnamefont {I.}~\bibnamefont
  {Kaminer}}, \bibinfo {author} {\bibfnamefont {J.}~\bibnamefont {Nemirovsky}},
  \bibinfo {author} {\bibfnamefont {M.}~\bibnamefont {Rechtsman}}, \bibinfo
  {author} {\bibfnamefont {R.}~\bibnamefont {Bekenstein}},\ and\ \bibinfo
  {author} {\bibfnamefont {M.}~\bibnamefont {Segev}},\ }\href
  {https://doi.org/10.1038/nphys3196} {\bibfield  {journal} {\bibinfo
  {journal} {Nature Phys.}\ }\textbf {\bibinfo {volume} {11}},\ \bibinfo
  {pages} {261} (\bibinfo {year} {2015})}\BibitemShut {NoStop}%
\bibitem [{\citenamefont {Hammad}\ \emph {et~al.}(2024)\citenamefont {Hammad},
  \citenamefont {Simard}, \citenamefont {Saadati},\ and\ \citenamefont
  {Landry}}]{DE15}%
  \BibitemOpen
  \bibfield  {author} {\bibinfo {author} {\bibfnamefont {F.}~\bibnamefont
  {Hammad}}, \bibinfo {author} {\bibfnamefont {M.}~\bibnamefont {Simard}},
  \bibinfo {author} {\bibfnamefont {R.}~\bibnamefont {Saadati}},\ and\ \bibinfo
  {author} {\bibfnamefont {A.}~\bibnamefont {Landry}},\ }\href
  {https://doi.org/10.1103/PhysRevD.110.065005} {\bibfield  {journal} {\bibinfo
   {journal} {Phys. Rev. D}\ }\textbf {\bibinfo {volume} {110}},\ \bibinfo
  {pages} {065005} (\bibinfo {year} {2024})}\BibitemShut {NoStop}%
\bibitem [{\citenamefont {Bialynicki-Birula}(2021)}]{DE16}%
  \BibitemOpen
  \bibfield  {author} {\bibinfo {author} {\bibfnamefont {I.}~\bibnamefont
  {Bialynicki-Birula}},\ }\href {https://doi.org/10.1103/PhysRevD.103.085001}
  {\bibfield  {journal} {\bibinfo  {journal} {Phys. Rev. D}\ }\textbf {\bibinfo
  {volume} {103}},\ \bibinfo {pages} {085001} (\bibinfo {year}
  {2021})}\BibitemShut {NoStop}%
\bibitem [{\citenamefont {Mishra}\ \emph {et~al.}(2024)\citenamefont {Mishra},
  \citenamefont {Kumar},\ and\ \citenamefont {Misra}}]{Amruta2024}%
  \BibitemOpen
  \bibfield  {author} {\bibinfo {author} {\bibfnamefont {A.}~\bibnamefont
  {Mishra}}, \bibinfo {author} {\bibfnamefont {A.}~\bibnamefont {Kumar}},\ and\
  \bibinfo {author} {\bibfnamefont {S.~P.}\ \bibnamefont {Misra}},\ }\href
  {https://doi.org/10.1103/PhysRevD.110.014003} {\bibfield  {journal} {\bibinfo
   {journal} {Phys. Rev. D}\ }\textbf {\bibinfo {volume} {110}},\ \bibinfo
  {pages} {014003} (\bibinfo {year} {2024})}\BibitemShut {NoStop}%
\bibitem [{\citenamefont {Alberto}\ \emph {et~al.}(2001)\citenamefont
  {Alberto}, \citenamefont {Fiolhais}, \citenamefont {Malheiro}, \citenamefont
  {Delfino},\ and\ \citenamefont {Chiapparini}}]{PS}%
  \BibitemOpen
  \bibfield  {author} {\bibinfo {author} {\bibfnamefont {P.}~\bibnamefont
  {Alberto}}, \bibinfo {author} {\bibfnamefont {M.}~\bibnamefont {Fiolhais}},
  \bibinfo {author} {\bibfnamefont {M.}~\bibnamefont {Malheiro}}, \bibinfo
  {author} {\bibfnamefont {A.}~\bibnamefont {Delfino}},\ and\ \bibinfo {author}
  {\bibfnamefont {M.}~\bibnamefont {Chiapparini}},\ }\href
  {https://doi.org/10.1103/PhysRevLett.86.5015} {\bibfield  {journal} {\bibinfo
   {journal} {Phys. Rev. Lett.}\ }\textbf {\bibinfo {volume} {86}},\ \bibinfo
  {pages} {5015} (\bibinfo {year} {2001})}\BibitemShut {NoStop}%
\bibitem [{\citenamefont {Ginocchio}(1997)}]{PS1}%
  \BibitemOpen
  \bibfield  {author} {\bibinfo {author} {\bibfnamefont {J.~N.}\ \bibnamefont
  {Ginocchio}},\ }\href {https://doi.org/10.1103/PhysRevLett.78.436} {\bibfield
   {journal} {\bibinfo  {journal} {Phys. Rev. Lett.}\ }\textbf {\bibinfo
  {volume} {78}},\ \bibinfo {pages} {436} (\bibinfo {year} {1997})}\BibitemShut
  {NoStop}%
\bibitem [{\citenamefont {Ginocchio}(1999)}]{PS2}%
  \BibitemOpen
  \bibfield  {author} {\bibinfo {author} {\bibfnamefont {J.~N.}\ \bibnamefont
  {Ginocchio}},\ }\href {https://doi.org/10.1016/S0370-1573(99)00021-6}
  {\bibfield  {journal} {\bibinfo  {journal} {Phys. Rep.}\ }\textbf {\bibinfo
  {volume} {315}},\ \bibinfo {pages} {231} (\bibinfo {year}
  {1999})}\BibitemShut {NoStop}%
\bibitem [{\citenamefont {Ginocchio}(2004)}]{PS3}%
  \BibitemOpen
  \bibfield  {author} {\bibinfo {author} {\bibfnamefont {J.~N.}\ \bibnamefont
  {Ginocchio}},\ }\href {https://doi.org/10.1103/PhysRevC.69.034318} {\bibfield
   {journal} {\bibinfo  {journal} {Phys. Rev. C}\ }\textbf {\bibinfo {volume}
  {69}},\ \bibinfo {pages} {034318} (\bibinfo {year} {2004})}\BibitemShut
  {NoStop}%
\bibitem [{\citenamefont {Ginocchio}(2005)}]{PS4}%
  \BibitemOpen
  \bibfield  {author} {\bibinfo {author} {\bibfnamefont {J.~N.}\ \bibnamefont
  {Ginocchio}},\ }\href {https://doi.org/10.1016/j.physrep.2005.04.003}
  {\bibfield  {journal} {\bibinfo  {journal} {Phys. Rep.}\ }\textbf {\bibinfo
  {volume} {414}},\ \bibinfo {pages} {165} (\bibinfo {year}
  {2005})}\BibitemShut {NoStop}%
\bibitem [{\citenamefont {Liang}\ \emph {et~al.}(2015)\citenamefont {Liang},
  \citenamefont {Meng},\ and\ \citenamefont {Zhou}}]{PS5}%
  \BibitemOpen
  \bibfield  {author} {\bibinfo {author} {\bibfnamefont {H.}~\bibnamefont
  {Liang}}, \bibinfo {author} {\bibfnamefont {J.}~\bibnamefont {Meng}},\ and\
  \bibinfo {author} {\bibfnamefont {S.-G.}\ \bibnamefont {Zhou}},\ }\href
  {https://doi.org/https://doi.org/10.1016/j.physrep.2014.12.005} {\bibfield
  {journal} {\bibinfo  {journal} {Phys. Rep.}\ }\textbf {\bibinfo {volume}
  {570}},\ \bibinfo {pages} {1} (\bibinfo {year} {2015})}\BibitemShut {NoStop}%
\bibitem [{\citenamefont {Zhou}\ \emph {et~al.}(2003)\citenamefont {Zhou},
  \citenamefont {Meng},\ and\ \citenamefont {Ring}}]{PS6}%
  \BibitemOpen
  \bibfield  {author} {\bibinfo {author} {\bibfnamefont {S.-G.}\ \bibnamefont
  {Zhou}}, \bibinfo {author} {\bibfnamefont {J.}~\bibnamefont {Meng}},\ and\
  \bibinfo {author} {\bibfnamefont {P.}~\bibnamefont {Ring}},\ }\href
  {https://doi.org/10.1103/PhysRevLett.91.262501} {\bibfield  {journal}
  {\bibinfo  {journal} {Phys. Rev. Lett.}\ }\textbf {\bibinfo {volume} {91}},\
  \bibinfo {pages} {262501} (\bibinfo {year} {2003})}\BibitemShut {NoStop}%
\bibitem [{\citenamefont {Page}\ \emph {et~al.}(2001)\citenamefont {Page},
  \citenamefont {Goldman},\ and\ \citenamefont {Ginocchio}}]{PS7}%
  \BibitemOpen
  \bibfield  {author} {\bibinfo {author} {\bibfnamefont {P.~R.}\ \bibnamefont
  {Page}}, \bibinfo {author} {\bibfnamefont {T.}~\bibnamefont {Goldman}},\ and\
  \bibinfo {author} {\bibfnamefont {J.~N.}\ \bibnamefont {Ginocchio}},\ }\href
  {https://doi.org/10.1103/PhysRevLett.86.204} {\bibfield  {journal} {\bibinfo
  {journal} {Phys. Rev. Lett.}\ }\textbf {\bibinfo {volume} {86}},\ \bibinfo
  {pages} {204} (\bibinfo {year} {2001})}\BibitemShut {NoStop}%
\bibitem [{\citenamefont {Arima}\ \emph {et~al.}(1969)\citenamefont {Arima},
  \citenamefont {Harvey},\ and\ \citenamefont {Shimizu}}]{PS8}%
  \BibitemOpen
  \bibfield  {author} {\bibinfo {author} {\bibfnamefont {A.}~\bibnamefont
  {Arima}}, \bibinfo {author} {\bibfnamefont {M.}~\bibnamefont {Harvey}},\ and\
  \bibinfo {author} {\bibfnamefont {K.}~\bibnamefont {Shimizu}},\ }\href
  {https://doi.org/10.1016/0370-2693(69)90443-2} {\bibfield  {journal}
  {\bibinfo  {journal} {Phys. Lett. B}\ }\textbf {\bibinfo {volume} {30}},\
  \bibinfo {pages} {517} (\bibinfo {year} {1969})}\BibitemShut {NoStop}%
\bibitem [{\citenamefont {Hecht}\ and\ \citenamefont {Adler}(1969)}]{PS9}%
  \BibitemOpen
  \bibfield  {author} {\bibinfo {author} {\bibfnamefont {K.}~\bibnamefont
  {Hecht}}\ and\ \bibinfo {author} {\bibfnamefont {A.}~\bibnamefont {Adler}},\
  }\href {https://doi.org/10.1016/0375-9474(69)90077-3} {\bibfield  {journal}
  {\bibinfo  {journal} {Nucl. Phys. A}\ }\textbf {\bibinfo {volume} {137}},\
  \bibinfo {pages} {129} (\bibinfo {year} {1969})}\BibitemShut {NoStop}%
\bibitem [{\citenamefont {de~Voigt}\ \emph {et~al.}(1983)\citenamefont
  {de~Voigt}, \citenamefont {Dudek},\ and\ \citenamefont
  {Szyma\ifmmode~\acute{n}\else \'{n}\fi{}ski}}]{PS10}%
  \BibitemOpen
  \bibfield  {author} {\bibinfo {author} {\bibfnamefont {M.~J.~A.}\
  \bibnamefont {de~Voigt}}, \bibinfo {author} {\bibfnamefont {J.}~\bibnamefont
  {Dudek}},\ and\ \bibinfo {author} {\bibfnamefont {Z.}~\bibnamefont
  {Szyma\ifmmode~\acute{n}\else \'{n}\fi{}ski}},\ }\href
  {https://doi.org/10.1103/RevModPhys.55.949} {\bibfield  {journal} {\bibinfo
  {journal} {Rev. Mod. Phys.}\ }\textbf {\bibinfo {volume} {55}},\ \bibinfo
  {pages} {949} (\bibinfo {year} {1983})}\BibitemShut {NoStop}%
\bibitem [{\citenamefont {Dudek}\ \emph {et~al.}(1987)\citenamefont {Dudek},
  \citenamefont {Nazarewicz}, \citenamefont {Szymanski},\ and\ \citenamefont
  {Leander}}]{PS11}%
  \BibitemOpen
  \bibfield  {author} {\bibinfo {author} {\bibfnamefont {J.}~\bibnamefont
  {Dudek}}, \bibinfo {author} {\bibfnamefont {W.}~\bibnamefont {Nazarewicz}},
  \bibinfo {author} {\bibfnamefont {Z.}~\bibnamefont {Szymanski}},\ and\
  \bibinfo {author} {\bibfnamefont {G.~A.}\ \bibnamefont {Leander}},\ }\href
  {https://doi.org/10.1103/PhysRevLett.59.1405} {\bibfield  {journal} {\bibinfo
   {journal} {Phys. Rev. Lett.}\ }\textbf {\bibinfo {volume} {59}},\ \bibinfo
  {pages} {1405} (\bibinfo {year} {1987})}\BibitemShut {NoStop}%
\bibitem [{\citenamefont {Nazarewicz}\ \emph {et~al.}(1990)\citenamefont
  {Nazarewicz}, \citenamefont {Twin}, \citenamefont {Fallon},\ and\
  \citenamefont {Garrett}}]{PS12}%
  \BibitemOpen
  \bibfield  {author} {\bibinfo {author} {\bibfnamefont {W.}~\bibnamefont
  {Nazarewicz}}, \bibinfo {author} {\bibfnamefont {P.~J.}\ \bibnamefont
  {Twin}}, \bibinfo {author} {\bibfnamefont {P.}~\bibnamefont {Fallon}},\ and\
  \bibinfo {author} {\bibfnamefont {J.~D.}\ \bibnamefont {Garrett}},\ }\href
  {https://doi.org/10.1103/PhysRevLett.64.1654} {\bibfield  {journal} {\bibinfo
   {journal} {Phys. Rev. Lett.}\ }\textbf {\bibinfo {volume} {64}},\ \bibinfo
  {pages} {1654} (\bibinfo {year} {1990})}\BibitemShut {NoStop}%
\bibitem [{\citenamefont {Zeng}\ \emph {et~al.}(1991)\citenamefont {Zeng},
  \citenamefont {Meng}, \citenamefont {Wu}, \citenamefont {Zhao}, \citenamefont
  {Xing},\ and\ \citenamefont {Chen}}]{PS13}%
  \BibitemOpen
  \bibfield  {author} {\bibinfo {author} {\bibfnamefont {J.~Y.}\ \bibnamefont
  {Zeng}}, \bibinfo {author} {\bibfnamefont {J.}~\bibnamefont {Meng}}, \bibinfo
  {author} {\bibfnamefont {C.~S.}\ \bibnamefont {Wu}}, \bibinfo {author}
  {\bibfnamefont {E.~G.}\ \bibnamefont {Zhao}}, \bibinfo {author}
  {\bibfnamefont {Z.}~\bibnamefont {Xing}},\ and\ \bibinfo {author}
  {\bibfnamefont {X.~Q.}\ \bibnamefont {Chen}},\ }\href
  {https://doi.org/10.1103/PhysRevC.44.R1745} {\bibfield  {journal} {\bibinfo
  {journal} {Phys. Rev. C}\ }\textbf {\bibinfo {volume} {44}},\ \bibinfo
  {pages} {R1745} (\bibinfo {year} {1991})}\BibitemShut {NoStop}%
\bibitem [{\citenamefont {Byrski}\ \emph {et~al.}(1990)\citenamefont {Byrski},
  \citenamefont {Beck}, \citenamefont {Curien}, \citenamefont {Schuck},
  \citenamefont {Fallon}, \citenamefont {Alderson}, \citenamefont {Ali},
  \citenamefont {Bentley}, \citenamefont {Bruce}, \citenamefont {Forsyth},
  \citenamefont {Howe}, \citenamefont {Roberts}, \citenamefont
  {Sharpey-Schafer}, \citenamefont {Smith},\ and\ \citenamefont {Twin}}]{PS14}%
  \BibitemOpen
  \bibfield  {author} {\bibinfo {author} {\bibfnamefont {T.}~\bibnamefont
  {Byrski}}, \bibinfo {author} {\bibfnamefont {F.~A.}\ \bibnamefont {Beck}},
  \bibinfo {author} {\bibfnamefont {D.}~\bibnamefont {Curien}}, \bibinfo
  {author} {\bibfnamefont {C.}~\bibnamefont {Schuck}}, \bibinfo {author}
  {\bibfnamefont {P.}~\bibnamefont {Fallon}}, \bibinfo {author} {\bibfnamefont
  {A.}~\bibnamefont {Alderson}}, \bibinfo {author} {\bibfnamefont
  {I.}~\bibnamefont {Ali}}, \bibinfo {author} {\bibfnamefont {M.~A.}\
  \bibnamefont {Bentley}}, \bibinfo {author} {\bibfnamefont {A.~M.}\
  \bibnamefont {Bruce}}, \bibinfo {author} {\bibfnamefont {P.~D.}\ \bibnamefont
  {Forsyth}}, \bibinfo {author} {\bibfnamefont {D.}~\bibnamefont {Howe}},
  \bibinfo {author} {\bibfnamefont {J.~W.}\ \bibnamefont {Roberts}}, \bibinfo
  {author} {\bibfnamefont {J.~F.}\ \bibnamefont {Sharpey-Schafer}}, \bibinfo
  {author} {\bibfnamefont {G.}~\bibnamefont {Smith}},\ and\ \bibinfo {author}
  {\bibfnamefont {P.~J.}\ \bibnamefont {Twin}},\ }\href
  {https://doi.org/10.1103/PhysRevLett.64.1650} {\bibfield  {journal} {\bibinfo
   {journal} {Phys. Rev. Lett.}\ }\textbf {\bibinfo {volume} {64}},\ \bibinfo
  {pages} {1650} (\bibinfo {year} {1990})}\BibitemShut {NoStop}%
\bibitem [{\citenamefont {Troltenier}\ \emph {et~al.}(1994)\citenamefont
  {Troltenier}, \citenamefont {Nazarewicz}, \citenamefont {Szymański},\ and\
  \citenamefont {Draayer}}]{PS15}%
  \BibitemOpen
  \bibfield  {author} {\bibinfo {author} {\bibfnamefont {D.}~\bibnamefont
  {Troltenier}}, \bibinfo {author} {\bibfnamefont {W.}~\bibnamefont
  {Nazarewicz}}, \bibinfo {author} {\bibfnamefont {Z.}~\bibnamefont
  {Szymański}},\ and\ \bibinfo {author} {\bibfnamefont {J.}~\bibnamefont
  {Draayer}},\ }\href {https://doi.org/10.1016/0375-9474(94)90026-4} {\bibfield
   {journal} {\bibinfo  {journal} {Nucl. Phys. A}\ }\textbf {\bibinfo {volume}
  {567}},\ \bibinfo {pages} {591} (\bibinfo {year} {1994})}\BibitemShut
  {NoStop}%
\bibitem [{\citenamefont {Stuchbery}(2002)}]{PS16}%
  \BibitemOpen
  \bibfield  {author} {\bibinfo {author} {\bibfnamefont {A.~E.}\ \bibnamefont
  {Stuchbery}},\ }\href {https://doi.org/10.1016/S0375-9474(01)01300-8}
  {\bibfield  {journal} {\bibinfo  {journal} {Nucl. Phys. A}\ }\textbf
  {\bibinfo {volume} {700}},\ \bibinfo {pages} {83} (\bibinfo {year}
  {2002})}\BibitemShut {NoStop}%
\bibitem [{\citenamefont {Nagai}\ \emph {et~al.}(1981)\citenamefont {Nagai},
  \citenamefont {Styczen}, \citenamefont {Piiparinen}, \citenamefont
  {Kleinheinz}, \citenamefont {Bazzacco}, \citenamefont {Brentano},
  \citenamefont {Zell},\ and\ \citenamefont {Blomqvist}}]{PS17}%
  \BibitemOpen
  \bibfield  {author} {\bibinfo {author} {\bibfnamefont {Y.}~\bibnamefont
  {Nagai}}, \bibinfo {author} {\bibfnamefont {J.}~\bibnamefont {Styczen}},
  \bibinfo {author} {\bibfnamefont {M.}~\bibnamefont {Piiparinen}}, \bibinfo
  {author} {\bibfnamefont {P.}~\bibnamefont {Kleinheinz}}, \bibinfo {author}
  {\bibfnamefont {D.}~\bibnamefont {Bazzacco}}, \bibinfo {author}
  {\bibfnamefont {P.~V.}\ \bibnamefont {Brentano}}, \bibinfo {author}
  {\bibfnamefont {K.~O.}\ \bibnamefont {Zell}},\ and\ \bibinfo {author}
  {\bibfnamefont {J.}~\bibnamefont {Blomqvist}},\ }\href
  {https://doi.org/10.1103/PhysRevLett.47.1259} {\bibfield  {journal} {\bibinfo
   {journal} {Phys. Rev. Lett.}\ }\textbf {\bibinfo {volume} {47}},\ \bibinfo
  {pages} {1259} (\bibinfo {year} {1981})}\BibitemShut {NoStop}%
\bibitem [{\citenamefont {Gaudefroy}\ \emph {et~al.}(2006)\citenamefont
  {Gaudefroy}, \citenamefont {Sorlin}, \citenamefont {Beaumel}, \citenamefont
  {Blumenfeld}, \citenamefont {Dombr\'adi}, \citenamefont {Fortier},
  \citenamefont {Franchoo}, \citenamefont {G\'elin}, \citenamefont {Gibelin},
  \citenamefont {Gr\'evy}, \citenamefont {Hammache}, \citenamefont {Ibrahim},
  \citenamefont {Kemper}, \citenamefont {Kratz}, \citenamefont {Lukyanov},
  \citenamefont {Monrozeau}, \citenamefont {Nalpas}, \citenamefont {Nowacki},
  \citenamefont {Ostrowski}, \citenamefont {Otsuka}, \citenamefont
  {Penionzhkevich}, \citenamefont {Piekarewicz}, \citenamefont {Pollacco},
  \citenamefont {Roussel-Chomaz}, \citenamefont {Rich}, \citenamefont
  {Scarpaci}, \citenamefont {St.~Laurent}, \citenamefont {Sohler},
  \citenamefont {Stanoiu}, \citenamefont {Suzuki}, \citenamefont {Tryggestad},\
  and\ \citenamefont {Verney}}]{PS18}%
  \BibitemOpen
  \bibfield  {author} {\bibinfo {author} {\bibfnamefont {L.}~\bibnamefont
  {Gaudefroy}}, \bibinfo {author} {\bibfnamefont {O.}~\bibnamefont {Sorlin}},
  \bibinfo {author} {\bibfnamefont {D.}~\bibnamefont {Beaumel}}, \bibinfo
  {author} {\bibfnamefont {Y.}~\bibnamefont {Blumenfeld}}, \bibinfo {author}
  {\bibfnamefont {Z.}~\bibnamefont {Dombr\'adi}}, \bibinfo {author}
  {\bibfnamefont {S.}~\bibnamefont {Fortier}}, \bibinfo {author} {\bibfnamefont
  {S.}~\bibnamefont {Franchoo}}, \bibinfo {author} {\bibfnamefont
  {M.}~\bibnamefont {G\'elin}}, \bibinfo {author} {\bibfnamefont
  {J.}~\bibnamefont {Gibelin}}, \bibinfo {author} {\bibfnamefont
  {S.}~\bibnamefont {Gr\'evy}}, \bibinfo {author} {\bibfnamefont
  {F.}~\bibnamefont {Hammache}}, \bibinfo {author} {\bibfnamefont
  {F.}~\bibnamefont {Ibrahim}}, \bibinfo {author} {\bibfnamefont {K.~W.}\
  \bibnamefont {Kemper}}, \bibinfo {author} {\bibfnamefont {K.-L.}\
  \bibnamefont {Kratz}}, \bibinfo {author} {\bibfnamefont {S.~M.}\ \bibnamefont
  {Lukyanov}}, \bibinfo {author} {\bibfnamefont {C.}~\bibnamefont {Monrozeau}},
  \bibinfo {author} {\bibfnamefont {L.}~\bibnamefont {Nalpas}}, \bibinfo
  {author} {\bibfnamefont {F.}~\bibnamefont {Nowacki}}, \bibinfo {author}
  {\bibfnamefont {A.~N.}\ \bibnamefont {Ostrowski}}, \bibinfo {author}
  {\bibfnamefont {T.}~\bibnamefont {Otsuka}}, \bibinfo {author} {\bibfnamefont
  {Y.-E.}\ \bibnamefont {Penionzhkevich}}, \bibinfo {author} {\bibfnamefont
  {J.}~\bibnamefont {Piekarewicz}}, \bibinfo {author} {\bibfnamefont {E.~C.}\
  \bibnamefont {Pollacco}}, \bibinfo {author} {\bibfnamefont {P.}~\bibnamefont
  {Roussel-Chomaz}}, \bibinfo {author} {\bibfnamefont {E.}~\bibnamefont
  {Rich}}, \bibinfo {author} {\bibfnamefont {J.~A.}\ \bibnamefont {Scarpaci}},
  \bibinfo {author} {\bibfnamefont {M.~G.}\ \bibnamefont {St.~Laurent}},
  \bibinfo {author} {\bibfnamefont {D.}~\bibnamefont {Sohler}}, \bibinfo
  {author} {\bibfnamefont {M.}~\bibnamefont {Stanoiu}}, \bibinfo {author}
  {\bibfnamefont {T.}~\bibnamefont {Suzuki}}, \bibinfo {author} {\bibfnamefont
  {E.}~\bibnamefont {Tryggestad}},\ and\ \bibinfo {author} {\bibfnamefont
  {D.}~\bibnamefont {Verney}},\ }\href
  {https://doi.org/10.1103/PhysRevLett.97.092501} {\bibfield  {journal}
  {\bibinfo  {journal} {Phys. Rev. Lett.}\ }\textbf {\bibinfo {volume} {97}},\
  \bibinfo {pages} {092501} (\bibinfo {year} {2006})}\BibitemShut {NoStop}%
\bibitem [{\citenamefont {Bastin}\ \emph {et~al.}(2007)\citenamefont {Bastin},
  \citenamefont {Gr\'evy}, \citenamefont {Sohler}, \citenamefont {Sorlin},
  \citenamefont {Dombr\'adi}, \citenamefont {Achouri}, \citenamefont
  {Ang\'elique}, \citenamefont {Azaiez}, \citenamefont {Baiborodin},
  \citenamefont {Borcea}, \citenamefont {Bourgeois}, \citenamefont {Buta},
  \citenamefont {B\"urger}, \citenamefont {Chapman}, \citenamefont {Dalouzy},
  \citenamefont {Dlouhy}, \citenamefont {Drouard}, \citenamefont {Elekes},
  \citenamefont {Franchoo}, \citenamefont {Iacob}, \citenamefont {Laurent},
  \citenamefont {Lazar}, \citenamefont {Liang}, \citenamefont {Li\'enard},
  \citenamefont {Mrazek}, \citenamefont {Nalpas}, \citenamefont {Negoita},
  \citenamefont {Orr}, \citenamefont {Penionzhkevich}, \citenamefont
  {Podoly\'ak}, \citenamefont {Pougheon}, \citenamefont {Roussel-Chomaz},
  \citenamefont {Saint-Laurent}, \citenamefont {Stanoiu}, \citenamefont
  {Stefan}, \citenamefont {Nowacki},\ and\ \citenamefont {Poves}}]{PS19}%
  \BibitemOpen
  \bibfield  {author} {\bibinfo {author} {\bibfnamefont {B.}~\bibnamefont
  {Bastin}}, \bibinfo {author} {\bibfnamefont {S.}~\bibnamefont {Gr\'evy}},
  \bibinfo {author} {\bibfnamefont {D.}~\bibnamefont {Sohler}}, \bibinfo
  {author} {\bibfnamefont {O.}~\bibnamefont {Sorlin}}, \bibinfo {author}
  {\bibfnamefont {Z.}~\bibnamefont {Dombr\'adi}}, \bibinfo {author}
  {\bibfnamefont {N.~L.}\ \bibnamefont {Achouri}}, \bibinfo {author}
  {\bibfnamefont {J.~C.}\ \bibnamefont {Ang\'elique}}, \bibinfo {author}
  {\bibfnamefont {F.}~\bibnamefont {Azaiez}}, \bibinfo {author} {\bibfnamefont
  {D.}~\bibnamefont {Baiborodin}}, \bibinfo {author} {\bibfnamefont
  {R.}~\bibnamefont {Borcea}}, \bibinfo {author} {\bibfnamefont
  {C.}~\bibnamefont {Bourgeois}}, \bibinfo {author} {\bibfnamefont
  {A.}~\bibnamefont {Buta}}, \bibinfo {author} {\bibfnamefont {A.}~\bibnamefont
  {B\"urger}}, \bibinfo {author} {\bibfnamefont {R.}~\bibnamefont {Chapman}},
  \bibinfo {author} {\bibfnamefont {J.~C.}\ \bibnamefont {Dalouzy}}, \bibinfo
  {author} {\bibfnamefont {Z.}~\bibnamefont {Dlouhy}}, \bibinfo {author}
  {\bibfnamefont {A.}~\bibnamefont {Drouard}}, \bibinfo {author} {\bibfnamefont
  {Z.}~\bibnamefont {Elekes}}, \bibinfo {author} {\bibfnamefont
  {S.}~\bibnamefont {Franchoo}}, \bibinfo {author} {\bibfnamefont
  {S.}~\bibnamefont {Iacob}}, \bibinfo {author} {\bibfnamefont
  {B.}~\bibnamefont {Laurent}}, \bibinfo {author} {\bibfnamefont
  {M.}~\bibnamefont {Lazar}}, \bibinfo {author} {\bibfnamefont
  {X.}~\bibnamefont {Liang}}, \bibinfo {author} {\bibfnamefont
  {E.}~\bibnamefont {Li\'enard}}, \bibinfo {author} {\bibfnamefont
  {J.}~\bibnamefont {Mrazek}}, \bibinfo {author} {\bibfnamefont
  {L.}~\bibnamefont {Nalpas}}, \bibinfo {author} {\bibfnamefont
  {F.}~\bibnamefont {Negoita}}, \bibinfo {author} {\bibfnamefont {N.~A.}\
  \bibnamefont {Orr}}, \bibinfo {author} {\bibfnamefont {Y.}~\bibnamefont
  {Penionzhkevich}}, \bibinfo {author} {\bibfnamefont {Z.}~\bibnamefont
  {Podoly\'ak}}, \bibinfo {author} {\bibfnamefont {F.}~\bibnamefont
  {Pougheon}}, \bibinfo {author} {\bibfnamefont {P.}~\bibnamefont
  {Roussel-Chomaz}}, \bibinfo {author} {\bibfnamefont {M.~G.}\ \bibnamefont
  {Saint-Laurent}}, \bibinfo {author} {\bibfnamefont {M.}~\bibnamefont
  {Stanoiu}}, \bibinfo {author} {\bibfnamefont {I.}~\bibnamefont {Stefan}},
  \bibinfo {author} {\bibfnamefont {F.}~\bibnamefont {Nowacki}},\ and\ \bibinfo
  {author} {\bibfnamefont {A.}~\bibnamefont {Poves}},\ }\href
  {https://doi.org/10.1103/PhysRevLett.99.022503} {\bibfield  {journal}
  {\bibinfo  {journal} {Phys. Rev. Lett.}\ }\textbf {\bibinfo {volume} {99}},\
  \bibinfo {pages} {022503} (\bibinfo {year} {2007})}\BibitemShut {NoStop}%
\bibitem [{\citenamefont {Troltenier}\ \emph {et~al.}(1995)\citenamefont
  {Troltenier}, \citenamefont {Bahri},\ and\ \citenamefont {Draayer}}]{PS20}%
  \BibitemOpen
  \bibfield  {author} {\bibinfo {author} {\bibfnamefont {D.}~\bibnamefont
  {Troltenier}}, \bibinfo {author} {\bibfnamefont {C.}~\bibnamefont {Bahri}},\
  and\ \bibinfo {author} {\bibfnamefont {J.}~\bibnamefont {Draayer}},\ }\href
  {https://doi.org/10.1016/0375-9474(94)00518-R} {\bibfield  {journal}
  {\bibinfo  {journal} {Nucl. Phys. A}\ }\textbf {\bibinfo {volume} {586}},\
  \bibinfo {pages} {53} (\bibinfo {year} {1995})}\BibitemShut {NoStop}%
\bibitem [{\citenamefont {Ahmadov}\ \emph {et~al.}(2022)\citenamefont
  {Ahmadov}, \citenamefont {Nagiyev}, \citenamefont {Aydin}, \citenamefont
  {Tarverdiyeva}, \citenamefont {Orujova},\ and\ \citenamefont
  {Badalov}}]{PS21}%
  \BibitemOpen
  \bibfield  {author} {\bibinfo {author} {\bibfnamefont {A.~I.}\ \bibnamefont
  {Ahmadov}}, \bibinfo {author} {\bibfnamefont {S.~M.}\ \bibnamefont
  {Nagiyev}}, \bibinfo {author} {\bibfnamefont {C.}~\bibnamefont {Aydin}},
  \bibinfo {author} {\bibfnamefont {V.~A.}\ \bibnamefont {Tarverdiyeva}},
  \bibinfo {author} {\bibfnamefont {M.~S.}\ \bibnamefont {Orujova}},\ and\
  \bibinfo {author} {\bibfnamefont {S.~V.}\ \bibnamefont {Badalov}},\ }\href
  {https://doi.org/10.1140/epjp/s13360-022-03255-9} {\bibfield  {journal}
  {\bibinfo  {journal} {Eur. Phys. J. Plus}\ }\textbf {\bibinfo {volume}
  {137}},\ \bibinfo {pages} {1075} (\bibinfo {year} {2022})}\BibitemShut
  {NoStop}%
\bibitem [{\citenamefont {de~Castro}\ and\ \citenamefont
  {Alberto}(2012)}]{DEA1}%
  \BibitemOpen
  \bibfield  {author} {\bibinfo {author} {\bibfnamefont {A.~S.}\ \bibnamefont
  {de~Castro}}\ and\ \bibinfo {author} {\bibfnamefont {P.}~\bibnamefont
  {Alberto}},\ }\href {https://doi.org/10.1103/PhysRevA.86.032122} {\bibfield
  {journal} {\bibinfo  {journal} {Phys. Rev. A}\ }\textbf {\bibinfo {volume}
  {86}},\ \bibinfo {pages} {032122} (\bibinfo {year} {2012})}\BibitemShut
  {NoStop}%
\bibitem [{\citenamefont {Xu}\ \emph {et~al.}(2015)\citenamefont {Xu},
  \citenamefont {Zhang}, \citenamefont {Signoracci}, \citenamefont {Smith},\
  and\ \citenamefont {Li}}]{DEA2}%
  \BibitemOpen
  \bibfield  {author} {\bibinfo {author} {\bibfnamefont {X.-D.}\ \bibnamefont
  {Xu}}, \bibinfo {author} {\bibfnamefont {S.-S.}\ \bibnamefont {Zhang}},
  \bibinfo {author} {\bibfnamefont {A.~J.}\ \bibnamefont {Signoracci}},
  \bibinfo {author} {\bibfnamefont {M.~S.}\ \bibnamefont {Smith}},\ and\
  \bibinfo {author} {\bibfnamefont {Z.~P.}\ \bibnamefont {Li}},\ }\href
  {https://doi.org/10.1103/PhysRevC.92.024324} {\bibfield  {journal} {\bibinfo
  {journal} {Phys. Rev. C}\ }\textbf {\bibinfo {volume} {92}},\ \bibinfo
  {pages} {024324} (\bibinfo {year} {2015})}\BibitemShut {NoStop}%
\bibitem [{\citenamefont {Fang}\ \emph {et~al.}(2017)\citenamefont {Fang},
  \citenamefont {Shi}, \citenamefont {Guo}, \citenamefont {Niu}, \citenamefont
  {Liang},\ and\ \citenamefont {Zhang}}]{DEA3}%
  \BibitemOpen
  \bibfield  {author} {\bibinfo {author} {\bibfnamefont {Z.}~\bibnamefont
  {Fang}}, \bibinfo {author} {\bibfnamefont {M.}~\bibnamefont {Shi}}, \bibinfo
  {author} {\bibfnamefont {J.-Y.}\ \bibnamefont {Guo}}, \bibinfo {author}
  {\bibfnamefont {Z.-M.}\ \bibnamefont {Niu}}, \bibinfo {author} {\bibfnamefont
  {H.}~\bibnamefont {Liang}},\ and\ \bibinfo {author} {\bibfnamefont {S.-S.}\
  \bibnamefont {Zhang}},\ }\href {https://doi.org/10.1103/PhysRevC.95.024311}
  {\bibfield  {journal} {\bibinfo  {journal} {Phys. Rev. C}\ }\textbf {\bibinfo
  {volume} {95}},\ \bibinfo {pages} {024311} (\bibinfo {year}
  {2017})}\BibitemShut {NoStop}%
\bibitem [{\citenamefont {Hartmann}\ and\ \citenamefont
  {Portnoi}(2014)}]{DEA4}%
  \BibitemOpen
  \bibfield  {author} {\bibinfo {author} {\bibfnamefont {R.~R.}\ \bibnamefont
  {Hartmann}}\ and\ \bibinfo {author} {\bibfnamefont {M.~E.}\ \bibnamefont
  {Portnoi}},\ }\href {https://doi.org/10.1103/PhysRevA.89.012101} {\bibfield
  {journal} {\bibinfo  {journal} {Phys. Rev. A}\ }\textbf {\bibinfo {volume}
  {89}},\ \bibinfo {pages} {012101} (\bibinfo {year} {2014})}\BibitemShut
  {NoStop}%
\bibitem [{\citenamefont {Ahmadov}\ \emph {et~al.}(2021)\citenamefont
  {Ahmadov}, \citenamefont {Dadashov}, \citenamefont {Huseynova},\ and\
  \citenamefont {Badalov}}]{GTHP1}%
  \BibitemOpen
  \bibfield  {author} {\bibinfo {author} {\bibfnamefont {H.~I.}\ \bibnamefont
  {Ahmadov}}, \bibinfo {author} {\bibfnamefont {E.~A.}\ \bibnamefont
  {Dadashov}}, \bibinfo {author} {\bibfnamefont {N.~S.}\ \bibnamefont
  {Huseynova}},\ and\ \bibinfo {author} {\bibfnamefont {V.~H.}\ \bibnamefont
  {Badalov}},\ }\href {https://doi.org/10.1140/epjp/s13360-021-01202-8}
  {\bibfield  {journal} {\bibinfo  {journal} {Eur. Phys. J. Plus}\ }\textbf
  {\bibinfo {volume} {136}},\ \bibinfo {pages} {244} (\bibinfo {year}
  {2021})}\BibitemShut {NoStop}%
\bibitem [{\citenamefont {Badalov}\ and\ \citenamefont
  {Badalov}(2023)}]{GTHP2}%
  \BibitemOpen
  \bibfield  {author} {\bibinfo {author} {\bibfnamefont {V.~H.}\ \bibnamefont
  {Badalov}}\ and\ \bibinfo {author} {\bibfnamefont {S.~V.}\ \bibnamefont
  {Badalov}},\ }\href {https://doi.org/10.1088/1572-9494/acd441} {\bibfield
  {journal} {\bibinfo  {journal} {Commun. Theor. Phys.}\ }\textbf {\bibinfo
  {volume} {75}},\ \bibinfo {pages} {075003} (\bibinfo {year}
  {2023})}\BibitemShut {NoStop}%
\bibitem [{\citenamefont {Woods}\ and\ \citenamefont {Saxon}(1954)}]{WS1}%
  \BibitemOpen
  \bibfield  {author} {\bibinfo {author} {\bibfnamefont {R.~D.}\ \bibnamefont
  {Woods}}\ and\ \bibinfo {author} {\bibfnamefont {D.~S.}\ \bibnamefont
  {Saxon}},\ }\href {https://doi.org/10.1103/PhysRev.95.577} {\bibfield
  {journal} {\bibinfo  {journal} {Phys. Rev.}\ }\textbf {\bibinfo {volume}
  {95}},\ \bibinfo {pages} {577} (\bibinfo {year} {1954})}\BibitemShut
  {NoStop}%
\bibitem [{\citenamefont {Badalov}\ \emph {et~al.}(2019)\citenamefont
  {Badalov}, \citenamefont {Baris},\ and\ \citenamefont {Uzun}}]{WS2}%
  \BibitemOpen
  \bibfield  {author} {\bibinfo {author} {\bibfnamefont {V.~H.}\ \bibnamefont
  {Badalov}}, \bibinfo {author} {\bibfnamefont {B.}~\bibnamefont {Baris}},\
  and\ \bibinfo {author} {\bibfnamefont {K.}~\bibnamefont {Uzun}},\ }\href
  {https://doi.org/10.1142/S0217732319501074} {\bibfield  {journal} {\bibinfo
  {journal} {Mod. Phys. Lett. A}\ }\textbf {\bibinfo {volume} {34}},\ \bibinfo
  {pages} {1950107} (\bibinfo {year} {2019})}\BibitemShut {NoStop}%
\bibitem [{\citenamefont {Rosen}\ and\ \citenamefont {Morse}(1932)}]{RM1}%
  \BibitemOpen
  \bibfield  {author} {\bibinfo {author} {\bibfnamefont {N.}~\bibnamefont
  {Rosen}}\ and\ \bibinfo {author} {\bibfnamefont {P.~M.}\ \bibnamefont
  {Morse}},\ }\href {https://doi.org/10.1103/PhysRev.42.210} {\bibfield
  {journal} {\bibinfo  {journal} {Phys. Rev.}\ }\textbf {\bibinfo {volume}
  {42}},\ \bibinfo {pages} {210} (\bibinfo {year} {1932})}\BibitemShut
  {NoStop}%
\bibitem [{\citenamefont {Manning}\ and\ \citenamefont {Rosen}(1933)}]{MR}%
  \BibitemOpen
  \bibfield  {author} {\bibinfo {author} {\bibfnamefont {M.~F.}\ \bibnamefont
  {Manning}}\ and\ \bibinfo {author} {\bibfnamefont {N.}~\bibnamefont
  {Rosen}},\ }\href {https://doi.org/10.1103/PhysRev.44.951} {\bibfield
  {journal} {\bibinfo  {journal} {Phys. Rev.}\ }\textbf {\bibinfo {volume}
  {44}},\ \bibinfo {pages} {951} (\bibinfo {year} {1933})}\BibitemShut
  {NoStop}%
\bibitem [{\citenamefont {Morse}(1929)}]{RM2}%
  \BibitemOpen
  \bibfield  {author} {\bibinfo {author} {\bibfnamefont {P.~M.}\ \bibnamefont
  {Morse}},\ }\href {https://doi.org/10.1103/PhysRev.34.57} {\bibfield
  {journal} {\bibinfo  {journal} {Phys. Rev.}\ }\textbf {\bibinfo {volume}
  {34}},\ \bibinfo {pages} {57} (\bibinfo {year} {1929})}\BibitemShut {NoStop}%
\bibitem [{\citenamefont {Jia}\ \emph {et~al.}(2012)\citenamefont {Jia},
  \citenamefont {Diao}, \citenamefont {Liu}, \citenamefont {Wang},
  \citenamefont {Liu},\ and\ \citenamefont {Zhang}}]{RM3}%
  \BibitemOpen
  \bibfield  {author} {\bibinfo {author} {\bibfnamefont {C.-S.}\ \bibnamefont
  {Jia}}, \bibinfo {author} {\bibfnamefont {Y.-F.}\ \bibnamefont {Diao}},
  \bibinfo {author} {\bibfnamefont {X.-J.}\ \bibnamefont {Liu}}, \bibinfo
  {author} {\bibfnamefont {P.-Q.}\ \bibnamefont {Wang}}, \bibinfo {author}
  {\bibfnamefont {J.-Y.}\ \bibnamefont {Liu}},\ and\ \bibinfo {author}
  {\bibfnamefont {G.-D.}\ \bibnamefont {Zhang}},\ }\href
  {https://doi.org/10.1063/1.4731340} {\bibfield  {journal} {\bibinfo
  {journal} {J. Chem. Phys.}\ }\textbf {\bibinfo {volume} {137}},\ \bibinfo
  {pages} {014101} (\bibinfo {year} {2012})}\BibitemShut {NoStop}%
\bibitem [{\citenamefont {Schiöberg}(1986)}]{SCH1}%
  \BibitemOpen
  \bibfield  {author} {\bibinfo {author} {\bibfnamefont {D.}~\bibnamefont
  {Schiöberg}},\ }\href {https://doi.org/10.1080/00268978600102631} {\bibfield
   {journal} {\bibinfo  {journal} {Mol. Phys.}\ }\textbf {\bibinfo {volume}
  {59}},\ \bibinfo {pages} {1123} (\bibinfo {year} {1986})}\BibitemShut
  {NoStop}%
\bibitem [{\citenamefont {Wang}\ \emph {et~al.}(2012)\citenamefont {Wang},
  \citenamefont {Zhang}, \citenamefont {Jia},\ and\ \citenamefont
  {Liu}}]{SCH2}%
  \BibitemOpen
  \bibfield  {author} {\bibinfo {author} {\bibfnamefont {P.-Q.}\ \bibnamefont
  {Wang}}, \bibinfo {author} {\bibfnamefont {L.-H.}\ \bibnamefont {Zhang}},
  \bibinfo {author} {\bibfnamefont {C.-S.}\ \bibnamefont {Jia}},\ and\ \bibinfo
  {author} {\bibfnamefont {J.-Y.}\ \bibnamefont {Liu}},\ }\href
  {https://doi.org/https://doi.org/10.1016/j.jms.2012.03.005} {\bibfield
  {journal} {\bibinfo  {journal} {J. Mol. Spectr.}\ }\textbf {\bibinfo {volume}
  {274}},\ \bibinfo {pages} {5} (\bibinfo {year} {2012})}\BibitemShut {NoStop}%
\bibitem [{\citenamefont {Chun-Sheng}\ \emph {et~al.}(2001)\citenamefont
  {Chun-Sheng}, \citenamefont {Ying}, \citenamefont {Xiang-Lin},\ and\
  \citenamefont {Liang-Tian}}]{FP1}%
  \BibitemOpen
  \bibfield  {author} {\bibinfo {author} {\bibfnamefont {J.}~\bibnamefont
  {Chun-Sheng}}, \bibinfo {author} {\bibfnamefont {Z.}~\bibnamefont {Ying}},
  \bibinfo {author} {\bibfnamefont {Z.}~\bibnamefont {Xiang-Lin}},\ and\
  \bibinfo {author} {\bibfnamefont {S.}~\bibnamefont {Liang-Tian}},\ }\href
  {https://doi.org/10.1088/0253-6102/36/6/641} {\bibfield  {journal} {\bibinfo
  {journal} {Commun. Theor. Phys.}\ }\textbf {\bibinfo {volume} {36}},\
  \bibinfo {pages} {641} (\bibinfo {year} {2001})}\BibitemShut {NoStop}%
\bibitem [{\citenamefont {Ikot}\ \emph {et~al.}(2016)\citenamefont {Ikot},
  \citenamefont {Lutfuoglu}, \citenamefont {Ngwueke}, \citenamefont {Udoh},
  \citenamefont {Zare},\ and\ \citenamefont {Hassanabadi}}]{FP2}%
  \BibitemOpen
  \bibfield  {author} {\bibinfo {author} {\bibfnamefont {A.~N.}\ \bibnamefont
  {Ikot}}, \bibinfo {author} {\bibfnamefont {B.~C.}\ \bibnamefont {Lutfuoglu}},
  \bibinfo {author} {\bibfnamefont {M.~I.}\ \bibnamefont {Ngwueke}}, \bibinfo
  {author} {\bibfnamefont {M.~E.}\ \bibnamefont {Udoh}}, \bibinfo {author}
  {\bibfnamefont {S.}~\bibnamefont {Zare}},\ and\ \bibinfo {author}
  {\bibfnamefont {H.}~\bibnamefont {Hassanabadi}},\ }\href
  {https://doi.org/10.1140/epjp/i2016-16419-5} {\bibfield  {journal} {\bibinfo
  {journal} {Eur. Phys. J. Plus}\ }\textbf {\bibinfo {volume} {131}},\ \bibinfo
  {pages} {419} (\bibinfo {year} {2016})}\BibitemShut {NoStop}%
\bibitem [{\citenamefont {Williams}\ and\ \citenamefont {Poulios}(1993)}]{WP1}%
  \BibitemOpen
  \bibfield  {author} {\bibinfo {author} {\bibfnamefont {B.~W.}\ \bibnamefont
  {Williams}}\ and\ \bibinfo {author} {\bibfnamefont {D.~P.}\ \bibnamefont
  {Poulios}},\ }\href {https://doi.org/10.1088/0143-0807/14/5/006} {\bibfield
  {journal} {\bibinfo  {journal} {Eur. J. Phys.}\ }\textbf {\bibinfo {volume}
  {14}},\ \bibinfo {pages} {222} (\bibinfo {year} {1993})}\BibitemShut
  {NoStop}%
\bibitem [{\citenamefont {Peña}\ \emph {et~al.}(2017)\citenamefont {Peña},
  \citenamefont {Morales},\ and\ \citenamefont {García-Ravelo}}]{WP2}%
  \BibitemOpen
  \bibfield  {author} {\bibinfo {author} {\bibfnamefont {J.~J.}\ \bibnamefont
  {Peña}}, \bibinfo {author} {\bibfnamefont {J.}~\bibnamefont {Morales}},\
  and\ \bibinfo {author} {\bibfnamefont {J.}~\bibnamefont {García-Ravelo}},\
  }\href {https://doi.org/10.1063/1.4979617} {\bibfield  {journal} {\bibinfo
  {journal} {J. Math. Phys.}\ }\textbf {\bibinfo {volume} {58}},\ \bibinfo
  {pages} {043501} (\bibinfo {year} {2017})}\BibitemShut {NoStop}%
\bibitem [{\citenamefont {Soto}\ and\ \citenamefont
  {Tom\`as~Valls}(2023)}]{Hadron_0}%
  \BibitemOpen
  \bibfield  {author} {\bibinfo {author} {\bibfnamefont {J.}~\bibnamefont
  {Soto}}\ and\ \bibinfo {author} {\bibfnamefont {S.}~\bibnamefont
  {Tom\`as~Valls}},\ }\href {https://doi.org/10.1103/PhysRevD.108.014025}
  {\bibfield  {journal} {\bibinfo  {journal} {Phys. Rev. D}\ }\textbf {\bibinfo
  {volume} {108}},\ \bibinfo {pages} {014025} (\bibinfo {year}
  {2023})}\BibitemShut {NoStop}%
\bibitem [{\citenamefont {Patel}\ and\ \citenamefont
  {Vinodkumar}(2009)}]{Hadron_1}%
  \BibitemOpen
  \bibfield  {author} {\bibinfo {author} {\bibfnamefont {B.}~\bibnamefont
  {Patel}}\ and\ \bibinfo {author} {\bibfnamefont {P.~C.}\ \bibnamefont
  {Vinodkumar}},\ }\href {https://doi.org/10.1088/0954-3899/36/3/035003}
  {\bibfield  {journal} {\bibinfo  {journal} {J. Phys. G: Nucl. Part. Phys.}\
  }\textbf {\bibinfo {volume} {36}},\ \bibinfo {pages} {035003} (\bibinfo
  {year} {2009})}\BibitemShut {NoStop}%
\bibitem [{\citenamefont {Sebastian}\ \emph {et~al.}(2023)\citenamefont
  {Sebastian}, \citenamefont {Jamal},\ and\ \citenamefont {Haque}}]{Hadron_2}%
  \BibitemOpen
  \bibfield  {author} {\bibinfo {author} {\bibfnamefont {J.}~\bibnamefont
  {Sebastian}}, \bibinfo {author} {\bibfnamefont {M.~Y.}\ \bibnamefont
  {Jamal}},\ and\ \bibinfo {author} {\bibfnamefont {N.}~\bibnamefont {Haque}},\
  }\href {https://doi.org/10.1103/PhysRevD.107.054040} {\bibfield  {journal}
  {\bibinfo  {journal} {Phys. Rev. D}\ }\textbf {\bibinfo {volume} {107}},\
  \bibinfo {pages} {054040} (\bibinfo {year} {2023})}\BibitemShut {NoStop}%
\bibitem [{\citenamefont {Brambilla}\ \emph {et~al.}(2005)\citenamefont
  {Brambilla}, \citenamefont {Pineda}, \citenamefont {Soto},\ and\
  \citenamefont {Vairo}}]{Hadron_3}%
  \BibitemOpen
  \bibfield  {author} {\bibinfo {author} {\bibfnamefont {N.}~\bibnamefont
  {Brambilla}}, \bibinfo {author} {\bibfnamefont {A.}~\bibnamefont {Pineda}},
  \bibinfo {author} {\bibfnamefont {J.}~\bibnamefont {Soto}},\ and\ \bibinfo
  {author} {\bibfnamefont {A.}~\bibnamefont {Vairo}},\ }\href
  {https://doi.org/10.1103/RevModPhys.77.1423} {\bibfield  {journal} {\bibinfo
  {journal} {Rev. Mod. Phys.}\ }\textbf {\bibinfo {volume} {77}},\ \bibinfo
  {pages} {1423} (\bibinfo {year} {2005})}\BibitemShut {NoStop}%
\bibitem [{\citenamefont {Purohit}\ \emph {et~al.}(2022)\citenamefont
  {Purohit}, \citenamefont {Jakhad},\ and\ \citenamefont {Rai}}]{purohit2022}%
  \BibitemOpen
  \bibfield  {author} {\bibinfo {author} {\bibfnamefont {K.~R.}\ \bibnamefont
  {Purohit}}, \bibinfo {author} {\bibfnamefont {P.}~\bibnamefont {Jakhad}},\
  and\ \bibinfo {author} {\bibfnamefont {A.~K.}\ \bibnamefont {Rai}},\ }\href
  {https://doi.org/10.1088/1402-4896/ac5bc2} {\bibfield  {journal} {\bibinfo
  {journal} {Phys. Scr.}\ }\textbf {\bibinfo {volume} {97}},\ \bibinfo {pages}
  {044002} (\bibinfo {year} {2022})}\BibitemShut {NoStop}%
\bibitem [{\citenamefont {Soni}\ \emph {et~al.}(2018)\citenamefont {Soni},
  \citenamefont {Joshi}, \citenamefont {Shah}, \citenamefont {Chauhan},\ and\
  \citenamefont {Randya}}]{soni2018}%
  \BibitemOpen
  \bibfield  {author} {\bibinfo {author} {\bibfnamefont {N.~R.}\ \bibnamefont
  {Soni}}, \bibinfo {author} {\bibfnamefont {B.~R.}\ \bibnamefont {Joshi}},
  \bibinfo {author} {\bibfnamefont {R.~P.}\ \bibnamefont {Shah}}, \bibinfo
  {author} {\bibfnamefont {H.~R.}\ \bibnamefont {Chauhan}},\ and\ \bibinfo
  {author} {\bibfnamefont {J.~N.}\ \bibnamefont {Randya}},\ }\href
  {https://doi.org/10.1140/epjc/s10052-018-6068-6} {\bibfield  {journal}
  {\bibinfo  {journal} {Eur. Phys. J. C}\ }\textbf {\bibinfo {volume} {78}},\
  \bibinfo {pages} {592} (\bibinfo {year} {2018})}\BibitemShut {NoStop}%
\bibitem [{\citenamefont {Bukor}\ and\ \citenamefont
  {Tekel}(2023)}]{bukor2023}%
  \BibitemOpen
  \bibfield  {author} {\bibinfo {author} {\bibfnamefont {B.}~\bibnamefont
  {Bukor}}\ and\ \bibinfo {author} {\bibfnamefont {J.}~\bibnamefont {Tekel}},\
  }\href {https://doi.org/10.1140/epjp/s13360-023-04049-3} {\bibfield
  {journal} {\bibinfo  {journal} {Eur. Phys. J. Plus}\ }\textbf {\bibinfo
  {volume} {138}},\ \bibinfo {pages} {499} (\bibinfo {year}
  {2023})}\BibitemShut {NoStop}%
\bibitem [{\citenamefont {Li}\ and\ \citenamefont {Chao}(2009)}]{Li2009}%
  \BibitemOpen
  \bibfield  {author} {\bibinfo {author} {\bibfnamefont {B.-Q.}\ \bibnamefont
  {Li}}\ and\ \bibinfo {author} {\bibfnamefont {K.-T.}\ \bibnamefont {Chao}},\
  }\href {https://doi.org/10.1103/PhysRevD.79.094004} {\bibfield  {journal}
  {\bibinfo  {journal} {Phys. Rev. D}\ }\textbf {\bibinfo {volume} {79}},\
  \bibinfo {pages} {094004} (\bibinfo {year} {2009})}\BibitemShut {NoStop}%
\bibitem [{\citenamefont {Ni}\ \emph {et~al.}(2022)\citenamefont {Ni},
  \citenamefont {Li},\ and\ \citenamefont {Zhong}}]{Ni2022}%
  \BibitemOpen
  \bibfield  {author} {\bibinfo {author} {\bibfnamefont {R.-H.}\ \bibnamefont
  {Ni}}, \bibinfo {author} {\bibfnamefont {Q.}~\bibnamefont {Li}},\ and\
  \bibinfo {author} {\bibfnamefont {X.-H.}\ \bibnamefont {Zhong}},\ }\href
  {https://doi.org/10.1103/PhysRevD.105.056006} {\bibfield  {journal} {\bibinfo
   {journal} {Phys. Rev. D}\ }\textbf {\bibinfo {volume} {105}},\ \bibinfo
  {pages} {056006} (\bibinfo {year} {2022})}\BibitemShut {NoStop}%
\bibitem [{\citenamefont {Molina}\ \emph {et~al.}(2017)\citenamefont {Molina},
  \citenamefont {De~Sanctis},\ and\ \citenamefont
  {Fern\'andez-Ram\'{\i}rez}}]{Molina2027}%
  \BibitemOpen
  \bibfield  {author} {\bibinfo {author} {\bibfnamefont {D.}~\bibnamefont
  {Molina}}, \bibinfo {author} {\bibfnamefont {M.}~\bibnamefont {De~Sanctis}},\
  and\ \bibinfo {author} {\bibfnamefont {C.}~\bibnamefont
  {Fern\'andez-Ram\'{\i}rez}},\ }\href
  {https://doi.org/10.1103/PhysRevD.95.094021} {\bibfield  {journal} {\bibinfo
  {journal} {Phys. Rev. D}\ }\textbf {\bibinfo {volume} {95}},\ \bibinfo
  {pages} {094021} (\bibinfo {year} {2017})}\BibitemShut {NoStop}%
\bibitem [{\citenamefont {Deng}\ \emph {et~al.}(2024)\citenamefont {Deng},
  \citenamefont {Ni}, \citenamefont {Li},\ and\ \citenamefont
  {Zhong}}]{Qian2024}%
  \BibitemOpen
  \bibfield  {author} {\bibinfo {author} {\bibfnamefont {Q.}~\bibnamefont
  {Deng}}, \bibinfo {author} {\bibfnamefont {R.-H.}\ \bibnamefont {Ni}},
  \bibinfo {author} {\bibfnamefont {Q.}~\bibnamefont {Li}},\ and\ \bibinfo
  {author} {\bibfnamefont {X.-H.}\ \bibnamefont {Zhong}},\ }\href
  {https://doi.org/10.1103/PhysRevD.110.056034} {\bibfield  {journal} {\bibinfo
   {journal} {Phys. Rev. D}\ }\textbf {\bibinfo {volume} {110}},\ \bibinfo
  {pages} {056034} (\bibinfo {year} {2024})}\BibitemShut {NoStop}%
\bibitem [{\citenamefont {Eichten}\ and\ \citenamefont
  {Quigg}(1995)}]{Quark_1}%
  \BibitemOpen
  \bibfield  {author} {\bibinfo {author} {\bibfnamefont {E.~J.}\ \bibnamefont
  {Eichten}}\ and\ \bibinfo {author} {\bibfnamefont {C.}~\bibnamefont
  {Quigg}},\ }\href {https://doi.org/10.1103/PhysRevD.52.1726} {\bibfield
  {journal} {\bibinfo  {journal} {Phys. Rev. D}\ }\textbf {\bibinfo {volume}
  {52}},\ \bibinfo {pages} {1726} (\bibinfo {year} {1995})}\BibitemShut
  {NoStop}%
\bibitem [{\citenamefont {Kawanai}\ and\ \citenamefont
  {Sasaki}(2014)}]{Quark_2}%
  \BibitemOpen
  \bibfield  {author} {\bibinfo {author} {\bibfnamefont {T.}~\bibnamefont
  {Kawanai}}\ and\ \bibinfo {author} {\bibfnamefont {S.}~\bibnamefont
  {Sasaki}},\ }\href {https://doi.org/10.1103/PhysRevD.89.054507} {\bibfield
  {journal} {\bibinfo  {journal} {Phys. Rev. D}\ }\textbf {\bibinfo {volume}
  {89}},\ \bibinfo {pages} {054507} (\bibinfo {year} {2014})}\BibitemShut
  {NoStop}%
\bibitem [{\citenamefont {Godfrey}\ and\ \citenamefont
  {Isgur}(1985)}]{Quark_3}%
  \BibitemOpen
  \bibfield  {author} {\bibinfo {author} {\bibfnamefont {S.}~\bibnamefont
  {Godfrey}}\ and\ \bibinfo {author} {\bibfnamefont {N.}~\bibnamefont
  {Isgur}},\ }\href {https://doi.org/10.1103/PhysRevD.32.189} {\bibfield
  {journal} {\bibinfo  {journal} {Phys. Rev. D}\ }\textbf {\bibinfo {volume}
  {32}},\ \bibinfo {pages} {189} (\bibinfo {year} {1985})}\BibitemShut
  {NoStop}%
\bibitem [{\citenamefont {Quigg}\ and\ \citenamefont {Rosner}(1979)}]{Quark_4}%
  \BibitemOpen
  \bibfield  {author} {\bibinfo {author} {\bibfnamefont {C.}~\bibnamefont
  {Quigg}}\ and\ \bibinfo {author} {\bibfnamefont {J.~L.}\ \bibnamefont
  {Rosner}},\ }\href
  {https://doi.org/https://doi.org/10.1016/0370-1573(79)90095-4} {\bibfield
  {journal} {\bibinfo  {journal} {Phys. Rep.}\ }\textbf {\bibinfo {volume}
  {56}},\ \bibinfo {pages} {167} (\bibinfo {year} {1979})}\BibitemShut
  {NoStop}%
\bibitem [{\citenamefont {Brambilla}\ \emph {et~al.}(2011)\citenamefont
  {Brambilla}, \citenamefont {Eidelman}, \citenamefont {Heltsley},
  \citenamefont {Vogt}, \citenamefont {Bodwin}, \citenamefont {Eichten},
  \citenamefont {Frawley}, \citenamefont {Meyer}, \citenamefont {Mitchell},
  \citenamefont {Papadimitriou}, \citenamefont {Petreczky}, \citenamefont
  {Petrov}, \citenamefont {Robbe}, \citenamefont {Vairo}, \citenamefont
  {Andronic}, \citenamefont {Arnaldi}, \citenamefont {Artoisenet},
  \citenamefont {Bali}, \citenamefont {Bertolin}, \citenamefont {Bettoni},
  \citenamefont {Brodzicka}, \citenamefont {Bruno}, \citenamefont {Caldwell},
  \citenamefont {Catmore}, \citenamefont {Chang}, \citenamefont {Chao},
  \citenamefont {Chudakov}, \citenamefont {Cortese}, \citenamefont {Crochet},
  \citenamefont {Drutskoy}, \citenamefont {Ellwanger}, \citenamefont
  {Faccioli}, \citenamefont {Gabareen~Mokhtar}, \citenamefont {Garcia~i Tormo},
  \citenamefont {Hanhart}, \citenamefont {Harris}, \citenamefont {Kaplan},
  \citenamefont {Klein}, \citenamefont {Kowalski}, \citenamefont {Lansberg},
  \citenamefont {Levichev}, \citenamefont {Lombardo}, \citenamefont
  {Lourenço}, \citenamefont {Maltoni}, \citenamefont {Mocsy}, \citenamefont
  {Mussa}, \citenamefont {Navarra}, \citenamefont {Negrini}, \citenamefont
  {Nielsen}, \citenamefont {Olsen}, \citenamefont {Pakhlov}, \citenamefont
  {Pakhlova}, \citenamefont {Peters}, \citenamefont {Polosa}, \citenamefont
  {Qian}, \citenamefont {Qiu}, \citenamefont {Rong}, \citenamefont
  {Sanchis-Lozano}, \citenamefont {Scomparin}, \citenamefont {Senger},
  \citenamefont {Simon}, \citenamefont {Stracka}, \citenamefont {Sumino},
  \citenamefont {Voloshin}, \citenamefont {Weiss}, \citenamefont {Wöhri},\
  and\ \citenamefont {Yuan}}]{Quark_4_1}%
  \BibitemOpen
  \bibfield  {author} {\bibinfo {author} {\bibfnamefont {N.}~\bibnamefont
  {Brambilla}}, \bibinfo {author} {\bibfnamefont {S.}~\bibnamefont {Eidelman}},
  \bibinfo {author} {\bibfnamefont {B.~K.}\ \bibnamefont {Heltsley}}, \bibinfo
  {author} {\bibfnamefont {R.}~\bibnamefont {Vogt}}, \bibinfo {author}
  {\bibfnamefont {G.~T.}\ \bibnamefont {Bodwin}}, \bibinfo {author}
  {\bibfnamefont {E.}~\bibnamefont {Eichten}}, \bibinfo {author} {\bibfnamefont
  {A.~D.}\ \bibnamefont {Frawley}}, \bibinfo {author} {\bibfnamefont {A.~B.}\
  \bibnamefont {Meyer}}, \bibinfo {author} {\bibfnamefont {R.~E.}\ \bibnamefont
  {Mitchell}}, \bibinfo {author} {\bibfnamefont {V.}~\bibnamefont
  {Papadimitriou}}, \bibinfo {author} {\bibfnamefont {P.}~\bibnamefont
  {Petreczky}}, \bibinfo {author} {\bibfnamefont {A.~A.}\ \bibnamefont
  {Petrov}}, \bibinfo {author} {\bibfnamefont {P.}~\bibnamefont {Robbe}},
  \bibinfo {author} {\bibfnamefont {A.}~\bibnamefont {Vairo}}, \bibinfo
  {author} {\bibfnamefont {A.}~\bibnamefont {Andronic}}, \bibinfo {author}
  {\bibfnamefont {R.}~\bibnamefont {Arnaldi}}, \bibinfo {author} {\bibfnamefont
  {P.}~\bibnamefont {Artoisenet}}, \bibinfo {author} {\bibfnamefont
  {G.}~\bibnamefont {Bali}}, \bibinfo {author} {\bibfnamefont {A.}~\bibnamefont
  {Bertolin}}, \bibinfo {author} {\bibfnamefont {D.}~\bibnamefont {Bettoni}},
  \bibinfo {author} {\bibfnamefont {J.}~\bibnamefont {Brodzicka}}, \bibinfo
  {author} {\bibfnamefont {G.~E.}\ \bibnamefont {Bruno}}, \bibinfo {author}
  {\bibfnamefont {A.}~\bibnamefont {Caldwell}}, \bibinfo {author}
  {\bibfnamefont {J.}~\bibnamefont {Catmore}}, \bibinfo {author} {\bibfnamefont
  {C.-H.}\ \bibnamefont {Chang}}, \bibinfo {author} {\bibfnamefont {K.-T.}\
  \bibnamefont {Chao}}, \bibinfo {author} {\bibfnamefont {E.}~\bibnamefont
  {Chudakov}}, \bibinfo {author} {\bibfnamefont {P.}~\bibnamefont {Cortese}},
  \bibinfo {author} {\bibfnamefont {P.}~\bibnamefont {Crochet}}, \bibinfo
  {author} {\bibfnamefont {A.}~\bibnamefont {Drutskoy}}, \bibinfo {author}
  {\bibfnamefont {U.}~\bibnamefont {Ellwanger}}, \bibinfo {author}
  {\bibfnamefont {P.}~\bibnamefont {Faccioli}}, \bibinfo {author}
  {\bibfnamefont {A.}~\bibnamefont {Gabareen~Mokhtar}}, \bibinfo {author}
  {\bibfnamefont {X.}~\bibnamefont {Garcia~i Tormo}}, \bibinfo {author}
  {\bibfnamefont {C.}~\bibnamefont {Hanhart}}, \bibinfo {author} {\bibfnamefont
  {F.~A.}\ \bibnamefont {Harris}}, \bibinfo {author} {\bibfnamefont {D.~M.}\
  \bibnamefont {Kaplan}}, \bibinfo {author} {\bibfnamefont {S.~R.}\
  \bibnamefont {Klein}}, \bibinfo {author} {\bibfnamefont {H.}~\bibnamefont
  {Kowalski}}, \bibinfo {author} {\bibfnamefont {J.-P.}\ \bibnamefont
  {Lansberg}}, \bibinfo {author} {\bibfnamefont {E.}~\bibnamefont {Levichev}},
  \bibinfo {author} {\bibfnamefont {V.}~\bibnamefont {Lombardo}}, \bibinfo
  {author} {\bibfnamefont {C.}~\bibnamefont {Lourenço}}, \bibinfo {author}
  {\bibfnamefont {F.}~\bibnamefont {Maltoni}}, \bibinfo {author} {\bibfnamefont
  {A.}~\bibnamefont {Mocsy}}, \bibinfo {author} {\bibfnamefont
  {R.}~\bibnamefont {Mussa}}, \bibinfo {author} {\bibfnamefont {F.~S.}\
  \bibnamefont {Navarra}}, \bibinfo {author} {\bibfnamefont {M.}~\bibnamefont
  {Negrini}}, \bibinfo {author} {\bibfnamefont {M.}~\bibnamefont {Nielsen}},
  \bibinfo {author} {\bibfnamefont {S.~L.}\ \bibnamefont {Olsen}}, \bibinfo
  {author} {\bibfnamefont {P.}~\bibnamefont {Pakhlov}}, \bibinfo {author}
  {\bibfnamefont {G.}~\bibnamefont {Pakhlova}}, \bibinfo {author}
  {\bibfnamefont {K.}~\bibnamefont {Peters}}, \bibinfo {author} {\bibfnamefont
  {A.~D.}\ \bibnamefont {Polosa}}, \bibinfo {author} {\bibfnamefont
  {W.}~\bibnamefont {Qian}}, \bibinfo {author} {\bibfnamefont {J.-W.}\
  \bibnamefont {Qiu}}, \bibinfo {author} {\bibfnamefont {G.}~\bibnamefont
  {Rong}}, \bibinfo {author} {\bibfnamefont {M.~A.}\ \bibnamefont
  {Sanchis-Lozano}}, \bibinfo {author} {\bibfnamefont {E.}~\bibnamefont
  {Scomparin}}, \bibinfo {author} {\bibfnamefont {P.}~\bibnamefont {Senger}},
  \bibinfo {author} {\bibfnamefont {F.}~\bibnamefont {Simon}}, \bibinfo
  {author} {\bibfnamefont {S.}~\bibnamefont {Stracka}}, \bibinfo {author}
  {\bibfnamefont {Y.}~\bibnamefont {Sumino}}, \bibinfo {author} {\bibfnamefont
  {M.}~\bibnamefont {Voloshin}}, \bibinfo {author} {\bibfnamefont
  {C.}~\bibnamefont {Weiss}}, \bibinfo {author} {\bibfnamefont {H.~K.}\
  \bibnamefont {Wöhri}},\ and\ \bibinfo {author} {\bibfnamefont {C.-Z.}\
  \bibnamefont {Yuan}},\ }\href
  {https://doi.org/10.1140/epjc/s10052-010-1534-9} {\bibfield  {journal}
  {\bibinfo  {journal} {Eur. Phys. J. C}\ }\textbf {\bibinfo {volume} {71}},\
  \bibinfo {pages} {1534} (\bibinfo {year} {2011})}\BibitemShut {NoStop}%
\bibitem [{\citenamefont {Eichten}\ \emph {et~al.}(1978)\citenamefont
  {Eichten}, \citenamefont {Gottfried}, \citenamefont {Kinoshita},
  \citenamefont {Lane},\ and\ \citenamefont {Yan}}]{Quark_5}%
  \BibitemOpen
  \bibfield  {author} {\bibinfo {author} {\bibfnamefont {E.}~\bibnamefont
  {Eichten}}, \bibinfo {author} {\bibfnamefont {K.}~\bibnamefont {Gottfried}},
  \bibinfo {author} {\bibfnamefont {T.}~\bibnamefont {Kinoshita}}, \bibinfo
  {author} {\bibfnamefont {K.~D.}\ \bibnamefont {Lane}},\ and\ \bibinfo
  {author} {\bibfnamefont {T.~M.}\ \bibnamefont {Yan}},\ }\href
  {https://doi.org/10.1103/PhysRevD.17.3090} {\bibfield  {journal} {\bibinfo
  {journal} {Phys. Rev. D}\ }\textbf {\bibinfo {volume} {17}},\ \bibinfo
  {pages} {3090} (\bibinfo {year} {1978})}\BibitemShut {NoStop}%
\bibitem [{\citenamefont {Sang}\ \emph {et~al.}(2023)\citenamefont {Sang},
  \citenamefont {Yang},\ and\ \citenamefont {Zhang}}]{Sang2023}%
  \BibitemOpen
  \bibfield  {author} {\bibinfo {author} {\bibfnamefont {W.-L.}\ \bibnamefont
  {Sang}}, \bibinfo {author} {\bibfnamefont {D.-S.}\ \bibnamefont {Yang}},\
  and\ \bibinfo {author} {\bibfnamefont {Y.-D.}\ \bibnamefont {Zhang}},\ }\href
  {https://doi.org/10.1103/PhysRevD.108.014021} {\bibfield  {journal} {\bibinfo
   {journal} {Phys. Rev. D}\ }\textbf {\bibinfo {volume} {108}},\ \bibinfo
  {pages} {014021} (\bibinfo {year} {2023})}\BibitemShut {NoStop}%
\bibitem [{\citenamefont {Particle{\,}Data{\,}Group}\ \emph
  {et~al.}(2022)\citenamefont {Particle{\,}Data{\,}Group}, \citenamefont
  {Workman}, \citenamefont {Burkert},\ and\ \citenamefont {\textit{et
  al.}}}]{workman2022}%
  \BibitemOpen
  \bibfield  {author} {\bibinfo {author} {\bibnamefont
  {Particle{\,}Data{\,}Group}}, \bibinfo {author} {\bibfnamefont {R.~L.}\
  \bibnamefont {Workman}}, \bibinfo {author} {\bibfnamefont {V.~D.}\
  \bibnamefont {Burkert}},\ and\ \bibinfo {author} {\bibnamefont {\textit{et
  al.}}},\ }\href {https://doi.org/10.1093/ptep/ptac097} {\bibfield  {journal}
  {\bibinfo  {journal} {Prog. Theor. Exp. Phys.}\ }\textbf {\bibinfo {volume}
  {2022}},\ \bibinfo {pages} {083C01} (\bibinfo {year} {2022})}\BibitemShut
  {NoStop}%
\bibitem [{\citenamefont {Godfrey}\ and\ \citenamefont
  {Moats}(2015)}]{Godfrey2015}%
  \BibitemOpen
  \bibfield  {author} {\bibinfo {author} {\bibfnamefont {S.}~\bibnamefont
  {Godfrey}}\ and\ \bibinfo {author} {\bibfnamefont {K.}~\bibnamefont
  {Moats}},\ }\href {https://doi.org/10.1103/PhysRevD.92.054034} {\bibfield
  {journal} {\bibinfo  {journal} {Phys. Rev. D}\ }\textbf {\bibinfo {volume}
  {92}},\ \bibinfo {pages} {054034} (\bibinfo {year} {2015})}\BibitemShut
  {NoStop}%
\bibitem [{\citenamefont {Greiner}(2001)}]{Greiner}%
  \BibitemOpen
  \bibfield  {author} {\bibinfo {author} {\bibfnamefont {W.}~\bibnamefont
  {Greiner}},\ }\href {https://doi.org/10.1007/978-3-642-56826-8} {\emph
  {\bibinfo {title} {Quantum Mechanics: An Introduction}}},\ \bibinfo {edition}
  {4th}\ ed.\ (\bibinfo  {publisher} {Springer Berlin Heidelberg},\ \bibinfo
  {year} {2001})\BibitemShut {NoStop}%
\bibitem [{\citenamefont {Nikiforov}\ and\ \citenamefont
  {Uvarov}(1988)}]{Nikiforov}%
  \BibitemOpen
  \bibfield  {author} {\bibinfo {author} {\bibfnamefont {A.~F.}\ \bibnamefont
  {Nikiforov}}\ and\ \bibinfo {author} {\bibfnamefont {V.~B.}\ \bibnamefont
  {Uvarov}},\ }\href {https://doi.org/10.1007/978-1-4757-1595-8} {\emph
  {\bibinfo {title} {Special Functions of Mathematical Physics}}}\ (\bibinfo
  {publisher} {Birkhäuser},\ \bibinfo {address} {Basel},\ \bibinfo {year}
  {1988})\BibitemShut {NoStop}%
\bibitem [{SM()}]{SM}%
  \BibitemOpen
  \href@noop {} {\bibinfo  {journal} {Additional details, theoretical
  derivations, and supplementary analyses can be found in the Supplementary
  Material, which also includes Ref.{\,}[\onlinecite{Greiner, PS4, PS5,
  Pekeris, GTHP1, GTHP2, Nikiforov, Abramowitz, workman2022, purohit2022,
  soni2018, bukor2023, Crater, 2BDE_2, 2BDE_3, Hilger}] to support this study}\
  }\BibitemShut {NoStop}%
\bibitem [{\citenamefont {Fioravanti}\ and\ \citenamefont
  {Gregori}(2020)}]{Fioravanti2020}%
  \BibitemOpen
\bibfield  {journal} {  }\bibfield  {author} {\bibinfo {author} {\bibfnamefont
  {D.}~\bibnamefont {Fioravanti}}\ and\ \bibinfo {author} {\bibfnamefont
  {D.}~\bibnamefont {Gregori}},\ }\href
  {https://doi.org/10.1016/j.physletb.2020.135376} {\bibfield  {journal}
  {\bibinfo  {journal} {Phys. Lett. B}\ }\textbf {\bibinfo {volume} {804}},\
  \bibinfo {pages} {135376} (\bibinfo {year} {2020})}\BibitemShut {NoStop}%
\bibitem [{\citenamefont {Fioravanti}\ and\ \citenamefont
  {Rossi}(2023)}]{Fioravanti2023}%
  \BibitemOpen
  \bibfield  {author} {\bibinfo {author} {\bibfnamefont {D.}~\bibnamefont
  {Fioravanti}}\ and\ \bibinfo {author} {\bibfnamefont {M.}~\bibnamefont
  {Rossi}},\ }\href {https://doi.org/10.1016/j.physletb.2023.137706} {\bibfield
   {journal} {\bibinfo  {journal} {Phys. Lett. B}\ }\textbf {\bibinfo {volume}
  {838}},\ \bibinfo {pages} {137706} (\bibinfo {year} {2023})}\BibitemShut
  {NoStop}%
\bibitem [{\citenamefont {Fioravanti}\ and\ \citenamefont
  {Gregori}(2021)}]{Fioravanti2021}%
  \BibitemOpen
  \bibfield  {author} {\bibinfo {author} {\bibfnamefont {D.}~\bibnamefont
  {Fioravanti}}\ and\ \bibinfo {author} {\bibfnamefont {D.}~\bibnamefont
  {Gregori}},\ }\href@noop {} {\bibinfo {title} {From quasinormal modes to
  exact wkb quantization in ads black holes}} (\bibinfo {year} {2021}),\
  \Eprint {https://arxiv.org/abs/2112.11434} {arXiv:2112.11434 [hep-th]}
  \BibitemShut {NoStop}%
\bibitem [{\citenamefont {Fioravanti}\ \emph {et~al.}(2022)\citenamefont
  {Fioravanti}, \citenamefont {Gregori},\ and\ \citenamefont
  {Shu}}]{Fioravanti2022}%
  \BibitemOpen
  \bibfield  {author} {\bibinfo {author} {\bibfnamefont {D.}~\bibnamefont
  {Fioravanti}}, \bibinfo {author} {\bibfnamefont {D.}~\bibnamefont
  {Gregori}},\ and\ \bibinfo {author} {\bibfnamefont {H.}~\bibnamefont {Shu}},\
  }\href@noop {} {\bibinfo {title} {Exact quantization of scalar quasinormal
  modes in ads black holes via $\mathcal{N}=2$ gauge theories}} (\bibinfo
  {year} {2022}),\ \Eprint {https://arxiv.org/abs/2208.14031} {arXiv:2208.14031
  [hep-th]} \BibitemShut {NoStop}%
\bibitem [{\citenamefont {Feizi}\ \emph {et~al.}(2012)\citenamefont {Feizi},
  \citenamefont {Shojaei},\ and\ \citenamefont {Rajabi}}]{Feizi}%
  \BibitemOpen
  \bibfield  {author} {\bibinfo {author} {\bibfnamefont {H.}~\bibnamefont
  {Feizi}}, \bibinfo {author} {\bibfnamefont {M.~R.}\ \bibnamefont {Shojaei}},\
  and\ \bibinfo {author} {\bibfnamefont {A.~A.}\ \bibnamefont {Rajabi}},\
  }\href {https://doi.org/10.1140/epjp/i2012-12041-y} {\bibfield  {journal}
  {\bibinfo  {journal} {Eur. Phys. J. Plus}\ }\textbf {\bibinfo {volume}
  {127}},\ \bibinfo {pages} {41} (\bibinfo {year} {2012})}\BibitemShut
  {NoStop}%
\bibitem [{\citenamefont {Isgur}\ and\ \citenamefont
  {Paton}(1985)}]{Isgur1985}%
  \BibitemOpen
  \bibfield  {author} {\bibinfo {author} {\bibfnamefont {N.}~\bibnamefont
  {Isgur}}\ and\ \bibinfo {author} {\bibfnamefont {J.}~\bibnamefont {Paton}},\
  }\href {https://doi.org/10.1103/PhysRevD.31.2910} {\bibfield  {journal}
  {\bibinfo  {journal} {Phys. Rev. D}\ }\textbf {\bibinfo {volume} {31}},\
  \bibinfo {pages} {2910} (\bibinfo {year} {1985})}\BibitemShut {NoStop}%
\bibitem [{\citenamefont {Bali}(2001)}]{Bali2001}%
  \BibitemOpen
  \bibfield  {author} {\bibinfo {author} {\bibfnamefont {G.~S.}\ \bibnamefont
  {Bali}},\ }\href {https://doi.org/10.1016/S0370-1573(00)00079-X} {\bibfield
  {journal} {\bibinfo  {journal} {Phys. Rep.}\ }\textbf {\bibinfo {volume}
  {343}},\ \bibinfo {pages} {1} (\bibinfo {year} {2001})}\BibitemShut {NoStop}%
\bibitem [{\citenamefont {Bali}\ \emph {et~al.}(2005)\citenamefont {Bali},
  \citenamefont {Neff}, \citenamefont {D{\"u}ssel}, \citenamefont {Lippert},\
  and\ \citenamefont {Schilling}}]{Bali2005}%
  \BibitemOpen
  \bibfield  {author} {\bibinfo {author} {\bibfnamefont {G.~S.}\ \bibnamefont
  {Bali}}, \bibinfo {author} {\bibfnamefont {H.}~\bibnamefont {Neff}}, \bibinfo
  {author} {\bibfnamefont {T.}~\bibnamefont {D{\"u}ssel}}, \bibinfo {author}
  {\bibfnamefont {T.}~\bibnamefont {Lippert}},\ and\ \bibinfo {author}
  {\bibfnamefont {K.}~\bibnamefont {Schilling}},\ }\href
  {https://doi.org/10.1103/PhysRevD.71.114513} {\bibfield  {journal} {\bibinfo
  {journal} {Phys. Rev. D}\ }\textbf {\bibinfo {volume} {71}},\ \bibinfo
  {pages} {114513} (\bibinfo {year} {2005})}\BibitemShut {NoStop}%
\bibitem [{\citenamefont {Bulava}\ \emph {et~al.}(2019)\citenamefont {Bulava},
  \citenamefont {H{\"o}rz}, \citenamefont {Knechtli}, \citenamefont {Koch},
  \citenamefont {Moir}, \citenamefont {Morningstar},\ and\ \citenamefont
  {Peardon}}]{Bulava2019}%
  \BibitemOpen
  \bibfield  {author} {\bibinfo {author} {\bibfnamefont {J.}~\bibnamefont
  {Bulava}}, \bibinfo {author} {\bibfnamefont {B.}~\bibnamefont {H{\"o}rz}},
  \bibinfo {author} {\bibfnamefont {F.}~\bibnamefont {Knechtli}}, \bibinfo
  {author} {\bibfnamefont {V.}~\bibnamefont {Koch}}, \bibinfo {author}
  {\bibfnamefont {G.}~\bibnamefont {Moir}}, \bibinfo {author} {\bibfnamefont
  {C.}~\bibnamefont {Morningstar}},\ and\ \bibinfo {author} {\bibfnamefont
  {M.}~\bibnamefont {Peardon}},\ }\href
  {https://doi.org/10.1016/j.physletb.2019.05.018} {\bibfield  {journal}
  {\bibinfo  {journal} {Phys. Lett. B}\ }\textbf {\bibinfo {volume} {793}},\
  \bibinfo {pages} {493} (\bibinfo {year} {2019})}\BibitemShut {NoStop}%
\bibitem [{\citenamefont {Bernard}\ \emph {et~al.}(2001)\citenamefont
  {Bernard}, \citenamefont {Burch}, \citenamefont {DeGrand}, \citenamefont
  {DeTar}, \citenamefont {Gottlieb}, \citenamefont {Heller}, \citenamefont
  {Lacock}, \citenamefont {Orginos}, \citenamefont {Sugar},\ and\ \citenamefont
  {Toussaint}}]{Bernard2000}%
  \BibitemOpen
  \bibfield  {author} {\bibinfo {author} {\bibfnamefont {C.}~\bibnamefont
  {Bernard}}, \bibinfo {author} {\bibfnamefont {T.}~\bibnamefont {Burch}},
  \bibinfo {author} {\bibfnamefont {T.}~\bibnamefont {DeGrand}}, \bibinfo
  {author} {\bibfnamefont {C.}~\bibnamefont {DeTar}}, \bibinfo {author}
  {\bibfnamefont {S.}~\bibnamefont {Gottlieb}}, \bibinfo {author}
  {\bibfnamefont {U.}~\bibnamefont {Heller}}, \bibinfo {author} {\bibfnamefont
  {P.}~\bibnamefont {Lacock}}, \bibinfo {author} {\bibfnamefont
  {K.}~\bibnamefont {Orginos}}, \bibinfo {author} {\bibfnamefont
  {R.}~\bibnamefont {Sugar}},\ and\ \bibinfo {author} {\bibfnamefont
  {D.}~\bibnamefont {Toussaint}},\ }\href
  {https://doi.org/https://doi.org/10.1016/S0920-5632(01)00863-5} {\bibfield
  {journal} {\bibinfo  {journal} {Nucl. Phys. B}\ }\textbf {\bibinfo {volume}
  {94}},\ \bibinfo {pages} {546} (\bibinfo {year} {2001})}\BibitemShut
  {NoStop}%
\bibitem [{\citenamefont {Crater}\ \emph {et~al.}(2009)\citenamefont {Crater},
  \citenamefont {Yoon},\ and\ \citenamefont {Wong}}]{Crater}%
  \BibitemOpen
  \bibfield  {author} {\bibinfo {author} {\bibfnamefont {H.~W.}\ \bibnamefont
  {Crater}}, \bibinfo {author} {\bibfnamefont {J.-H.}\ \bibnamefont {Yoon}},\
  and\ \bibinfo {author} {\bibfnamefont {C.-Y.}\ \bibnamefont {Wong}},\ }\href
  {https://doi.org/10.1103/PhysRevD.79.034011} {\bibfield  {journal} {\bibinfo
  {journal} {Phys. Rev. D}\ }\textbf {\bibinfo {volume} {79}},\ \bibinfo
  {pages} {034011} (\bibinfo {year} {2009})}\BibitemShut {NoStop}%
\bibitem [{\citenamefont {Scott}\ \emph {et~al.}(1992)\citenamefont {Scott},
  \citenamefont {Shertzer},\ and\ \citenamefont {Moore}}]{2BDE_2}%
  \BibitemOpen
  \bibfield  {author} {\bibinfo {author} {\bibfnamefont {T.~C.}\ \bibnamefont
  {Scott}}, \bibinfo {author} {\bibfnamefont {J.}~\bibnamefont {Shertzer}},\
  and\ \bibinfo {author} {\bibfnamefont {R.~A.}\ \bibnamefont {Moore}},\ }\href
  {https://doi.org/10.1103/PhysRevA.45.4393} {\bibfield  {journal} {\bibinfo
  {journal} {Phys. Rev. A}\ }\textbf {\bibinfo {volume} {45}},\ \bibinfo
  {pages} {4393} (\bibinfo {year} {1992})}\BibitemShut {NoStop}%
\bibitem [{\citenamefont {Ferreira}(1988)}]{2BDE_3}%
  \BibitemOpen
  \bibfield  {author} {\bibinfo {author} {\bibfnamefont {P.~L.}\ \bibnamefont
  {Ferreira}},\ }\href {https://doi.org/10.1103/PhysRevD.38.2648} {\bibfield
  {journal} {\bibinfo  {journal} {Phys. Rev. D}\ }\textbf {\bibinfo {volume}
  {38}},\ \bibinfo {pages} {2648} (\bibinfo {year} {1988})}\BibitemShut
  {NoStop}%
\bibitem [{\citenamefont {Hilger}\ \emph {et~al.}(2015)\citenamefont {Hilger},
  \citenamefont {Popovici}, \citenamefont {G\'omez-Rocha},\ and\ \citenamefont
  {Krassnigg}}]{Hilger}%
  \BibitemOpen
  \bibfield  {author} {\bibinfo {author} {\bibfnamefont {T.}~\bibnamefont
  {Hilger}}, \bibinfo {author} {\bibfnamefont {C.}~\bibnamefont {Popovici}},
  \bibinfo {author} {\bibfnamefont {M.}~\bibnamefont {G\'omez-Rocha}},\ and\
  \bibinfo {author} {\bibfnamefont {A.}~\bibnamefont {Krassnigg}},\ }\href
  {https://doi.org/10.1103/PhysRevD.91.034013} {\bibfield  {journal} {\bibinfo
  {journal} {Phys. Rev. D}\ }\textbf {\bibinfo {volume} {91}},\ \bibinfo
  {pages} {034013} (\bibinfo {year} {2015})}\BibitemShut {NoStop}%
\bibitem [{\citenamefont {Pekeris}(1934)}]{Pekeris}%
  \BibitemOpen
  \bibfield  {author} {\bibinfo {author} {\bibfnamefont {C.~L.}\ \bibnamefont
  {Pekeris}},\ }\href {https://doi.org/10.1103/PhysRev.45.98} {\bibfield
  {journal} {\bibinfo  {journal} {Phys. Rev.}\ }\textbf {\bibinfo {volume}
  {45}},\ \bibinfo {pages} {98} (\bibinfo {year} {1934})}\BibitemShut {NoStop}%
\bibitem [{\citenamefont {Abramowitz}\ and\ \citenamefont
  {Stegun}(2012)}]{Abramowitz}%
  \BibitemOpen
  \bibfield  {author} {\bibinfo {author} {\bibfnamefont {M.}~\bibnamefont
  {Abramowitz}}\ and\ \bibinfo {author} {\bibfnamefont {I.~A.}\ \bibnamefont
  {Stegun}},\ }\href {https://books.google.de/books?id=KiPCAgAAQBAJ} {\emph
  {\bibinfo {title} {Handbook of Mathematical Functions with Formulas, Graphs,
  and Mathematical Tables}}}\ (\bibinfo  {publisher} {Dover Publications},\
  \bibinfo {address} {New York},\ \bibinfo {year} {2012})\BibitemShut {NoStop}%
\end{thebibliography}%
\end{document}